**Estimating data center water use: Best practices and critical questions**

Eric Masanet[1,2,*]

[1] Bren School of Environmental Science and Management, University of California, Santa Barbara, USA

[2] Department of Mechanical Engineering, University of California, Santa Barbara, USA

*corresponding author: emasanet@ucsb.edu

# Abstract

Concerns are growing about the water use of data centers. To quantify the scales, drivers, and possible trajectories of this water use, analytical estimation is required. However, current estimation methods and results vary widely due, in part, to a lack of established best practices. This variance is illustrated with a U.S. case study demonstrating a roughly 24-fold difference in possible outcomes based on observed U.S. study design variations. To address these challenges, this review synthesizes 14 best practices for advancing the science and utility of data center water use estimation. These proposed best practices can also be used as critical questions for stakeholders when evaluating the quality of any study. Using a structured rubric, it further assesses the literature to date (31 studies) against these best practices. The assessment reveals important opportunities for improvement and standardization in future study scopes, clarity, methods, and scientific knowledge building utility. It further finds that, while many studies provide reasonable transparency, scope definitions, replicability, and uncertainty treatment, many fall short in scientific justification, data representativeness, transparency on hydropower assumptions, terminological consistency, and meaningful discussions of limitations and future work. Recommendations are proposed for analysts, stakeholders, and policymakers to leverage these findings to help build a vibrant data center water use research community. Finally, this review can also be used as a basic primer for understanding data center cooling configurations and water use among analysts, policymakers, the media, and the public.

# 1. Introduction

Since the artificial intelligence (AI) “boom” began in the early 2020s (1), the electricity consumption of the world’s data centers has been rising rapidly (2,3). Under current projections, this trend may continue into the next decade with data centers comprising 3% of global electricity demand by 2030, up from 1.7% in 2025 (4).

Data center energy studies have a long history in the literature, extending back to the first major studies in 2007-2008 (5–9). However, their recent publication pace has increased dramatically, reflecting growing concerns about the climate and energy infrastructure implications of rapid data center growth. For example, recent reviews have identified well over 100 data center energy studies published since 2007, with nearly one-half of these published since 2024 alone (10,11).

In parallel, concerns are also growing about the water resource implications of this rapid growth. These concerns are rooted in the fact that many data centers have traditionally relied on evaporative methods for data center cooling, primarily via cooling towers, adiabatic coolers, or their combinations (12–14). In the United States, which is home to the world’s largest data center fleet (4), many of these data centers are also located in water-stressed regions (15).

Concerns also extend to the water use associated with generating the electricity needed to power data centers, both now and in the future (16,17). Such “indirect” water consumption can be associated with thermoelectric power plant cooling, hydroelectric reservoir evaporation, and the supply chains needed for power plant fuels and technologies (18–21).

Unlike energy studies, data center water use studies are a more recent phenomenon. Since the first elaboration of the data center water “footprint” in 2015 (22), this review identified 31 subsequent macro-level studies that have estimated the past, present, and/or future water use of data centers in different regions (Supplementary Information, Table S1). Ninety percent of these studies have been published since 2023, coinciding with the current “AI boom,” but also with rapidly growing attention to water use among the media and policymakers (Section S2).

For example, global English language news stories mentioning data centers and water use tripled from around 8,000 in 2021 to more than 24,000 in 2025, while more than 50,000 appeared in the first half of 2026 alone (Figure S1) (23). In the United States, where data center electricity growth is particularly pronounced (3,4,24), more than 120 local, state, and federal policies have been introduced related to data center water use since January 1st, 2025 (25). This exploding media and regulatory attention has likely been both in reaction to, and a strong motivator for, the water use estimates published to date. This “push-pull” dynamic between researchers and stakeholders is likely to continue, especially due to ongoing community concerns around data center water use (26).

## 1.1 Why are estimates needed?

Water use estimation is important for several reasons. First, many large operators do not disclose their water use, especially at the level of individual data centers (27,28). Therefore, analysts must often make estimates to fill data gaps. However, data availability may improve as governments require disclosure, for example, in the European Union (29). Second, even if large operator data gaps are eventually addressed, there are many thousands of enterprise data centers run by non-tech companies globally that may never disclose their water use. For example, Lei et al. (30) estimated that there may be around 13,000 midsize enterprise data centers in the United States alone with onsite water consumption intensities similar to larger data centers.

Third, even when reported data are available, the models used for estimation can reveal important underlying drivers of that water use, including the effects of climate variables, cooling technologies, cooling strategies such as “free cooling,” and power sources (12,13,31). Fourth, these models will undoubtedly remain important for both projecting future water use and conducting “what if” evaluations of how various technology, operations, and policy interventions might reduce that future water use (12).

## 1.2 The case for best practices

Despite the recent surge of studies, data center water use estimation remains a nascent domain without a well-established research community. Such communities typically share datasets, compare methods, collaborate, develop standards, and enable collective learning; in short, they conduct the critical activities needed to advance the science of any domain (32,33). A necessary early step in community building is to establish best practices, which has a dual benefit.

First, best practices provide analysts with common standards for ensuring the rigor, clarity, and intercomparability of studies. Second, they provide stakeholders–whether they be policymakers, the public, the media, or other researchers–with a trusted means of evaluating the quality of any study and its suitability for their needs. These two benefits promote virtuous cycles, wherein stakeholders are empowered with critical questions to rate the utility of studies and researchers respond with continuously improved outputs.

While best practices have been proposed for data center energy analysis (28,34–36), no analogous guidance has been proposed for the data center water analyst. This review aims to address this important need by synthesizing 14 best practices for data center water use estimation related to ensuring clarity, scientific diligence, quality, replicability, and community knowledge building.

Section 2 reviews data center water uses and terminology to provide a common taxonomy for both analysts and their stakeholders. Section 3 proposes best practices by reviewing how water use estimates are calculated, illustrating the relevance and importance of best practices at each analysis step. In Section 4, the 31 compiled water use studies are evaluated against these best practices, revealing specific opportunities for community advancement. Section 5 evaluates emerging topics for future consideration. Section 6 proposes next steps for developing a robust research community.

# 2. Data center water use: components, metrics, and estimation

Figure 1 depicts the three stages of data center water use and the key terms and metrics that are most relevant to analysts when making estimates (2,22,37). These stages have been labeled direct, indirect, and embodied water use according to standard terminology observed across the literature (16,38).

Importantly, when analyzing each stage, analysts must distinguish between four distinct components of water “use:” withdrawals, inputs, consumption, and discharges (39). Withdrawals refer to quantities of water extracted directly from natural resources, such as surface waters or groundwater. Inputs refer to the total quantities of water taken into a facility from a water supply system, which can include local water utilities or direct withdrawals from nearby natural resources. In the case of direct withdrawals, inputs equal withdrawals. In the case of water utilities, treated water can often be lost due to leaks in distribution systems (40), in which case withdrawal quantities would be greater than input quantities. Direct withdrawals also typically require onsite water treatment at the data center (41). Discharges refer to the input quantities returned to local wastewater treatment systems, returned directly to the original water sources, or discharged elsewhere. Consumption is defined as the difference between facility inputs and discharges.

## 2.1 Direct water use

Direct water use occurs at the data center property, where water inputs can be divided into five main categories (Table 1). The presence and scale of each type of water input depend highly on the data center type and configuration. For example, a small server room located within an office building may have none of these inputs dedicated to its operation, while a large data center may require all of these inputs in various data halls and/or cooling system modes.

Water inputs for personnel and sanitation are driven by employees rather than equipment operations. Thus, their quantities are expected to be similar to those at typical office buildings with comparable occupancies (43). Water inputs for dedicated data hall air humidifiers are also generally small at most data centers (12,13). Compared to these two categories, the water inputs associated with adiabatic air cooling and evaporative heat rejection are typically much greater at data centers that employ these two cooling system technologies (12). Therefore, accurate estimation of direct water use requires a thorough understanding of data cooling systems and how they operate, as discussed in the next section. This includes consideration of water use for initial filling of cooling system fluid loops, which is rarely considered by analysts but should be for completeness.

**Table 1:** Five main categories of direct data center water inputs

| Water input category | Description |
|---|---|
| Personnel and sanitation | Water use in kitchens, restrooms, sinks, drinking fountains, and facility sanitation |
| Dedicated data hall air humidification | Water used by conventional data center humidification units, such as steam canisters, infrared humidifiers, or ultrasonic humidifiers that keep data hall air humidity above minimum thresholds to prevent static discharge in IT equipment (42) |
| Evaporative heat rejection | Water that is evaporated in open-loop or closed-loop cooling towers to move heat out of the data center and into the atmosphere |
| Initial filling of fluid loops | Water used to initially fill closed and recirculating IT cooling, primary cooling, and/or heat rejection fluid loops at the data center, depending on their design, and when aqueous fluids are required |
| Adiabatic air cooling | Water used in adiabatic coolers, which inject water into internal or external data center airstreams to reduce airstream temperature |

## 2.11 Data center cooling systems: the basics

The goal of any data center cooling system is to remove the heat generated within IT racks and to expel it outside the data center. This process must occur continuously to keep IT equipment within acceptable operating temperature ranges (44).

Figure 2a depicts three major cooling subsystems that are commonly used by midsize to large data centers to perform this service (3). In practice, these subsystems are often referred to as "loops." A simple way to understand how these loops work together is to "follow the heat" in Figure 2a, starting with the heat generated by the IT equipment (point A).  Similar simplified "follow the heat" schematics of other cooling system design variations are provided in Section S4.

### 2.111 The IT cooling loop

The first loop in Fig. 2a is the IT cooling loop. Its job is to move heat out of the IT racks within the data hall. It does so by introducing a cool fluid into the rack (point B), which absorbs the heat generated by the IT equipment (Point A). When it exits the rack as a warm fluid (point C), it takes this heat with it.

IT cooling loops that use air as the heat transfer fluid are known as "air-cooled" systems, whereas those that use a liquid are known as "liquid-cooled" systems (45,46). Air-cooled systems require fans and blowers for air circulation within the data hall and IT racks, whereas liquid-cooled systems require pumps for liquid circulation. All of these components consume electricity (Table 2) (47).

When liquid-cooled systems are used, the IT cooling loop is "closed loop," given that it contains a fixed amount of fluid that circulates continuously. Typically, this fluid is either highly-treated water or a water/glycol mixture, but water-free dielectric liquids or two-phase fluids can also be used (46,48). Liquid-cooled IT cooling loops are sometimes referred to as the "technology cooling loop" or the "secondary loop" (49). Liquid-cooling solutions include direct-to-chip and immersion cooling systems, the latter of which immerses servers in water-free dielectric fluids (50).

Lastly, IT racks can also be cooled using "hybrid" combinations of air and liquid cooling. One common variation is the use of direct-to-chip liquid cooling for high-power server components (e.g., CPUs and GPUs) alongside air cooling for the remaining components and devices within the rack; another is the use of rear-door heat exchangers, which further cool the rack exit air using a primary coolant loop mounted on the rear side of the rack (45,51).

### 2.112 The primary cooling loop

The second loop in Figure 2a is the primary cooling loop. This loop is also sometimes referred to as the "primary loop" or the "facility loop" (49). Its job is to sufficiently cool down the warm fluid that exits the racks (point C) so that it can be recirculated back to the rack (point B). This process requires a heat exchanger (52) and a primary cooling liquid. The primary cooling liquid enters the heat exchanger at a cool temperature (point D), allowing it to absorb heat from the warm fluid (point E) that exited the racks. The primary cooling liquid exits the heat exchanger as a warm fluid (point F), taking the heat from the IT cooling fluid with it. The IT cooling fluid is now cool and returns to the rack (point B), where it can once again enter as a cool inlet fluid for another heat removal cycle.

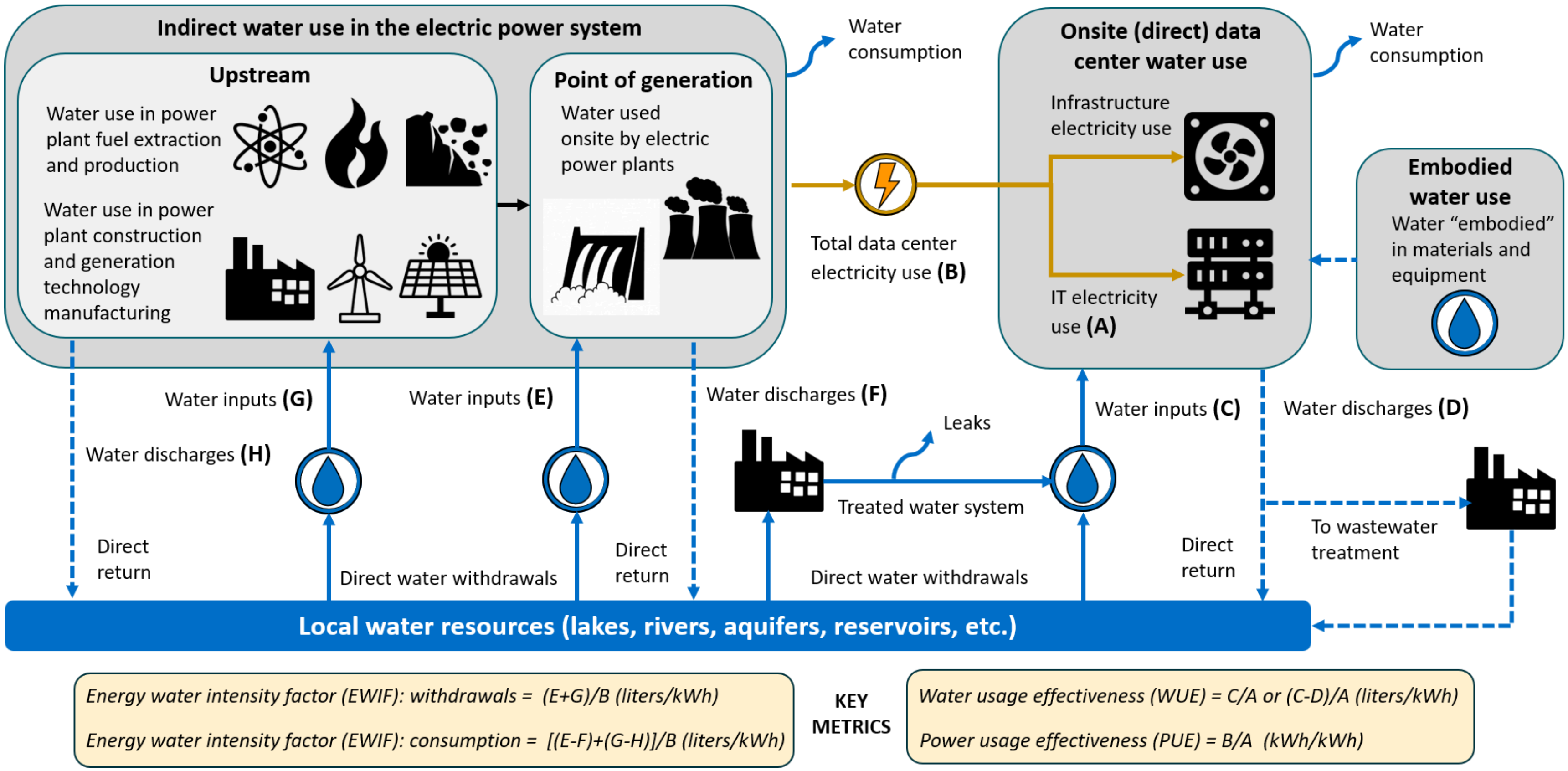

**Figure 1:** Stages and components of data center water use and related metrics**.** Note: icons courtesy of https://uxwing.com/.

At many data centers, the primary cooling fluid is chilled water or a chilled water/glycol mixture (48), both of which are made cool by a mechanical water "chiller" (53). The chiller is often located in a mechanical equipment room that is separated from the data hall for climate control, leak isolation, noise, and maintenance reasons. The chiller takes in the warm primary cooling fluid (point F) and makes it cold again so that it can return to the heat exchanger (point D) for another heat removal cycle. To do this, the chiller employs a mechanical vapor compression (MVC) refrigeration cycle, which requires electricity to power the compressor (47,53). The primary cooling loop also requires pumps for fluid movement, which also consume electricity (47,54). While chillers are used in many data centers globally, compressor-based direct expansion systems are sometimes used in small data centers to provide cool primary cooling fluid (point D) (Table 2).

The primary cooling loop is "closed loop," given that it contains a fixed amount of liquid that circulates continuously. For air-cooled IT racks, an air-to-liquid heat exchanger is required (Fig 2a, point E), such as a computer room air handler (CRAH) (48). For liquid-cooled IT racks, a liquid-to-liquid heat exchanger is required (see, e.g., Fig S4, point E), such as a cooling distribution unit (CDU) (49).

### 2.113 The outdoor heat rejection loop

The third loop in Figure 2a is the outdoor heat rejection loop, which can also be referred to as the "building heat rejection loop." Its job is to take the heat contained in the warm primary cooling fluid (point F) and permanently expel that heat into the outside environment. The medium into which this heat is expelled is called the exterior "heat sink." At most data centers, the heat sink is outdoor ambient air. However, data centers can also use surface waters, the Earth's subsurface, or heat recovery/export systems as the heat sink. In the last option, the heat moved out of the data center is transferred to another end use, such as a nearby building for space heating (55).

When the outdoor heat rejection loop is associated with a chiller, as in Figure 2a, it is typically referred to as the "condenser water loop" (49). This is because the chiller's MVC refrigeration cycle moves heat from its evaporator to its condenser (point G) (53). The former is the "cold side," which cools the primary cooling fluid, while the latter is the "hot side," which contains the heat to be rejected from the building.

The outdoor heat rejection loop requires a heat exchanger, a heat rejection device, and a heat sink (Table 2). In the heat exchanger, the heat from the chiller's condenser is transferred (point G) to the heat rejection fluid, which enters as a cool fluid (point H) and exits as a warm fluid (point I), taking the heat with it. Typically, treated water is used as the heat rejection circuit fluid, but glycol/water mixtures or other liquid coolants can be used, depending on the heat rejection device.

The job of the heat rejection device is twofold. First, it must sufficiently cool down the warm fluid that exits the condenser heat exchanger (point I) so that it can be recirculated back as a cool fluid (point H) for another cooling cycle. Second, it must reject that heat to the heat sink outside the building, permanently and completely removing it from the data center.

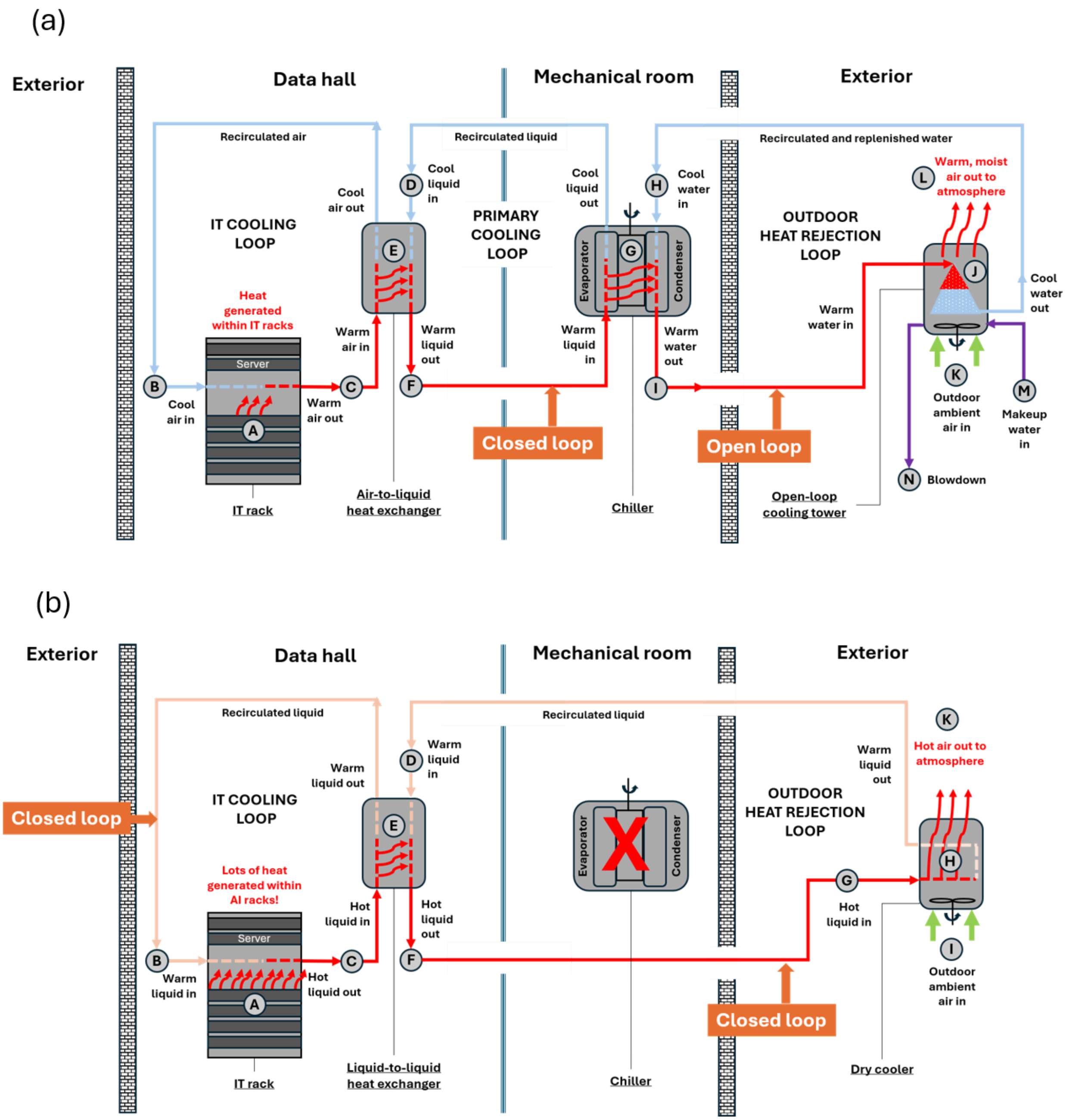

**Figure 2:** Follow the heat: simplified cooling loop configurations in a data center with (a) mechanical chilling and open-loop cooling tower outdoor heat rejection running at cool rack inlet temperatures and (b) no mechanical chilling and dry cooler outdoor heat rejection running at warm rack inlet temperatures. Note: refer to the text for an explanation of lettered symbols (A) through (P).

Figure 2a depicts an open-loop cooling tower, which is a heat rejection device used by many conventional data centers (56). Cooling towers are located outside the data center, typically on its roof or on the building perimeter grounds. The warm heat rejection fluid enters the top of the cooling tower (point J), where it is sprayed and falls downward. Ambient air acts as the heat sink and is drawn (point K) into and up through the tower, cooling and evaporating portions of the falling water along the way. The air exits the cooling tower more warm and humid than it entered, taking the heat and some vapor from the falling water with it (point L). Cool water exits the bottom of the cooling tower, where it returns to the condenser heat exchanger (point H) for another heat removal cycle.

### 2.114 Overall system and design variations

All three loops in Figure 2a work synergistically to continuously transfer the heat generated at the IT racks (point A) to the exterior heat sink (point L). In practice, different combinations of loops, fluids, heat exchangers, heat rejection devices, and heat sinks exist across data centers (see Table 2 and Section S3). Design choices are primarily driven by the local climate patterns, necessary heat removal quantities, available space, available heat sinks, and energy and water availability and costs (56).

For example, Figure 2b summarizes a high rack temperature “chillerless” design that involves only two fluid loops (57). It features closed-loop liquid cooling in the IT cooling loop that is directly coupled to a closed-loop, “dry” heat rejection circuit. In this design, heat is removed from the IT equipment using liquid-cooled cold plates attached directly to server components, whereas a “dry” cooler is used as the heat rejection device. A “dry” cooler works by forcing outside air (Fig 2b, point G) over a heat exchanger that is fed with the hot heat rejection fluid (Fig 2b, point H), which subsequently exits as a warm fluid (Fig 2b, point D) to be recirculated for another cooling cycle. The heat is transferred to the ambient air, which acts as the heat sink, and exits at a higher temperature (Fig 2b, point I), moving heat into the atmosphere.

Such designs may become increasingly common for power-dense AI computing racks operating at high temperatures that may make efficient “dry” cooling possible in many climates (58). For example, if the “warm” fluid design temperature (Fig 2b, point B) were to be 45 C, then the outside air could be as warm as 40 C for the dry cooler to operate reliably, using a 5 C approach temperature rule of thumb. Such outdoor conditions can occur in most climates at most times of the year globally (59). However, when such outdoor conditions do not occur, operation of the chiller becomes necessary (Figure 2a).

## 2.12 Open loops, closed loops, and water use

Estimating the water use of a data center cooling system thus requires knowledge of: (a) which cooling loop design(s) are, or will be, present in the data centers under study, (b) whether and when those loops require water, and, (c) if so, how much. Significant variations can exist, as shown in Table 2. When water is required, the analyst should distinguish between “open loop” and “closed loop” water uses within the relevant loops. The simplified cooling system depictions in Section S4 can provide guidance.

### 2.121 IT cooling loop water use

In the IT cooling loop, when liquid-cooling systems utilize water, they are “closed-loop” as described in Section 2.111. However, “closed-loop” does not necessarily mean “zero water.” This is because these systems will still require water inputs for initial filling when water or water/glycol mixtures are used, as discussed below (60).

When “air cooled” systems are used, two types of “open-loop” water use can occur. The first is for humidification of air, in which all water inputs are equal to consumption (61,62). In fact, even data halls with closed-loop, liquid-cooled racks often require air humidification.

The second is when adiabatic air coolers are used to cool the incoming rack air, whether directly or indirectly. In direct adiabatic coolers (Figure S15), water injection lowers the incoming airstream temperature to improve heat transfer and also humidifies the data hall air (13,63). In indirect adiabatic coolers, water is injected into outside air, which cools incoming air indirectly via a heat exchanger (Figure S16). Thus, both types of adiabatic cooler consume water. Both types may also require “bleeding” and flushing to remove contaminants, leading to water discharges (64).

### 2.122 Primary cooling loop water use

When present, the primary cooling loops (Figure 2a) are always “closed loop.” However, water inputs are still required for initial filling when water or water/glycol mixtures are used. Primary cooling loops may also require infrequent water discharges, and subsequent water inputs, to facilitate any needed system flushing and liquid refresh cycles to remove contaminants (54,65,66).

### 2.123 Outdoor heat rejection loop water use

In the outdoor heat rejection loop, both “open loop” and “closed loop” heat rejection devices exist. As discussed above, open-loop cooling towers (Figure 2a) facilitate building heat rejection through the evaporation of water, leading to substantial water consumption when they are operated (12,13,53). These cooling towers must also be regularly bled or “blown down” to remove minerals that are concentrated due to evaporation, leading to regular water discharges (53,67). Treated water inputs to to replace water lost to both evaporation and discharges is known as cooling tower “makeup” water.

Although currently less common, “once-through” heat rejection is also open loop (Figs S17, S18). In such systems, water continuously withdrawn from water bodies (e.g., rivers or lakes) is used for building heat rejection, wherein all water withdrawals are immediately returned as discharges, similar to once-through power plant cooling (68). Examples include terrestrial data centers cooled by seawater (14), floating data centers (69), and underwater data centers (70).

Closed-loop heat rejection devices require water inputs for initial filling when water or water/glycol mixtures are used as the heat transfer fluid. Closed-loop “dry” coolers, such as the one depicted in Figure 2b, require no further water use. They reject heat by drawing outside air across a heat exchanger that contains the outdoor heat rejection liquid in a closed circuit (48). No evaporation is involved (Fig S19).

However, some closed-loop heat rejection devices still consume water. For example, some “dry” coolers can operate with adiabatic cooling assistance, which injects water into the incoming outdoor airstream (i.e., the heat sink) to lower its temperature and improve cooling performance (Fig S20). Additionally, closed-loop cooling towers rely on evaporative cooling via a water spray, even though the heat rejection fluid itself recirculates within a closed circuit (71). Therefore, it is critical for analysts to understand that “closed-loop” heat rejection doesn’t mean “zero-water” heat rejection.

**Table 2:** Common technology variations in data center cooling loops and their uses of water. Note: see Section S4 for simplified schematics and descriptions of overall system design variations.

| Loop | Technology | Loop fluid | Typical fluid type(s) | Loop type and water use | | | | Fluid-related electricity use |
|---|---|---|---|---|---|---|---|---|
| | | | | **Type** | **Inputs** | **Consumption** | **Discharges** | |
| **Data hall IT cooling loop** | Recirculated air | Air | Indoor air | Open-loop | Humidifying water mists or steam as needed | Humidified data hall air exhausted from building | None | Fans/blowers |
| | Airside economizer | | Filtered outdoor air, humidified as necessary | | | | | Fans/blowers, pumps |
| | Airside economizer with direct adiabatic assist | | Filtered, cooled, and humidified outdoor air | | Water spray for wetted media; makeup water | | Bleed-off or blowdown | |
| | Airside economizer with indirect adiabatic assist | | Filtered and cooled outdoor air | | | Water vapor into external air | | |
| | Rear door heat exchange[1] | Liquid[2] | Treated water, water/glycol | Closed-loop | For treated water or water/glycol mixtures, initial filling of closed-loop IT cooling circuit, refilling after any system flushes | None beyond any consumption associated with indoor air humidification | Infrequent system flushes of treated water, water/glycol systems | Fans/blowers, pumps |
| | Direct-to-chip liquid cooling | Liquid/ fluid[2] | Treated water, water/glycol, dielectric fluid (single- or two-phase) | | | | | Pumps |
| | Immersion cooling | | Dielectric fluid (single- or two-phase) for immersion bath; treated water, water/glycol, or dielectric fluid (single- or two-phase) for IT coolant circuit | | | | | |
| **Primary cooling loop** | Direct expansion system | Liquid | Refrigerant | None | | | | Compressors and pumps |
| | Chiller | | Treated water or water/glycol mixture | Closed-loop | For treated water or water/glycol, initial filling of primary coolant circuit, refilling after any system flushes; makeup water | None | Infrequent system flushes of treated water, water/glycol systems | |

| Loop | Technology | Loop fluid | Typical fluid type(s) | Loop type and water use | | | | Fluid-related electricity use |
|---|---|---|---|---|---|---|---|---|
| | | | | Type | Inputs | Consumption | Discharges | |
| **Outdoor heat rejection loop[3] and outdoor heat sink** | Dry cooler for heat rejection to ambient air | Liquid | Refrigerant (DX systems); treated water, water/glycol | Closed-loop | For treated water or water/glycol, initial filling of closed-loop heat rejection liquid circuit, refilling after any system flushes, makeup water | None | Infrequent system flushes of treated water, water/glycol systems | Fans and pumps |
| | Dry cooler with direct adiabatic assist for heat rejection to ambient air | | Refrigerant (DX systems); treated water, water/glycol | Closed-loop and open-loop[4] | | Water vapor into external air | Bleed-off or blowdown of open-loop cooling water | |
| | Closed-circuit cooling tower for heat rejection to ambient air | | Treated water, water/glycol | | | | | |
| | Open-loop cooling tower for heat rejection to ambient air | | Treated water | Open-loop | Water spray for wetted media; makeup water | | | |
| | Once-through cooling for heat rejection to water bodies | | Surface water | | Direct withdrawals from cooling water body | None | Direct discharges to cooling water body | Pumps |

[1] provides supplemental cooling to air-cooled racks by passing exit air over an air-to-liquid heat exchanger that contains IT cooling or primary cooling liquid

[2] even fully closed-loop, liquid-cooled racks can require data hall air humidification; see the "recirculated air" water and electricity use entry

[3] all heat rejection devices can be operated as "waterside" economizers for "free cooling" to bypass the need for primary mechanical cooling

[4] open-loop water use occurs due to the addition of evaporative water to the heat sink (outdoor ambient air) that passes over the closed-loop heat rejection loop

### 2.124 Initial filling of closed-loop fluid systems

None of the 31 studies assessed in this review considered the water use associated with initial filling of aqueous fluid loops. However, water use for initial filling is likely to be relatively small, even for large data centers. For example, when the primary cooling loop uses chilled water, one conventional rule of thumb is that they should be filled with 3 to 10 gallons (11.4 to 37.9 liters) of water per installed ton of cooling (65,72). Accordingly, a large data center with 100 MW of peak IT capacity would require around 34 thousand tons of installed chilled water capacity. This estimate assumes that 1 MW of heat removal requires 284 tons of chilling, a 1.2 safety factor, and data hall power chain losses of 10% (3,73). Under these assumptions, the initial filling would require 468 thousand to 1.5 million liters of water input. Assuming that the average U.S. home requires 138 gallons per day (522 liters) for indoor use (74), this filling quantity is equivalent to the annual indoor water inputs of 3 to 8 average U.S. homes. Even if initial filling quantities are small, the analyst should ideally estimate them for IT cooling, primary cooling, and outdoor heat rejection loops as relevant for completeness.

### 2.125 Free cooling and its effects on water use

Lastly, the analyst must be aware that the use of “free cooling” (75) can substantially affect a data center’s water needs, even when the cooling loop types have been established. Data centers with air-cooled IT racks can employ air-side economizers, which are devices that simply route outside air into the data center when its temperature is cooler than the required rack air supply temperature (Fig S14). These conditions can occur in many hours of the year in favorable climate zones (76). Air-side economizers require fans and air filters to ensure sufficient airflow rates and quality, which consumes power, but their operation saves substantial electricity compared to using mechanically-chilled water for air cooling (5). Avoiding chiller operation can also save water when those chillers utilize outdoor heat rejection devices with open-loop water use.

The term “free cooling” also applies to data centers that bypass, or altogether eliminate, the use of chillers or direct expansion systems for the primary cooling loop. They do so by coupling the IT cooling loop and outdoor heat rejection loop as shown in Figure 2b, for example. This configuration is possible when outdoor air conditions (i.e., temperature and/or humidity) are such that operating either a closed-loop “dry,” or an open-loop evaporative, outdoor heat rejection device is sufficient for providing a cool enough liquid back to the IT cooling loop heat exchanger (point D in Figs 2a and 2b) such that the chiller can be turned off. These outdoor air conditions can also occur in many hours of the year in favorable climate zones (59). The term “water-side” economizer refers to using any heat rejection device (Table 2) for “chillerless” cooling of the IT cooling loop fluid (5).

## 2.13 Cooling system efficiency metrics

In addition to understanding the cooling loops of the data centers under study, analysts must also consider how these systems will perform under different conditions. These include different rack temperature setpoints, local climate variations, and the extent to which “free cooling” may be used.

To do so, analysts typically use two standard metrics for assessing the water and energy efficiency of different cooling system designs under different conditions: water usage effectiveness (WUE) and power usage effectiveness (PUE) (13). As shown in Figure 1, and as discussed in Section 3.31, WUE is expressed as the ratio of either total data center operational water inputs or total data center operational water consumption to total IT electricity use (77,78). Thus, WUE accounts for water use for personnel and sanitation, humidification, adiabatic cooling, and evaporative heat rejection (Table 1), although their individual contributions to WUE can vary based on whether water input or water consumption is considered in the WUE numerator.  Water use for initial filling of aqueous cooling loops is excluded from WUE. PUE is expressed as the ratio of total data center electricity use to IT electricity use (79,80).

### 2.131 Effects of cooling system designs and climate

Figure 3 depicts simulated WUE and PUE ranges for different data center types and cooling system configurations (employing combinations from Table 2) across nineteen different U.S. climate zones.

The ranges in Figure 3 illustrate the wide possible variations in water and energy efficiency that can occur across different types of data centers and climates depending on their cooling system designs and operating strategies (3,13). In particular, Figure 3 illustrates the substantial reductions in PUE and WUE that can occur when “free cooling” air-side or water-side economizers are utilized compared to continuous chiller operation (75).

In general, the highest WUEs are associated with the use of air-cooled IT cooling loops at low rack air supply temperatures, mechanically-chilled water primary cooling loops, minimal to no use of “free cooling,” and open-loop cooling towers in the outdoor heat rejection loop (81). Systems with “dry” coolers for outdoor heat rejection can require minimal water inputs (i.e., just for initial filling and infrequent refreshing of closed-loop water) or even no water inputs (in the case of direct-expansion systems (48) that use non-aqueous refrigerants).  However, “dry” operations have historically come at the expense of higher power use (i.e., high PUEs) compared to evaporative heat rejection systems (i.e., the so-called “energy-water tradeoff” as seen in Figure 3) (3,13,77).  The classical “energy-water tradeoff” for a chiller with dry heat rejection (low WUE, high PUE) versus a chiller with open-loop cooling tower heat rejection (low PUE, high WUE) can also be seen in Figure 4.

However, recent innovations enabling future shifts toward liquid-cooled, power-dense AI racks with high rack supply temperatures (e.g, 45C or higher) may enable “dry” cooling that is also highly energy efficient (i.e., low-to-zero WUEs with low PUEs) in many locations, reducing the “energy-water tradeoff” effect (59).

### 2.132 Temporal variations and “peak” WUE

Significant temporal variations in WUE and PUE can also occur due to diurnal, seasonal, and annual changes in outdoor climate conditions. Figure 4 plots such temporal variations on hourly and monthly bases for an illustrative data center in El Paso, Texas. These variations, simulated by Weller and Nichols (82), highlight the potentially large differences between peak and average WUE and PUE values depending on the cooling system configuration, time of day, and time of year as simulated by

Peak periods can place stresses on local water resources and power systems, depending on the data center location and design (83).  Quantification of peak WUE can be important for planning future data center water infrastructure requirements, but is a topic that has rarely been considered in data center water use estimates to date (see Section 5) (83).

However, data centers can also reduce peak power grid and water utility demand through the use of onsite water and/or chilled water storage tanks, which can be filled and discharged strategically to minimize stress on external grids and water systems (84,85). Therefore, the real-world implications of peak WUE will also depend on specific data center designs, their locations, and their water resilience strategies (85).

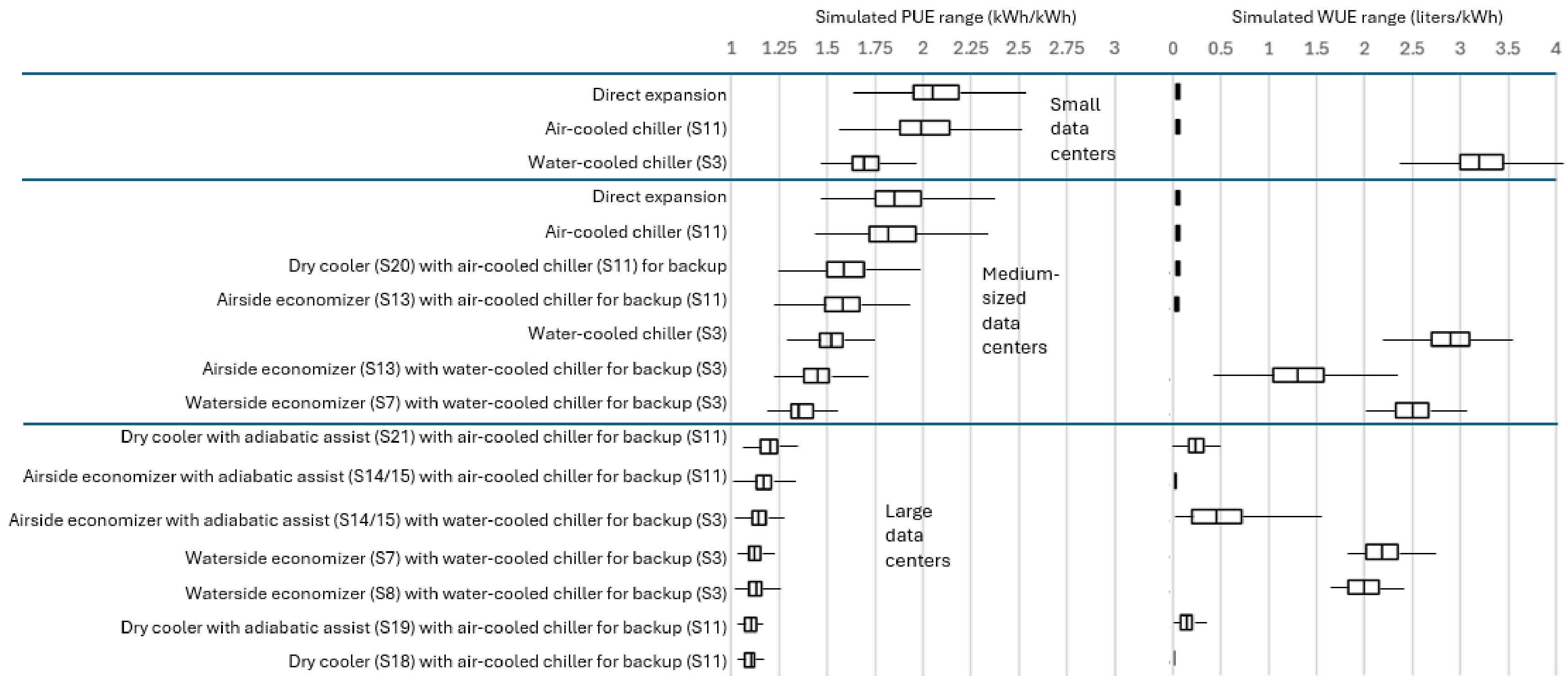

**Figure 3:** Simulated PUE and WUE ranges over 19 different U.S. climate zones for small, medium, and large data centers and by cooling system configuration (3,13,31). Note: terms in parentheses refer to the associated figure(s) in the supplementary information file.

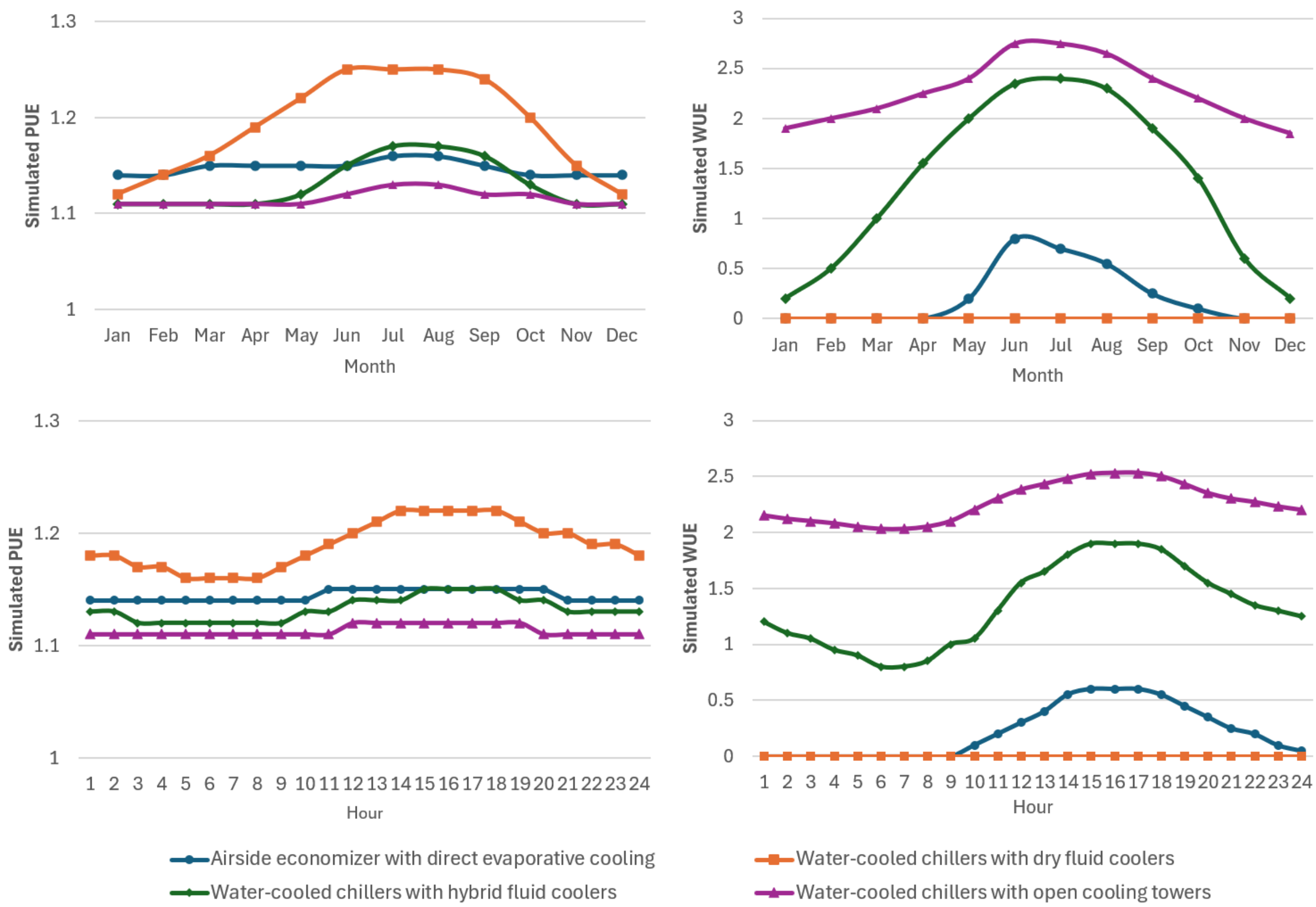

**Figure 4:** Illustrative annual average hourly variations in PUE (top left) and WUE (top right) and monthly average variations in PUE (bottom left) and WUE (bottom right) for a data center in El Paso, Texas as simulated by Weller and Nichols (82).

## 2.2 Indirect water use

Indirect water use is attributable to power generation, where water inputs can occur at the point of generation (POG) and within upstream processes.  Here, distinguishing between water withdrawals, consumption, and discharges is also very important, because their relative scales can vary significantly.

### 2.2.1 Upstream vs. POG water uses

Upstream water use relates to: (a) the water inputs required during the manufacturing of generation technologies such as photovoltaic panels, wind turbines, and steam turbines; (b) hydrofracturing water use when creating geothermal heat wells; (c) biocrop irrigation; and (d) other fuel cycle processes, such as oil refining and coal cleaning (18,20,86).

POG water use is associated with: (a) the operation of cooling towers or once-through cooling systems at thermoelectric power plants; (b) cleaning and maintenance of power generation equipment such as solar panels; and (c) water evaporation at hydroelectric dam reservoirs (18,87). Similar to the patterns seen in Figure 4, the POG water use of thermoelectric power plants and hydroelectric reservoirs can also vary by hour, day, season, and year due to local climate variations.

### 2.2.2 Quantifying indirect water use intensities

Analysts typically quantify indirect water use using an energy water intensity factor (EWIF), which can either quantify the indirect water withdrawals or indirect water consumption, in liters, associated with a kilowatt-hour of electricity consumed by a data center as depicted in Figure 1 (77). Because deriving original EWIFs can be a massive research undertaking, data center water use analysts typically don't derive original EWIFs when estimating indirect water use, although exceptions exist (15).

Instead, they often rely on EWIFs that are available for the regions and grid technologies under study, sourced from public datasets (88) and/or previous literature (18,20,86,87,89,90). The scales of available EWIFs can exhibit large variations within and across power generation technologies and regional grid mixes depending on the source's underlying data origins, ages, locations, power plant technology mixes, and EWIF derivation methods. In general, withdrawal-based factors exhibit (often much) higher ranges of values than consumption-based factors. For example, frequently cited withdrawal-based EWIF values summarized in Figure 5 for different combinations of power plant technologies and cooling systems range from 0 to more than 200 liters per kWh of generation, whereas the consumption-based EWIF values for those and similar combinations (Figure 6) range from 0 to less than 5 liters per kWh (89). In light of all this variability, analysts must choose their EWIFs carefully.

### 2.2.3 Hydroelectric power EWIFs: variations and cautions

In particular, EWIF values for hydroelectric power sources must be viewed with caution. First, different arbitrary conventions for estimating hydropower water use can lead to massive differences in published EWIF values, as shown in Figure 7. Second, many published hydropower EWIFs can be much higher (Figure 7) than EWIFs for other common forms of power generation (Figure 6). Therefore, when hydropower EWIFs are employed to estimate data center indirect water use, results can be much higher than estimates made without hydropower EWIFs.

Three accounting conventions strongly affect the scale of a published hydropower EWIF value. The first accounting decision is whether hydroelectric reservoir evaporation, which occurs at all reservoirs, was estimated on a gross or a net basis. Gross evaporation considers only the current evaporation of the reservoir, which is typically calculated as a function of its surface area and climate. Net evaporation subtracts away from the gross value any evaporation and evapotranspiration that was estimated to occur before the reservoir's dam was built (19,91).

The second accounting decision is what fraction of the estimated evaporation, whether on a gross or net basis, was allocated to hydroelectricity at reservoirs that are also used for other purposes, such as irrigation, flood control, and recreation (19,87,91). Different combinations of evaporation accounting and allocation choices can lead to substantial differences in the resulting consumption-based EWIFs available to analysts for hydroelectric power. Using the data source in Figure 7a as an example, the choice of a U.S. consumption-based hydropower EWIF based on gross reservoir evaporation that is completely allocated to hydroelectric generation would lead to a value (84 liters/kWh) that is around twenty times higher than one that is based on net evaporation with equal allocation to all reservoir uses (4.32 liters/kWh) (19).

The first two accounting decisions relate to consumption-based hydropower EWIFs. The third relates to withdrawal-based hydropower EWIFs, which are rare in the literature. Specifically, how a water "withdrawal" is defined is highly consequential. For example, Reig et al. (2020) treats the "sum of the water that is needed for turbine operation of the hydropower plant" as water that is "withdrawn" (90), even when that water might flow downstream for subsequent use. Under this accounting convention, withdrawal-based hydropower EWIFs can be more than 1000-times greater than consumption-based hydropower EWIFs within the same geographic region (Figure 7b).

The effect of hydroelectric power EWIFs on data center water use estimates is further explored in Section 3.4.

## 2.3 Embodied water use

Embodied water use refers to the "embedded" or "cradle-to-gate" water use associated with data center equipment and materials manufacturing and their upstream supply chains (27,92). This stage of water use is beyond the scope of this review, due in part to its scarcity of coverage in the literature to date (see Section 5.5). Embodied water use is also excluded due to its reliance on life-cycle assessment (LCA) data and methodologies, which have their own established standards and communities of practice (93).

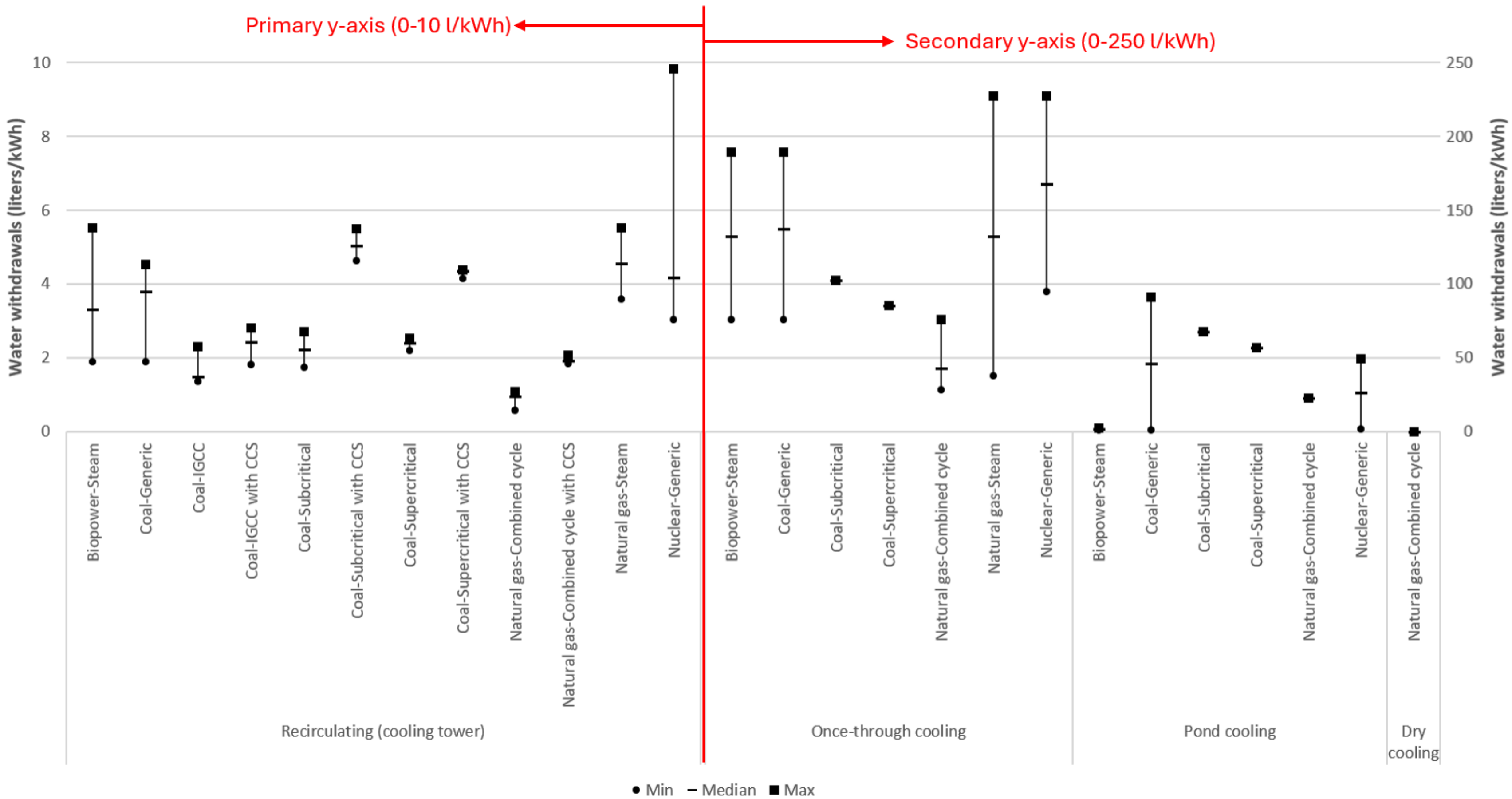

**Figure 5:** Literature ranges of water withdrawal EWIFs for different fuels and power plant cooling technologies (89)

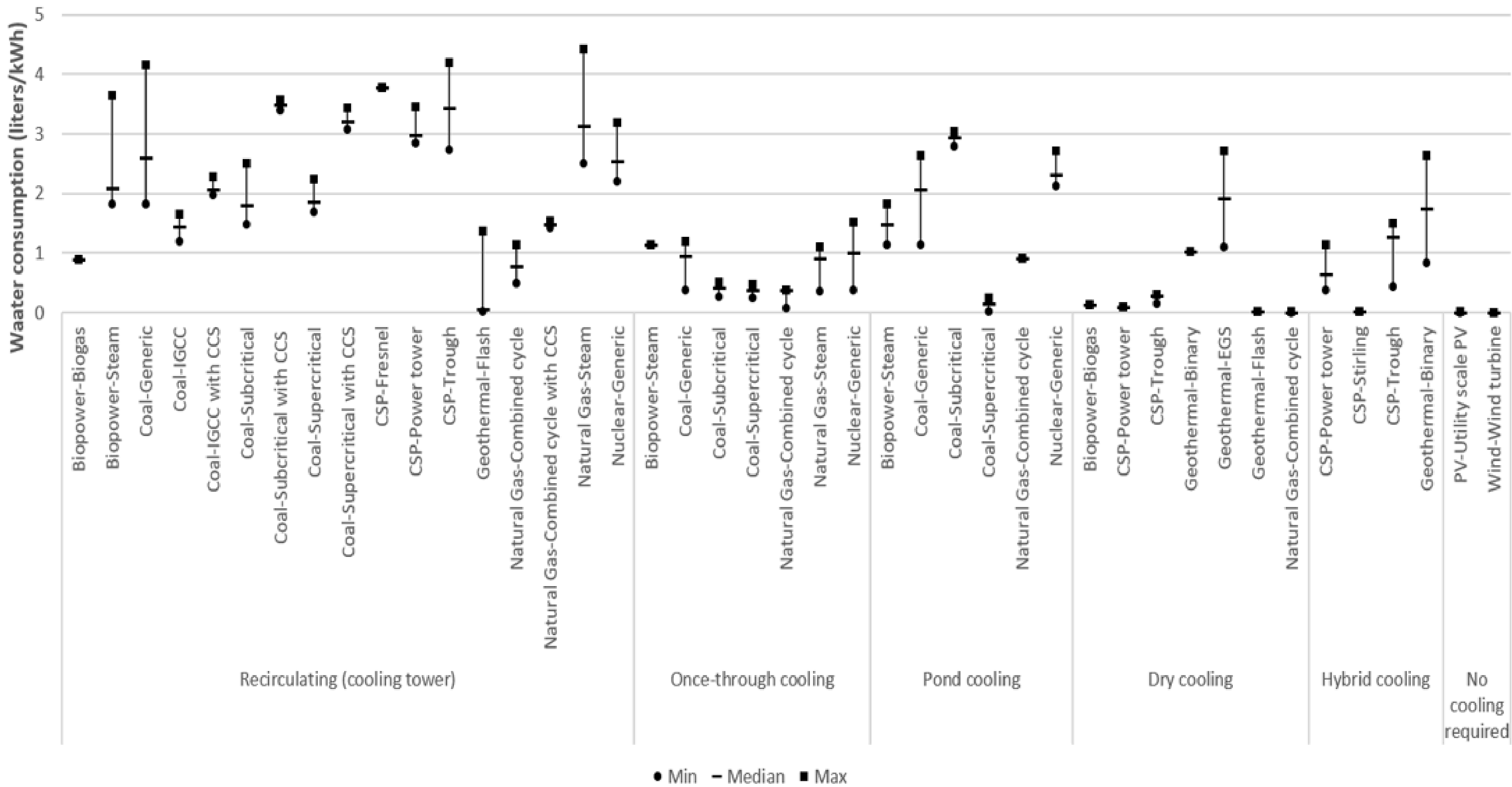

**Figure 6:** Literature ranges of water consumption EWIFs for different fuels and power plant cooling technologies (89)

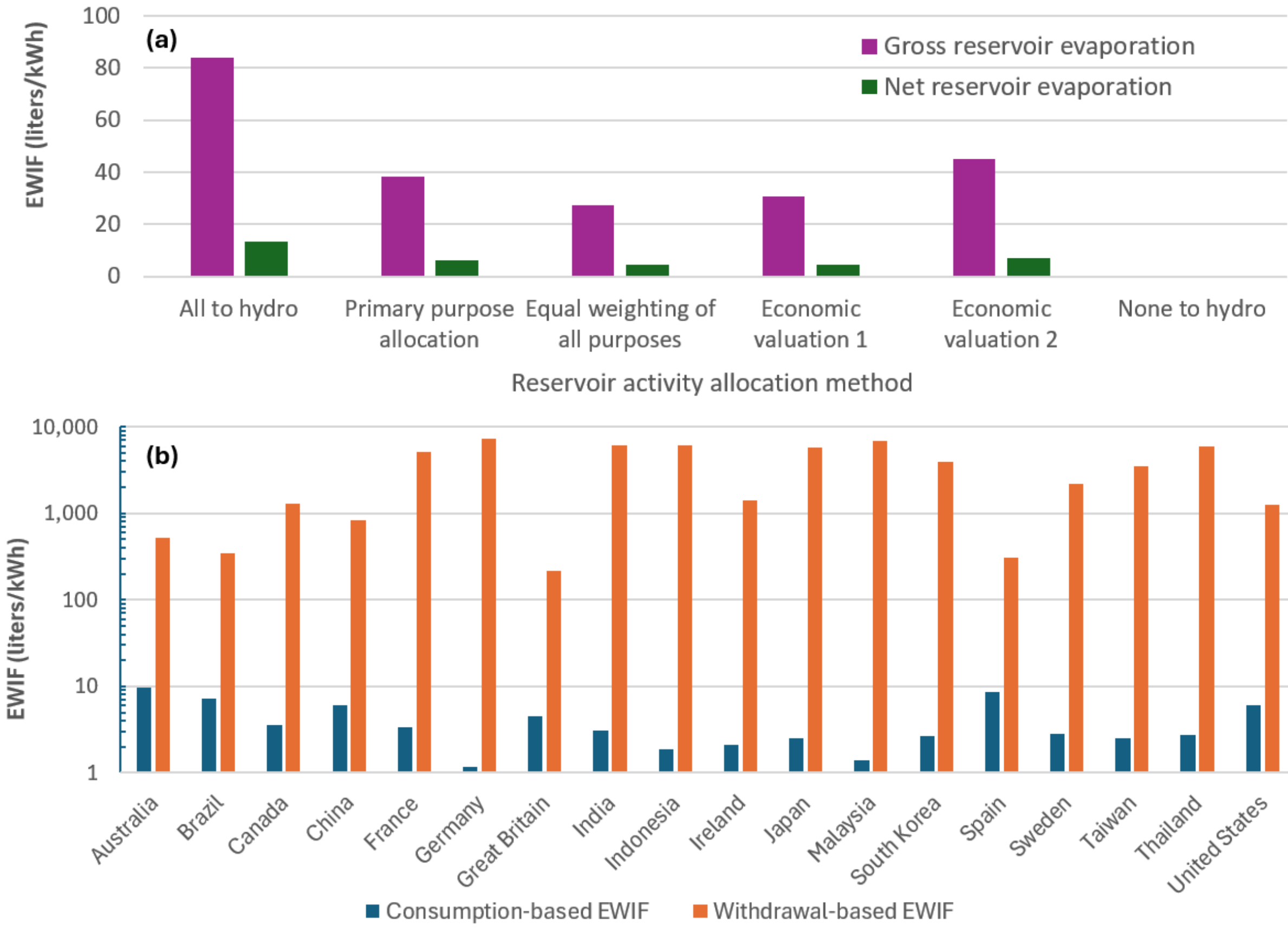

**Figure 7:** Effect of (a) evaporation and allocation accounting bases on the magnitude of consumption-based hydropower EWIF values for an illustrative case in the United States as presented in (19); and (b) withdrawal-based versus consumption-based EWIF values for major data center countries as presented in (90).

## 2.4 Direct and indirect water use estimation

To estimate direct and indirect data center water use, analysts generally rely on the approach depicted in Figure 8, based on the intrinsic relationships and metrics depicted in Figure 1.

**Direct water consumption/inputs (liters)** =
Data center IT electricity use (kWh) * WUE (liters/kWh)

**+**

**Indirect water consumption/inputs (liters)** =
Data center IT electricity use (kWh) * PUE (kWh/kWh) * EWIF (liters/kWh)

**=**

**Total (direct + indirect) water consumption/inputs (liters)**

**Figure 8**: Steps in the estimation of data center direct and indirect water use

However, Equation 1 provides a more precise mathematical expression of this approach. Importantly, the indices on each variable reinforce the critical need for consistency of assumptions related to the spatial, temporal, technology, and water use components under study. As discussed in Section 3, when analysts "mix and match" inconsistent assumptions in Equation 1, such as WUE values that are not valid for the timeframes, regions, and data center types under study, large errors can be introduced.

**Equation 1:** $$W_{i,j} = \sum_{1}^{t}\sum_{1}^{k}([IT_{j,k,t} * WUE_{i,j,k,t}] + [IT_{j,k,t} * PUE_{j,k,t} * EWIF_{i,j,k,t}])$$

where W is the total (direct and indirect) water use in liters of type *i* (inputs, withdrawals, or consumption), in region *j*, IT is the data center IT electricity use in kilowatt-hours (kWh) for data center type *k* over time period *t*; WUE is water usage effectiveness in liters per kWh of IT electricity use; PUE is power usage effectiveness in kWh per kWh of IT electricity use, and EWIF is the energy water intensity factor in liters per kWh of total data center electricity use. The first and second bracketed terms in the summation quantify direct and indirect water use, respectively.

While Equation 1 is mathematically straightforward, in reality the analyst has to contend with numerous technology, climate, and temporal variations, data gaps, and differences in data sources and conventions when applying Eq 1. The next section proposes 14 best practices for navigating these challenges by stepping through an illustrative application of Eq. 1, illustrating the relevance and importance of each best practice along the way.

# 3. Best practices for data center water use estimation

Table 3 summarizes the 14 best practices (BPs) proposed in this review. These proposed best practices are drawn from analogous best practices for data center energy analysis, water studies, and open science as well as from examples of best practices observed across the assessed studies (28,33–36,39). These best practices are referenced in this section according to the labels in column 2. To motivate each best practice, this section steps through the analyst's process for estimating projected data center water use via Eq. 1, using an illustrative case of the United States and drawing on various data from the 31 studies compiled for this review. The United States is chosen because it accounts for the largest share of current and projected data center electricity use globally (2–4) and has been the subject of more than half of the identified water estimation studies to date. Figure 9 depicts schematically how numerical results can diverge at each step based on analyst choices, which helps to illustrate the importance of best practices. Assumptions for the illustrative case are further summarized in the Supplementary Information Excel file.

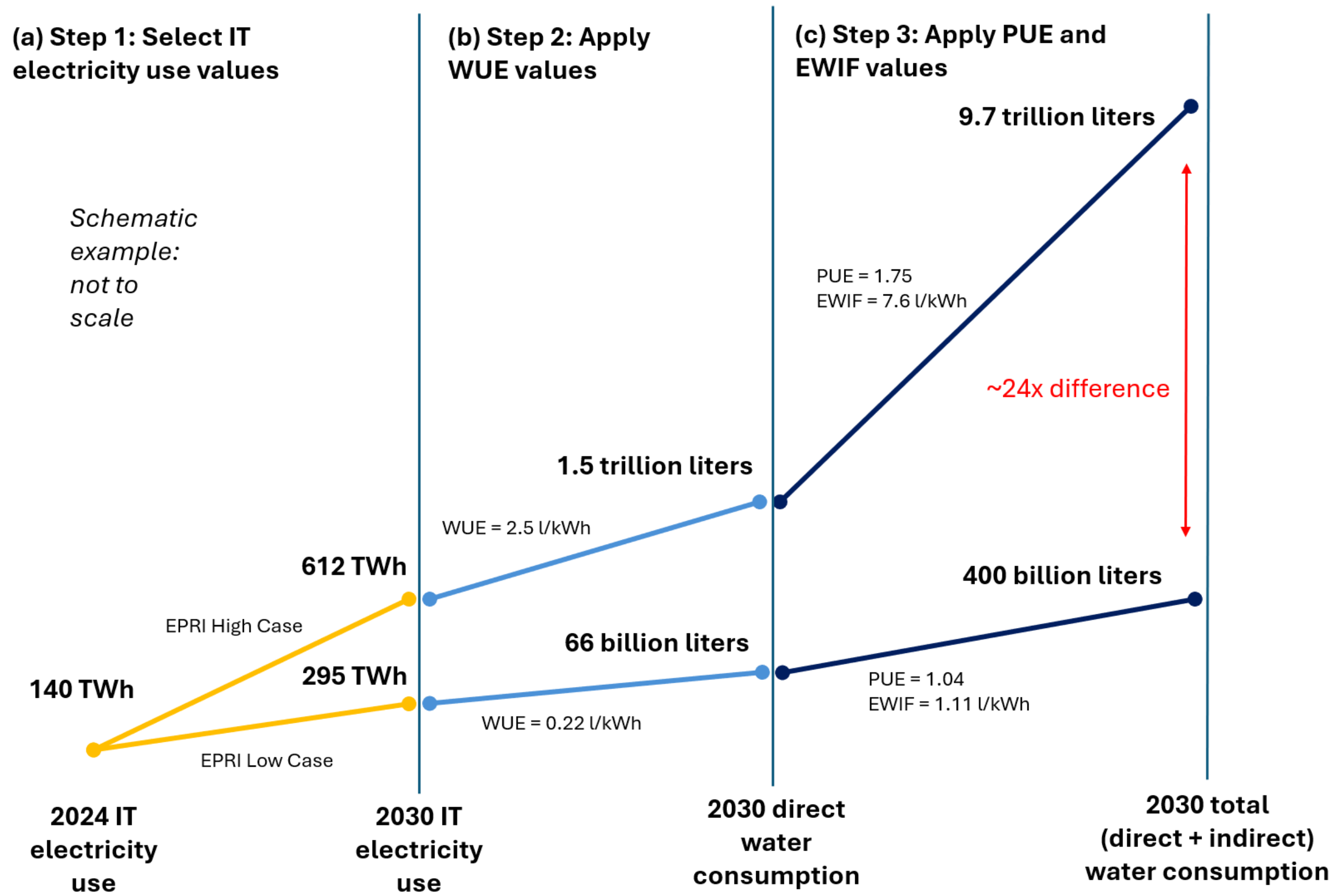

**Figure 9:** Illustrative example of two different estimation pathways for U.S. data center water consumption in 2030 using observed values from the literature (not to scale)

**Table 3:** Proposed best practices and critical questions for data center water use estimation

| Category | Label | Best practice for analysts | Critical questions for stakeholders |
|---|---|---|---|
| Clarity | 1 | Clearly state the water use components (withdrawals, inputs, consumption, and/or discharges) and water use stages included in the study scope | Can the specific data center water use components and scopes included in the study be easily determined? |
| | 2 | Precisely define all included water use components to eliminate confusion | Is it clear how all water use components are defined and do those definitions align with common conventions? |
| | 3 | Use consistent terminology throughout the study when describing water use data, methods, and results; avoid competing and vague terms | In all graphs, tables, and discussions, is it clear which water use components are being referred to? |
| Scientific due diligence | 4 | Transparency: provide all input data and important contextual information for assumptions related to the selected: (a) IT electricity demand values; (b) WUE values; (c) PUE values; and (d) EWIF values. | Does the study clearly document all input data values and sources and provide important context, such as whether the values relate to conservative or aggressive future outlooks? |
| | 5 | Justification: offer scientific justification, including acknowledgement of any limitations, related to the selected: (a) IT electricity demand values; (b) WUE values; (c) PUE values; and (d) EWIF values. | Does the study convince the reader that all input data values are well researched and are the most defensible and appropriate choices? |
| | 6 | Explicitly state the reservoir evaporation accounting and allocation methods used in any EWIF values that include hydroelectric power | If the study region includes hydroelectric power sources, are the water accounting bases for the hydroelectric power EWIFs clearly stated? |
| | 7 | Uncertainty: acknowledge and quantitatively assess uncertainties in study results through parametric uncertainty, sensitivity, and/or scenario analysis | Does the study provide ranges, scenarios, sensitivity analysis, or other uncertainty quantifications instead of only point values? |
| Scientific quality and replicability | 8 | Representativeness: ensure that all input data are representative of the spatial, temporal, and technological aspects of the systems under study for the selected: (a) WUE values; (b) PUE values; and (c) EWIF values. | Do the study's input data values make sense for the specific locations, time frames, and types of data centers included in the study, including any expected changes over time? |

| Category | Label | Best practice for analysts | Critical questions for stakeholders |
|---|---|---|---|
| | 9 | Replicability: ensure that all input data and related calculations are made publicly available to enable independent replicability of results | Can all input data and calculations be accessed such that the study's results can be fully replicated? |
| | 10 | Results granularity: present results with sufficient granularity to reflect estimates by water use component, water use scope, regions, and/or data center types for maximum utility and comparability | Are results presented separately for all included water use components, water use stages, and data center types? |
| | 11 | Subject the study to peer review by independent experts to help ensure study clarity, scientific quality, and stakeholder utility; clearly indicate peer review status and methods | Has the study been peer reviewed in a meaningful way? |
| Community knowledge building | 12 | Offer meaningful acknowledgements and discussions of study limitations related to data, methods, results, and intended applications | Are study limitations related to data, methods, results, and/or applications sufficiently discussed? |
| | 13 | Offer actionable suggestions for future work to address study limitations, expand study scope, and apply study contributions to new research questions | Does the study offer actionable suggestions for future work to address limitations and expand the domain? |
| Communi-cations | 14 | When offering water use equivalency comparisons, ensure that they are scientifically valid by only comparing inputs to inputs, etc. | If the study offers comparisons to other water using activities, are the comparisons truly "apples to apples?" |

## 3.1 Laying a solid foundation

Before starting any analysis, the analyst must decide which water use components (i.e., inputs, withdrawals, consumption, and/or discharges) and stages (i.e., direct, indirect, or both) to estimate. While these choices depend on study objectives, three best practices are critical for ensuring clarity and consistency.

First, the analyst must ***clearly state (BP1)*** the water use component(s) and water use stages included in the study, such that any reader can accurately interpret the study's results. For example, a study that focuses on water withdrawals will typically yield higher water estimates than one that focuses on water consumption, particularly for indirect water use estimates. Studies that only use vague terms such as water "use," "demand," or "requirements" without clearly stating the included components muddy the study's scope and preclude independent verification of whether Eq. 1 was applied consistently across water use stages. Studies that only include direct water use will typically yield lower estimates than those that include both direct and indirect water use.

Second, the analyst should offer ***precise definitions (BP2)*** of all included water use component(s) for maximum clarity. The need for absolute clarity has been stressed by Grubert et al. (2020), who describe how water use assessments are commonly hindered by the use of inconsistent terminology and reporting standards (39). For example, if a study includes withdrawals, are they defined as direct withdrawals from water sources or does this actually refer to water inputs from municipal supplies, which may not account for withdrawn quantities lost as leaks (Figure 1) (40)? Is consumption defined as the difference between input and discharge quantities, or is the term "consumption" being used casually?

Third, even after clearly stating and defining the included water use components, analysts should ***avoid using vague and competing terms (BP3)*** when subsequently discussing methods and presenting and interpreting study results. For example, figures labeled with general terms like water "use" or "demand" can inject confusion and enable future misinterpretations by stakeholders, even when a study has clearly stated and defined its water use components somewhere in the text.

In the illustrative case (Figure 9), water consumption is chosen and defined as the difference between withdrawals and discharges in the direct and indirect stages.

As discussed in Section 2, there are numerous technology, climate, and temporal variations, data gaps, and differences in data sources and conventions that the analyst will encounter when populating Eq. 1. Therefore, the analyst must provide full ***transparency (BP4)*** and ***scientific justification (BP5)*** for all independent variable choices and use the most ***representative values (BP8)*** whenever possible. Considerations for BP4, BP5, and BP8, which can differ based on each variable in Eq. 1, are discussed in more detail in Sections 3.2 to 3.6.

## 3.2 Charting the energy trajectory

Next, the analyst must derive or choose IT electricity demand values, which will drive the study's water use estimates. These values will depend on the region and time horizon of the study and the types of IT equipment, workloads, and data centers under consideration.

Wide variance can exist among data center electricity demand projections, owing to differences in methods, data sources, assumptions, and scenario narratives (10,11,36). For example, EPRI (2026) summarized nine recent projections of the total electricity use of U.S. data centers, with 2030 results ranging from 235 TWh to over 1,000 TWh – a five-fold difference in published outcomes only four years into the future (24).

While the analyst must derive or choose the IT electricity demand estimates that are most appropriate to a given water use study, this wide variance warrants the following considerations.

First, when providing ***transparency*** on the IT electricity demand value(s) chosen for a study, the analyst should ensure that the underlying data are explicitly declared and made publicly available. Furthermore, the analyst must clearly describe the forecast or scenario narratives associated with the chosen IT electricity demand values, for example, whether they correspond to aggressive (i.e., bearish) or conservative (i.e., bullish) growth assumptions. The analyst should also state clearly the types of data centers or IT hardware associated with the projections; for example, do the projections include all types of data centers (3), only those that have been publicly identified (94), or only those associated with AI computing (16)? Doing so provides important context for stakeholders when interpreting and using a study's results and can enable greater incomparability of results between studies.

Second, when offering ***scientific justification***, the analyst should clearly defend the choice of IT electricity demand values(s) compared to other options available in the literature. Doing so helps the analyst convince both themselves and the study's stakeholders that the chosen values are representative and defensible for the spatial, temporal, and technological scope of the study. Conversely, merely stating chosen values with no discussion can erode credibility and limit confidence in a study.

The analyst should also ensure that the spatial, temporal, and technological scope of the chosen IT electricity demand values will match those of the subsequently chosen WUE, PUE, and EWIF values (Eq 1), as discussed later in Sections 3.3 to 3.4.

### 3.21 Illustrative case: The effect of different IT electricity projections

The illustrative case (Figure 9, Panel a) presents upper and lower scenario results for U.S. data center IT electricity demand from 2025 to 2030, derived from recent low and high total electricity demand scenarios from EPRI (2026) and their associated PUE values (24,95). The 2030 IT electricity demand results of these two scenarios, 295 TWh and 612 TWh, differ by a factor of two. Therefore, from Eq. 1, it is evident that the high scenario values will lead to 2030 water use estimates that would be at least twice the quantities of the low scenario values. In practice, this difference could be far greater (e.g., 5x) considering the highest and lowest electricity demand values available for selection from the broader literature. Whatever the analyst's choice(s), providing both transparency and scientific justification is essential for responsible analysis and proper interpretation of ultimate water use results.

## 3.3 Cooling system assumptions: Handling with care

Next, the analyst must derive or choose the WUE and PUE values that will drive the direct and indirect water use estimates, respectively, in Eq. 1. These values should be consistent with the region and time horizon of the study and cooling systems associated with the IT equipment, workloads, and data centers implied by the analyst's chosen IT electricity use values.

### 3.31 WUE

Most commonly, the analyst will utilize reported or simulated onsite WUE values. However, these must be used carefully due to variations in how WUE may be measured, as summarized in Table 4. For example, the ISO standard 30134-9 defines two different measurement categories for onsite WUE as summarized in (78). A Category 1 WUE considers water inputs only, whereas a Category 2 WUE subtracts from inputs the water quantity leaving the data center boundary for non-industrial reuse. Thus, a Category 1 WUE always quantifies water inputs in Eq. 1, but it only truly quantifies water consumption if discharges for non-industrial use are negligible. Conversely, a Category 2 WUE always quantifies water consumption when discharges are for non-industrial use, but it doesn't quantify water inputs in Eq. 1. Additionally, neither measurement level accounts for any water supply leaks that may occur between source water withdrawals and data center input quantities.

The original "site" WUE definition from the Green Grid (2011) is somewhat vague in its construction, stating that "annual water usage" should be used in the metric's numerator, as opposed to the more precise concepts of withdrawals, inputs, or consumption (77). This leaves WUE measurement and reporting assumptions open to different interpretations. Other differences in WUE definitions are evident in Table 4.

All of these variations mean that the analyst must take care to verify the measurement definitions and/or categories associated with any WUE values used in Eq. 1 and to ensure that they are consistent with the stated water use components under study. For example, if the analyst uses a Category 1 WUE to estimate water consumption, they must first verify that discharges are negligible, otherwise water consumption will be overestimated. Conversely, if the analyst uses a Category 2 WUE to estimate water inputs or withdrawals, those quantities may be underestimated. If the analyst uses a reported WUE value without knowledge of which water use components are contained in the numerator, the analyst may introduce errors with no ability to identify or communicate them.

Considerations of data center types, locations, and technology changes over time are also critical. As summarized in Table 2 and Figure 3, different types of data centers can utilize different cooling loop configurations whose water use intensities will be affected by local climates and operational choices, such as rack temperatures. Additionally, trends toward liquid-cooled racks and dry coolers for outdoor heat rejection that are enabled by high temperature set points may also substantially lower the typical WUE values of future data centers with AI racks. This is an especially important consideration for IT electricity demand projections that include substantial AI workload growth (59).

### 3.32 PUE

As discussed in Section 2, and as seen in Figure 3, a data center's WUE and PUE values are intrinsically linked by its cooling system configuration, operating practices, and local climate patterns (12,13). Thus, it is important that the analyst chooses PUE values in Eq. 1 that are based on the same spatial, temporal, and technological assumptions as the chosen WUE values for a given data center type. If PUE values are chosen independently of WUE values, the analyst risks using incorrect or improbable combinations of these two interrelated metrics in Eq.1, leading to analytical inconsistencies.

Furthermore, like WUE, there are different measurement protocols associated with the PUE metric, as summarized in Table 4. Specifically, the ISO standard 30134-3 defines three measurement categories for PUE, which are based on the Green Grid's original PUE proposal in 2007 (79,80). Mathematically, only a Category 3 PUE is strictly compatible with the calculation in Eq. 1. When Category 1 or Category 2 PUE values are used, the analyst will underestimate the total electricity use at the data center level, leading to underestimates of indirect water use. This is because Category 1 and Category 2 PUE values are always lower than Category 3 values for the same system, given that they include data center power chain electricity losses in their denominators in addition to IT electricity use, and potentially even non-IT devices like rack-mounted fans as well.

### 3.33 Best practice considerations for WUE and PUE

In light of the above nuances and considerations, providing transparency and justification and ensuring representativeness is essential.

When striving for ***representativeness*** in the choice all WUE and PUE values, the analyst should ensure that they reflect the observed and expected data center types, cooling system configurations, operating practices, and climate effects of the region under study as well as how these factors may change over the study's time horizon consistent with the chosen IT electricity demand projections. Importantly, this includes choosing congruent WUE and PUE combinations that are always representative of the exact same conditions, including when conditions change over time, given their intrinsic interrelationships. If the analyst simply chooses WUE and PUE values that are convenient without verifying representativeness, large errors may be introduced.

When providing ***transparency*** on the chosen WUE and PUE values, the analyst should clearly state all actual values used, their data sources, and the measurement categories associated with all chosen values. The analyst should also clearly state the data center types, cooling system configurations, climate conditions, and ages and congruence of the chosen WUE and PUE values, even when these characteristics may not be fully representative of the systems under study (e.g., due to data limitations).

Finally, whatever the choices of WUE and PUE values, the analyst should offer clear ***scientific justification*** for these choices. This would include not only defending the chosen values and demonstrating due diligence, but also acknowledging any unknowns (e.g., unknown PUE or WUE measurement categories) and/or study limitations that may be related to data gaps, assumptions, or unavoidable mathematical inconsistencies in Eq. 1 and explaining why results are still useful despite such limitations. Conversely, when no scientific justification is offered, neither the analyst nor their stakeholders can be sure that results are truly defensible.

### 3.34 Illustrative case: The effects of different WUE and PUE values

For the United States, the highest and lowest 2030 national-average WUE values proposed in the reviewed studies are 2.5 l/kWh (96,97) and 0.22 l/kWh (38), respectively. The former assumes mostly air-cooled racks with evaporative heat rejection in 2030 while the latter considers substantial shifts to advanced liquid-cooled racks by 2030. Both values refer to water consumption in the original sources.

Figure 9b applies these two extremes in the illustrative case, demonstrating how different combinations of chosen IT electricity demand projections and WUE values can lead to onsite water consumption quantity estimates (1.5 trillion liters vs. 66 billion liters) that differ by a factor of more than 20 for the same region and time horizon. Whatever the analyst's choices may be, providing transparency, offering scientific justification, and ensuring WUE representativeness are crucial for providing stakeholders with needed context for interpreting onsite water use results.

Next, the highest and lowest 2030 U.S. national average PUE values found in the reviewed studies are 1.75 (96,97) and 1.04 (38), respectively. Importantly, these PUE values are presented as congruent with the highest and lowest WUE values from these same original sources, making them reasonable selections in the illustrative case. The combined effects of PUE and EWIF choices on the indirect water use estimate are discussed in the next section.

**Table 4: Common variations in WUE and PUE measurement standards**

| Metric | Variant | Numerator | Considerations | Denominator | Considerations |
|---|---|---|---|---|---|
| WUE | Green Grid site WUE (2011) (77) | Annual site water usage (liters) | The original proposal of the WUE metric by the Green Grid (Azevedo et al. 2011), this document defines WUE as “a site-based metric that is an assessment of the water used on-site for operation of the data center. This includes water used for humidification and water evaporated on-site for energy production or cooling of the data center and its support systems (similar to carbon Scope 1).” It further defines onsite water use as including “humidification” and “water consumed for cooling the data center or data center–associated power-generating equipment (including cooling tower evaporation, blowdown, and drift).” The inclusion of blowdown discharges implies that water usage refers to data center water inputs rather than the strict definition of consumption, which is the difference between inputs and discharges. Thus, the analyst should seek to understand how “usage” was defined when using any WUE values that are based on the Green Grid 2011 definition. | IT equipment energy (kWh) | The document states “Like PUE and CUE, the WUE metric uses the familiar value of IT Equipment Energy as its denominator. Once determined for PUE or CUE, the same value should be used as the denominator for this new metric as well.” Thus, the analyst should seek to determine whether a Level 1, 2, or 3 measurement is associated with any WUE values that are based on the Green Grid 2011 definition for mathematical consistency with chosen PUE values. See PUE considerations below. |
| | Green Grid WUE source (2011) (77) | Annual site water usage + annual source water usage (liters) | The Green Grid WUEsource metric further considers the indirect water usage of power generation in the numerator, which is derived from the data center’s PUE value and the energy water intensity factor (EWIF) of its utility grid. Because a WUEsource value includes indirect water use, it should never be used by the analyst to represent onsite water intensity alone. However, it can be used by the analyst to estimate total data center water usage if the calculation methodology is understood. | | |
| | ISO/IEC 30134-9 (2022) - Formula 1 (WUE) and 2 (Uw) (78) | Uw = Iw - Ow (m3) | Formula 2 defines annual water use (Uw) as the difference between “total water input from outside the data centre boundaries (Iw)” and “total water returned out of the data centre boundaries (Ow).” Thus, the formula 2 construction is based on the strict definition of water consumption, and is not valid for estimating water input intensities unless discharges are zero. | $E_{IT}$ (MWh) | The variable EIT is defined as “IT equipment energy consumption (annual) in MWh.” A related standard for PUE (ISO/IEC 30134-2:2016) states that EIT can be determined based on Level 1, Level 2, or Level 3 measurements. See PUE considerations below. Thus, the analyst should seek to determine whether a Level 1, 2, or 3 measurement is associated with any WUE values that are based on the Formula 1 definition for mathematical consistency with chosen PUE values. |
| | ISO/IEC 30134-9 (2022) - Category 1 ($WUE_1$) (78) | Uw = Iw (m3) | A Category 1 WUE defines water use (Uw) as the “physical water input of the data center.” The document further states that “WUE1 requires only basic measurements for the water input” and that “measurements for the water output are not necessary.” Thus, a Category 1 WUE can only be used definitively by the analyst to quantify water input intensity. For use as a water consumption intensity, the analyst must verify that WUE1 water outputs (i.e., discharges) are zero. | | |
| | ISO/IEC 30134-9 | Uw = Iw − | A Category 2 WUE defines water use (Uw) as the “physical water input of the | | |

| Metric | Variant | Numerator | Considerations | Denominator | Considerations |
|---|---|---|---|---|---|
| | (2022) - Category 2 ($WUE_2$) (78) | Ur,w,NI (m3) | data center (Iw)" minus "water output of the DC for non-industrial reuse (Ur,w,NI)." The document further states that WUE2 requires "basic measurements for the water input" and that "the non-industrial water reuse shall be measured." In order to definitively use a Category 2 WUE to quantify water consumption intensity, the analyst should seek to determine if all water discharges are categorized as non-industrial water use. If this is not possible, the analyst risks overestimation of the water consumption intensity when using WUE2 values when non-industrial reuse is a fraction of total water discharges. | | |
| | ISO/IEC 30134-9 (2022) - Category 3 ($WUE_3$) (78) | Uw = (Iw − Iw,rw + EDC × fEWI) − (Ur,w,I + Ur,w,IN) (m3) | A Category 3 WUE is the most comprehensive, as it includes indirect water use in power generation as calculated by multiplying total data center electricity use (EDC) by an energy water intensity factor (fEWI). It further subtracts from water inputs (Iw) any "water input from rainwater (Iw,rw)" and water outputs from the data center for non-industrial (Ur,W, NI) and industrial (Ur,W,I) reuse. Because a WUE3 value includes indirect water use, it should never be used by the analyst to represent onsite water intensity alone. However, it can be used by the analyst to estimate total data center water usage if the calculation methodology is understood. | | |
| PUE | ISO/IEC 30134-2 (2016) - Category 1 (80) | EDC (kWh) | The variable is defined as "the total data centre energy consumption (annual), in kWh." It should be "measured at or near the data centre's utility meter(s) to accurately reflect the energy entering the data centre." The value and measurement protocol for EDC is the same irrespective of PUE category. | $E_{IT}$ (kWh) | The variable EIT is defined as "the IT equipment energy consumption (annual), in kWh." A Category 1 EIT value is measured at the uninterruptible power supply (UPS) output. This means that, in addition to IT electricity load, a Category 1 EIT measurement includes electricity losses associated with power distribution unit (PDU) transformers, static switches, electrical distribution components and non-IT related devices. |
| | ISO/IEC 30134-2 (2016) - Category 2 (80) | | | | A Category 2 EIT value is measured at the PDU output. This means that, in addition to IT electricity load, a Category 2 EIT measurement includes electricity losses associated with electrical distribution components and non-IT related devices. |
| | ISO/IEC 30134-2 | | | | A Category 3 EIT value is measured at the IT equipment input. Therefore, a Category 3 EIT measurement includes |

| Metric | Variant | Numerator | Considerations | Denominator | Considerations |
|---|---|---|---|---|---|
| | (2016) - Category 3<br>(80) | | | | only IT electricity use. |

## 3.4 Locking in the “thirstiness” of the power grid

As discussed in Section 2.2, large variances in EWIF values exist across the literature depending on the power generation technologies included, whether withdrawals or consumption are considered, the ages and types of data available, the water accounting principles employed (particularly for hydroelectric power), and other methodological differences. Thus, EWIF values must be chosen and applied with care, especially since indirect water use typically dominates data center water use estimates (17).

First, when striving for ***representativeness*** of EWIF values, the analysts should choose values that are consistent with the water use component(s) under study (i.e., withdrawals and/or consumption) and the observed and expected power mixes, power plant cooling technologies, and their changes for the region(s) and time period(s) under study. This may include consideration of trends toward onsite generation technologies at large data centers, where EWIF values may differ significantly from those of local utility grids (20,89).

However, the age of available EWIF values can often be a barrier to temporal representativeness (see Section S8). For example, U.S. data center water use projections to 2030 by Jha et al. (2025) relied on an EWIF value from Shehabi et al. (2016), which is based on Torcellini et al. (2003), which itself is based on data from 1995-2000, which is up to three decades older than the Jha et al. (2025) study time horizon (96–98). The 2022 ISO WUE standard 30134-3 (78), another potential source of EWIF data, provides example EWIF values for different generation technologies that were published in 2002 (99).

Even when EWIF values are reasonably representative, the analyst must sometimes choose between available EWIF values that differ substantially based on their underlying methods, data sources, and water accounting assumptions. For example, Figure 10 compares three different U.S. EWIF data sources at the eGrid subregion grid level (100), any of which could be chosen for a U.S. study. Figure 10 demonstrates the substantial differences in available EWIF values that can occur due to differences in indirect system boundaries (e.g., POG vs. upstream water consumption) and allocation and evaporation accounting decisions for hydroelectricity (18,87,90). Specifically, the EWIFs from Reig et al. (90) utilize a gross evaporation basis with all evaporation allocated to hydropower, whereas the EWIFs from Peer et al. (18) and Lee et al. (87) utilize a net evaporation basis but with differing allocation procedures. These accounting conventions play out differently in different eGrid subregions, with smaller differences in some and much greater differences in others.

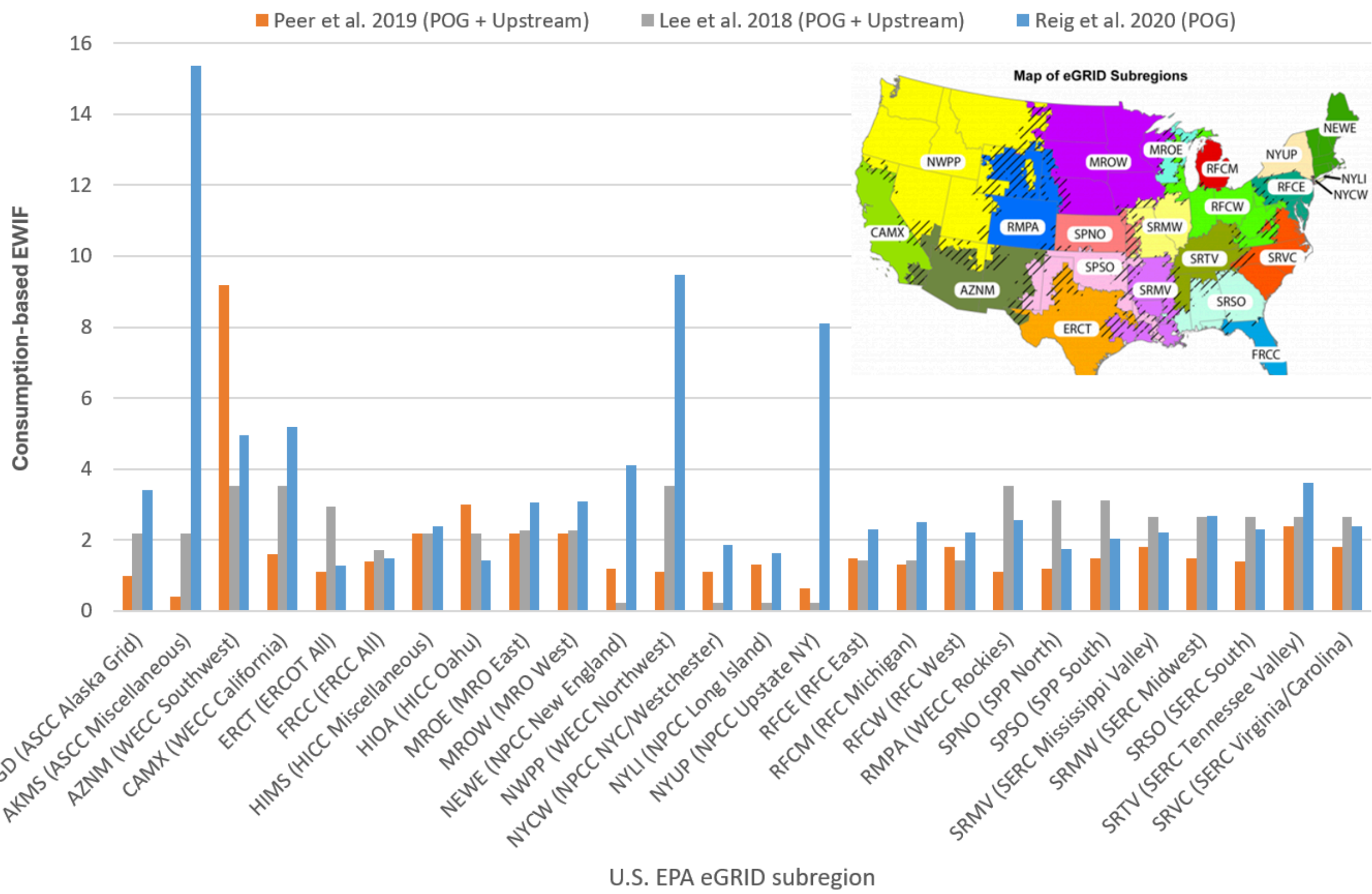

**Figure 10:** Differences in consumption-based EWIF values for the same U.S. grid subregions from three different published sources (18,87,90).

Thus, when providing ***transparency***, the analyst should ideally disclose which EWIF values are chosen, which power grid mixes and locations they reflect, the age of the data (which may be older than the age of the studies from which they were sourced), and whether the EWIF values reflect water withdrawals, water consumption, and POG and/or upstream accounting scopes.

Additionally, given the influence of allocation and evaporation accounting decisions for hydroelectricity EWIFs, the analyst should further disclose the ***water accounting bases (BP6)*** used for hydroelectric power when it is present in the study's grid mix(es) for maximum transparency (see Section 2.23). Such a declaration will ensure that both the analyst and their stakeholders are aware of how these accounting choices affect both the scale of final indirect water estimates and their comparability to other studies that might use different hydroelectric water accounting bases.

When providing ***scientific justification*** for chosen EWIF values, the analyst should discuss why certain values were chosen compared to others available in the literature, especially when it comes to the ages of the selected data and the choice of hydroelectricity EWIFs.  Further, the analyst should explain any study limitations that may be related to data gaps, assumptions, or unavoidable mathematical inconsistencies in Eq. 1 and convince themselves and their stakeholders why results are still useful despite such limitations.

### 3.41 Illustrative case: EWIFs and a 24-fold difference in outcomes

For the United States, the highest and lowest assumed EWIF values found in the reviewed studies are 7.6 l/kWh (96,97) and 1.11 (101) respectively. Both values are national average, consumption-based EWIFs. The former is a projected 2030 national value. The latter is a 2026 value that excludes hydroelectric power, and is extended to 2030 as a representative lower extreme in the illustrative example.

Per Equation 1, the product of EWIF and PUE drive indirect water use estimation. Therefore, these options for high and low EWIF values are multiplied by the options for high and low PUE values (Section 3.4) in Figure 9c, leading to products of 13.3 and 1.15, respectively, in the illustrative case.

With this final step, the illustrative case is completed. In Figure 9c it can be seen that a roughly 24-fold difference exists between the 2030 water consumption estimates that are theoretically possible in this illustrative case, based on observed assumptions in U.S. studies.

While the analyst ultimately decides which path to take, adherence to best practices will ensure diligence in documenting, justifying, and communicating why a particular path was taken. In turn, this will enable stakeholders to better assess the methods and quality of studies and to better compare results and approaches across studies.

## 3.5 Embracing, not ignoring, uncertainties

Given the many spatial, temporal, and technological variabilities, data gaps, and potential data inconsistencies that can be encountered, it is clear that data center water use estimation is not an exact science. Therefore, analysts should strive to ***quantify the uncertainties (BP7)*** inherent in their water use results, rather than simply presenting point estimates. Uncertainty analysis can involve generating results ranges using parametric uncertainty analysis, conducting sensitivity analysis to identify the most important assumptions, and/or constructing multiple scenarios to explore different outcomes based on different assumption sets.

Besides signaling to stakeholders that all estimates come with uncertainties, which is important for proper interpretation, uncertainty analysis can help identify insights that are robust even in light of uncertainties. For example, by identifying the actions that can help reduce water use (e.g., higher rack set points) across all scenarios even if the total scenario quantities are uncertain. Embracing uncertainty is also crucial for building research communities and advancing the state of science, by shining a light on where data and methods need to be improved to reduce uncertainties, which can inform research agendas moving forward.

At bare minimum, uncertainties should always be discussed qualitatively to signal to readers that water use estimation is not a precise science.

## 3.6 Building community through open science

To build a more cohesive and collaborative research community, data center water use analysts must commit to some basic tenets of open science (33,102). First, all methods and results must be ***fully replicable (BP9)*** by others. This means that all input data, mathematical equations, and supporting calculations should be made clear and available to stakeholders. Without this, the credibility and scientific utility of studies are diminished.

When it comes to data center water use estimation, a key component of scientific openness and replicability is reporting results with ***sufficient granularity (BP10)***. This means that when a study includes different water use components, stages, regions, and/or data center types, the analysts should report separate results for each category. For example, when both direct and indirect water use are estimated, they should be reported separately. Similarly, when traditional data centers and AI data centers are assessed, their water use should also be reported separately. Doing so will maximize the utility and replicability of studies far more than when only highly-aggregated results are reported. Moreover, reporting with maximum granularity will enable comparisons between studies that share common categories, but may differ in their overall scopes.

All studies should be ***peer reviewed (BP11)*** by independent experts to help ensure study clarity, scientific quality, and stakeholder utility.

Finally, due to the many uncertainties, variabilities, and data gaps that have been discussed, analysts should always ***acknowledge and discuss the key limitations (BP12)*** of their data and methods. They should also offer meaningful ***suggestions for future work (BP13)*** that would address limitations, expand study scopes, and leverage results in novel ways. These reflective and constructive steps are fundamental to the scientific method and will help the domain advance much more quickly that it can without them.

## 3.7 Comparing apples to apples

When communicating environmental results to non-experts, many analysts choose to make comparisons to everyday activities. For example, the U.S. Environmental Projection Agency offers equivalency factors used by many energy analysts to convert quantities of greenhouse gases into activities that anyone can understand, such as the equivalent number of cars driven for one year (103). Google provides one example related to data center water use by equating their reported onsite water consumption to an equivalent number of irrigated golf courses (104).

When data center analysts choose to use such analogies, it is important to use ***valid comparisons (BP14)*** so as not to mislead or confuse stakeholders about the scales of estimated water use. This means being careful to compare withdrawals to withdrawal activities, consumption to consumption activities, and so on. For example, the average U.S. home requires 138 gallons of water inputs per day (522 liters) for indoor use (68), but much of this quantity is later discharged into water treatment from sinks, showers, and toilets (105). Therefore, comparing the onsite water consumption of U.S. data centers to an equivalent number of U.S. homes assuming 522 liters per day is misleading, since the latter refers to water inputs, not water consumption. The number of equivalent homes on a consumption basis would be much higher than on an input basis. Given how important such equivalencies are for communicating study results in the media, and to policymakers and the public, it is critical to get them right.

# 4. Assessing the current state of practice

This section assesses how well the 31 studies published to date have adhered to the proposed best practices in Table 3. The goal is to identify collective strengths that can be replicated across future studies as well as needed improvements to help inform research community priorities moving forward. To do so, a three-point scoring rubric was developed to assess alignment with each best practice. Scores of 3, 2, and 1 correspond to strong, moderate, and weak/no alignment, respectively. The scoring rubric was designed to be as objective and replicable as possible. The full rubric can be found in Table S3, whereas the rationale for each score assigned to a study can be reviewed in the Supplementary Information Excel file. Assessment results are summarized via a heat map in Figure 11 with studies arranged in chronological order from top to bottom.

| | Best practices alignment score | | | | | | | | | | | | | | | | | | | | | |
|---|---|---|---|---|---|---|---|---|---|---|---|---|---|---|---|---|---|---|---|---|---|---|
| | | | | Scientific due diligence | | | | | | | | | | Scientific quality & replicability | | | | | | | | |
| | Clarity | | | Transparency | | | | | Justification | | | | | Represent-ativeness | | | | | | Knowledge building | | |
| | Scope | Definitions | Terms | Demand | WUE | PUE | EWIF | Hydro | Demand | WUE | PUE | EWIF | Uncertainty | WUE | PUE | EWIF | Replicability | Granularity | Peer review | Limitations | Future work | Comparisons |
| BP# | 1 | 2 | 3 | 4a | 4b | 4c | 4d | 6 | 5a | 5b | 5c | 5d | 7 | 8a | 8b | 8c | 9 | 10 | 11 | 12 | 13 | 14 |
| 2016-1 | 3 | 2 | 2 | 3 | 1 | 2 | 2 | 1 | 3 | 1 | 2 | 1 | 2 | 1 | 2 | 1 | 1 | 2 | 3 | 1 | 1 | N/A |
| 2021-1 | 1 | 1 | 1 | 1 | 2 | 1 | 2 | 1 | 1 | 1 | 1 | 1 | 3 | 1 | 1 | 2 | 2 | 1 | 3 | 3 | 1 | 1 |
| 2021-2 | 3 | 1 | 1 | 2 | 2 | 2 | 3 | 2 | 1 | 2 | 1 | 3 | 1 | 1 | 1 | 3 | 2 | 3 | 3 | 1 | 1 | N/A |
| 2023-1 | 3 | 1 | 1 | 2 | 1 | 1 | 3 | 1 | 1 | 1 | 1 | 2 | 3 | 1 | 1 | 1 | 1 | 2 | 3 | 1 | 1 | 1 |
| 2024-1 | 1 | 1 | 1 | 1 | 1 | 1 | 1 | 1 | 1 | 1 | 1 | 1 | 1 | 1 | 1 | 1 | 3 | 1 | 1 | 1 | 1 | 1 |
| 2024-2 | 3 | 2 | 3 | 3 | 2 | 2 | 3 | 1 | 3 | 1 | 1 | 2 | 1 | 1 | 1 | 2 | 2 | 1 | 3 | 1 | 2 | 2 |
| 2024-3 | 3 | 3 | 1 | 3 | 3 | 3 | 3 | 1 | 3 | 3 | 3 | 2 | 2 | 3 | 3 | 2 | 3 | 2 | 3 | 1 | 2 | N/A |
| 2025-1 | 2 | 1 | 3 | 2 | 1 | 1 | 3 | 1 | 3 | 1 | 1 | 1 | 2 | 1 | 1 | 1 | 2 | 1 | 3 | 1 | 1 | 1 |
| 2025-2 | 3 | 3 | 3 | 3 | 3 | 3 | 3 | 1 | 2 | 2 | 2 | 3 | 3 | 2 | 1 | 2 | 3 | 3 | 1 | 2 | 1 | 3 |
| 2025-3 | 3 | 3 | 2 | 3 | 2 | 3 | 1 | 1 | 3 | 1 | 2 | 1 | 1 | 1 | 2 | 1 | 1 | 3 | 3 | 1 | 1 | 1 |
| 2025-4 | 3 | 1 | 2 | 1 | 3 | 3 | 2 | 1 | 3 | 1 | 2 | 1 | 3 | 1 | 1 | 1 | 1 | 3 | 3 | 1 | 1 | N/A |
| 2025-5 | 2 | 1 | 1 | 1 | 1 | 1 | 1 | 1 | 3 | 1 | 1 | 1 | 3 | 1 | 1 | 1 | 1 | 2 | 1 | 1 | 1 | N/A |
| 2025-6 | 3 | 3 | 1 | 3 | 3 | 3 | 3 | 1 | 1 | 1 | 1 | 1 | 3 | 1 | 1 | 1 | 3 | 1 | 3 | 2 | 3 | N/A |
| 2025-7 | 3 | 1 | 1 | 2 | 1 | 1 | 1 | 1 | 1 | 1 | 1 | 1 | 1 | 1 | 1 | 1 | 1 | 3 | 1 | 1 | 1 | 1 |
| 2025-8 | 2 | 1 | 1 | 3 | 3 | 3 | 3 | 1 | 3 | 3 | 3 | 3 | 3 | 3 | 3 | 3 | 3 | 1 | 3 | 3 | 3 | N/A |
| 2025-9 | 3 | 1 | 1 | 3 | 3 | 3 | N/A | N/A | 3 | 2 | 1 | N/A | 1 | 1 | 1 | N/A | 3 | 3 | 1 | 1 | 1 | 1 |
| 2025-10 | 3 | 3 | 3 | 3 | 3 | 3 | 3 | 1 | 3 | 2 | 2 | 3 | 3 | 1 | 1 | 1 | 3 | 3 | 1 | 2 | 1 | 1 |
| 2025-11 | 2 | 2 | 1 | 3 | 3 | 3 | 3 | 3 | 1 | 2 | 2 | 1 | 3 | 1 | 1 | 1 | 3 | 1 | 3 | 1 | 3 | N/A |
| 2026-1 | 3 | 3 | 3 | 3 | 3 | 3 | 3 | 1 | 1 | 3 | 1 | 2 | 1 | 1 | 2 | 1 | 3 | 3 | 3 | 3 | 1 | 3 |
| 2026-2 | 2 | 2 | 1 | 1 | 3 | 1 | 1 | 1 | 1 | 3 | 1 | 1 | 1 | 3 | 1 | 1 | 1 | 2 | 1 | 2 | 1 | 1 |
| 2026-3 | 3 | 3 | 2 | 2 | 2 | 1 | 3 | 1 | 2 | 1 | 1 | 1 | 3 | 2 | 1 | 2 | 2 | 3 | 1 | 2 | 1 | N/A |
| 2026-4 | 2 | 1 | 1 | 2 | 2 | 1 | 3 | 1 | 2 | 3 | 1 | 2 | 3 | 1 | 1 | 2 | 3 | 3 | 1 | 1 | 2 | 2 |
| 2026-5 | 3 | 3 | 3 | 3 | 3 | 3 | N/A | N/A | 2 | 3 | 2 | N/A | 3 | 3 | 2 | N/A | 3 | 3 | 1 | 2 | 2 | 3 |
| 2026-6 | 3 | 2 | 2 | 2 | 3 | 3 | 3 | 3 | 1 | 1 | 1 | 2 | 3 | 1 | 1 | 1 | 2 | 2 | 3 | 2 | 2 | 1 |
| 2026-7 | 1 | 1 | 1 | 3 | 3 | 1 | 1 | 1 | 1 | 2 | 1 | 2 | 2 | 1 | 1 | 1 | 2 | 2 | 1 | 3 | 1 | N/A |
| 2026-8 | 1 | 1 | 1 | 2 | 1 | 2 | 2 | 1 | 1 | 1 | 1 | 1 | 1 | 1 | 1 | 1 | 2 | 1 | 1 | 1 | 1 | 1 |
| 2026-9 | 3 | 3 | 1 | 2 | 3 | 3 | 2 | 2 | 2 | 2 | 2 | 1 | 3 | 1 | 1 | 1 | 3 | 3 | 1 | 3 | 3 | N/A |
| 2026-10 | 3 | 1 | 3 | 2 | 1 | 2 | 1 | 1 | 1 | 1 | 1 | 1 | 2 | 1 | 2 | 1 | 1 | 2 | 3 | 3 | 3 | N/A |
| 2026-11 | 3 | 3 | 3 | 1 | 1 | N/A | N/A | N/A | 1 | 1 | N/A | N/A | 1 | 1 | N/A | N/A | 1 | 3 | 1 | 2 | 2 | 3 |
| 2026-12 | 3 | 3 | 1 | 3 | N/A | 1 | 1 | 1 | 1 | N/A | 1 | 2 | 3 | N/A | 1 | 2 | 2 | 3 | 3 | 3 | 1 | 2 |
| 2026-13 | 3 | 1 | 3 | 1 | 1 | N/A | N/A | N/A | 1 | 1 | N/A | N/A | 1 | 1 | N/A | N/A | 1 | 3 | 1 | 1 | 1 | N/A |
| AVG | 2.5 | 1.9 | 1.7 | 2.2 | 1.8 | 2.1 | 2.2 | 1.2 | 1.8 | 1.6 | 1.4 | 1.6 | 2.1 | 1.3 | 1.3 | 1.4 | 2.1 | 2.2 | 2.0 | 1.7 | 1.5 | 1.6 |
| | 2.05 | | | 1.91 | | | | | 1.61 | | | | | 1.35 | | | | | | 1.61 | | |

**Figure 11:** Summary of alignment scores by best practice, where strong alignment (green) is indicated by a 3, moderate alignment (yellow) is indicated by a 2, and weak/no alignment (red) is indicated by a 2, with averages presented by column and category.

## 4.1 Patterns, strengths, and weaknesses

Three broad conclusions can be drawn from observing the overall color patterns in Figure 11. First, organic growth in the adoption of best practices does not seem to be evident across the board. If increased adoption were occurring naturally over time, it would appear as clear shifts away from red/yellow, and toward green, in any given column from top to bottom. Second, while some studies demonstrated more best practices adherence than others, every study showed at least some opportunities for improvement. Third, as a whole, studies showed higher adherence to best practices related to clarity, transparency, and replicability and lower adherence to best practices related to scientific justification, use of representative data, community knowledge building, and communications categories. However, even within higher-performing categories, assessment results were uneven. These three patterns suggest a need for deliberate efforts by analysts and the broader research community to standardize and reinforce best practices to more quickly advance the field, especially in underperforming categories.

### 4.11 Clear scopes but fuzzy terms?

In the clarity category, while most studies provided explicit (68%) or adequate (19%) descriptions of their included water use stages and components (BP1), fewer (35%) offered explicit and unambiguous definitions of the included water use component(s) (BP2) and even fewer (29%) carefully avoided the use of competing, imprecise terms (BP3) to minimize potential misinterpretations and misuses of study results. In fact, Figure 12 shows the frequency of competing terms such as water "footprint," "demand," and "usage" exceeded that of the more precise terms "withdrawal" and "consumption" by a factor of nearly 1.5 within the body of assessed studies (see Section S6 for further details).

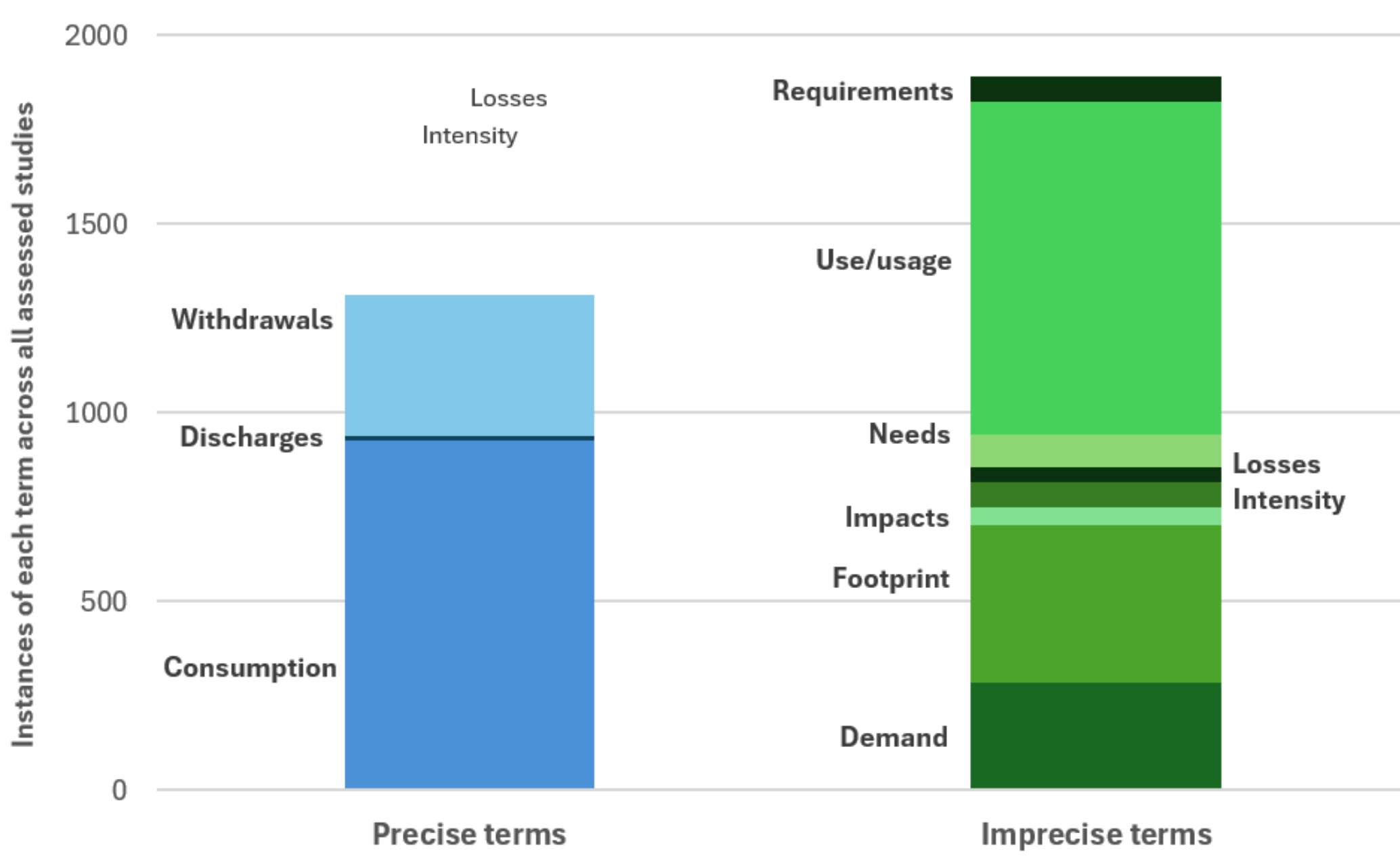

**Figure 12:** Instances of precise and imprecise water terms across the assessed studies

## 4.12 Mostly transparent, but not always

In the transparency category (BP4), well over half of the studies either explicitly or indirectly provided their assumed values for electricity demand (77%), WUE (67%), PUE (62%), and EWIF (70%). In stark contrast, however, the vast majority of studies (85%) that included indirect water use estimates neither acknowledged the presence of, nor explained the evaporation accounting bases associated with, hydroelectric power that may contribute to their assumed EWIF values (BP6). Only one study (106) explicitly declared the absence of hydroelectricity in its assumed power mix and only three studies (15,101,107) explicitly discussed the allocation approach used to assign reservoir evaporation to hydroelectric power. However, two of the latter three studies didn't elaborate on whether that evaporation was calculated on a gross or net basis. In fact, hydroelectric power transparency (BP6) received the lowest alignment score of any best practice, with a mean score of only 1.2. Given how hydroelectric power EWIFs can substantially affect indirect water estimates (see Section 2.2.3), increased compliance with BP6 should be a key community priority.

## 4.13 Uncertainty and replicability

Regarding uncertainty (BP7), around one-half (48%) of the assessed studies tackled uncertainty head on by providing results ranges or scenarios, while a smaller number (16%) mentioned uncertainty qualitatively but still provided only point estimates as results. This leaves around one-third (36%) of studies that neither analyzed nor discussed the substantial uncertainties associated with data center water use estimation, as evident in Figure 11.

In the replicability category (BP9), less than half (39%) of the assessed studies provided the necessary input data, equations, and/or access to modeling code needed for easy results replication, whereas around one-third (29%) should be technically, although not necessarily easily, replicable based on presented information. Therefore, roughly four-fifths of the studies should be replicable, although ease of replication should be a future priority.

Only around half (48%) of the assessed studies presented results with full granularity (BP10), meaning that a full half of the studies may be difficult to fully interpret and/or compare to other studies simply due to overly-aggregated results presentation. As such, this lack of granularity can be a barrier to cross-learning and cross-leveraging among the studies, which hampers scientific progress.

## 4.14 Peer review versus the need for speed?

In the peer review category (BP11), alignment results were mixed. On the positive side, 12 of 18 studies that appeared prior to 2026 were peer reviewed, indicating a reasonable commitment to independent review among the growing number of data center water analysts. However, of the 13 studies that appeared in 2026, only four (27,107–109) have been peer reviewed while two (83,94) are preprint articles concurrently under peer review. It should be noted that although the results of preprint articles can change due to peer review, preprint versions can sometimes be cited heavily by both researchers and the media in the meantime, due to high interest in data center water use.

Preprints are also becoming increasingly common due to slow journal peer review processes and the desire to release results on fast-moving topics for which they can be helpful. On the precautionary side, preprints risk “locking in” early, but perhaps not final or robust, estimates in the public consciousness on a very “hot” topic. Therefore, strategies for balancing the scientific benefits of preprints against the need for rigorously-scientific clarity among stakeholders on data center water use should be a key community priority moving forward, as should the promotion of peer review as a best practice standard. In particular, many of the non peer-reviewed studies in Table S1 were published as reports or articles by consultancies, think tanks, and research institutions. While peer review may delay the release of such publications, evidence of peer-reviewed reports in Table S1 (3,92,97,106,109) suggests that independent peer review is not impossible, but must be pursued deliberately.

## 4.15 More justification needed

While most studies were reasonably transparent (BP4) about their chosen electricity demand, WUE, EWIF, and PUE value(s), far fewer studies accompanied these disclosures with meaningful scientific justifications (BP5). In fact, more than one-half of the assessed studies provided no justification for their chosen electricity demand (52%) or WUE (57%) values, while around two-thirds provided no justification for their chosen PUE (66%) or EWIF (56%) values. Instead, these studies tended to merely state the values and sources for their assumptions and move on without any discussion to convince stakeholders that those choices were appropriate and defensible.

## 4.16 Data representativeness: the weakest link?

The general lack of scientific justification becomes more problematic when the representativeness (BP8) of the chosen Equation 1 input values is considered. More than three-quarters of the assessed studies utilized WUE (80%) or PUE (76%) values that were not strictly representative of the technologies, geographies, and/or time horizons associated with the studies or they did not present sufficient information for representativeness to be confidently established. Similarly, around two-thirds (67%) of the studies utilized EWIF values with low or unsubstantiated representativeness. After lack of clarity on hydroelectric power assumptions (BP6), the lack of representativeness (BP8) of chosen WUE, PUE, and/or EWIF values was the next most deficient area of best practice adherence identified in this review.

Importantly, more than one-third of direct water use studies strayed from using an ISO 30134-compliant WUE value (78) when estimating direct water use (see Table 4). Instead, such studies employed an onsite water intensity assumption, which was expressed in liters of water inputs per kWh of total data center electricity use. As shown and discussed in Section S7, these values can often be traced back to a 2014 onsite water intensity value introduced by Shehabi et al. (97), which used a similar convention. Doing so complicates comparability to other studies that use WUE values, given the difference in water intensity denominators. Doing so can also decouple a study's onsite water intensity assumptions from its PUE values–again, due to the difference in denominators–making it difficult or impossible to determine if a study's WUE and PUE assumptions are congruent.

Additionally, 40% of the studies that included indirect water use estimates didn't disclose their PUE values, which makes verification of congruence with any utilized WUE values impossible (see Section S7). Non-disclosure of PUE values also obscures assumptions related to the cooling technologies and efficiencies underlying the analyses, which is a barrier to stakeholder transparency.

The use of non-representative WUE and/or PUE values is undoubtedly related to a general lack of operator-reported data (27), and especially reported WUE and PUE values for different individual data center locations, cooling technology types, and climates. This lack of empirical data forces the analyst to make assumptions, either by choosing WUE and PUE values from previous studies or by simulating PUE and/or WUE from available models (12,13,31). As evident in the rationale for each representativeness score (see the Supplementary Information Excel file), many studies employed WUE and/or PUE values that were not strictly representative of the locations, potential cooling technologies, and time horizons under study. Additionally, in some studies, assumed WUE and PUE values remained static over multi-year time horizons, which can render them less representative with each passing year. However, nondisclosure of WUE and/or PUE sources and values is a practice that impedes transparency, and shouldn't be done even in light of persistent data gaps.

Notably, none of the assessed studies indicated the WUE or PUE measurement category associated with their assumed values (Table 4). As discussed in Sections 3.31 and 3.32, ensuring that measurement categories are consistent with other terms in Equation 1, or at least acknowledging when they aren't, are important for reducing errors and ensuring transparency.

Finally, as discussed in Section 2, most analysts must rely on publicly-available EWIF databases or studies for indirect water use estimates. However, the availability and age of published EWIF values can vary widely in the literature. In fact, the majority of low representativeness scores for EWIF assumptions related to the use of values that were outdated, were held static in forward-looking analyses, or both. As discussed further in Section S8, in some studies, the age of applied EWIF values could precede a given year within the analysis time period by decades. Therefore, advocating for, or generating, more timely, location-specific, and technology-specific EWIF values in future studies should be major priorities.

## 4.16 Comparisons: clear or confusing?

Regarding comparisons (BP14), only 18 of 31 studies (58%) included comparisons to equivalent water-using activities when communicating study results. However, only 7 of these 18 studies utilized comparisons that were made on reasonably "apples to apples" bases with sources cited for comparative data. Eleven of these 18 studies included comparisons that were technically incorrect (e.g., comparing consumption to withdrawals), vague (e.g., comparing to an equivalent number of Olympic swimming pools without stating their volume), and/or not associated with a source citation for independent verification.

### 4.17 Community building: a missed opportunity?

Finally, the heatmap in Figure 11 indicates low overall compliance with two best practices that are critical for scientific community building: clearly communicating study limitations (BP12) and providing suggestions for future work (BP13). This low compliance is evidence of missed opportunities for knowledge building within the broader research community. Only around one-quarter (23%) of the assessed studies meaningfully discussed the limitations of their analyses, which is important for informing readers about how results should and shouldn't be interpreted. Similarly, only 16% of the assessed studies offered meaningful and actionable suggestions for future work, as opposed to providing overly-general suggestions (19%) or no suggestions at all (65%).

# 5. Emerging analysis priorities

In addition to assessing alignment with the proposed best practices, this review also assessed coverage of selected emerging issues across the included studies. Emerging issues are defined as those that are relevant to broader impacts or scopes associated with data center water use quantities, and which may become important in future water use estimation studies. The goal was to quickly identify whether a given emerging issue was included within each study's scope, and, if so, whether it was quantitatively analyzed, only qualitatively discussed (even if briefly), or not included in the study's scope. This review considered the following emerging issues, which are further discussed below: water stress, wastewater discharges, peak water demand, water source types, and embodied water use. The results of this high-level evaluation are presented in Figure 13 and summarized in the Supplementary Information Excel file.

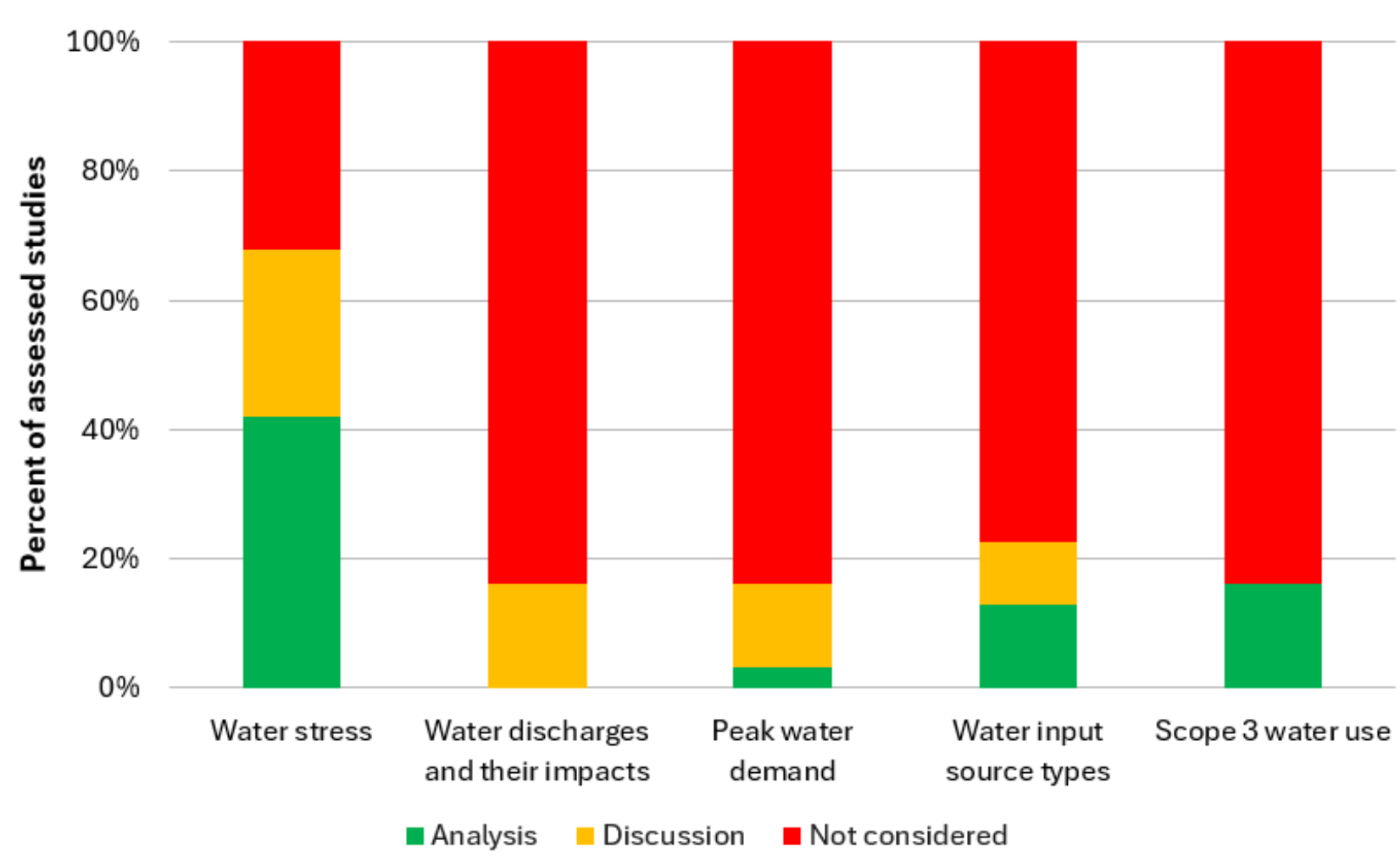

**Figure 13:** Coverage of emerging data center water issues in the assessed studies (n=31)

## 5.1 Quantifying water stress

This review focused on estimation of water use quantities alone, whether they be withdrawals, inputs, consumption, and/or discharges. However, the water stress and scarcity risks that those quantities pose to local watersheds can be very important, especially in water-stressed or water-constrained regions (15,110). Assessing water stress requires geographically locating estimated water use quantities, ideally at the local water supply level, which is a finer level of spatial resolution than was observed in most of the assessed studies. Still, two-fifths (40%) of the assessed studies presented some sort of water stress or scarcity analysis, whereas around one-quarter (26%) offered a qualitative discussion or at least a brief acknowledgement of water stress as a potential data center water use concern. Future water use studies should consider including water stress as an additional metric, given its relevance to understanding how water use quantities relate to potential water scarcity in very local contexts.

## 5.2 Quantifying water discharges and quality

As described in Section 2.12, many data centers can discharge wastewater, whether continuously due to blowdown/bleeding of evaporative cooling towers and adiabatic coolers or due to infrequent flushing and refilling of water-based “closed-loop” systems. Discharged water can contain concentrated minerals, water treatment chemicals, and biological contaminants that can require subsequent removal by onsite and/or offsite treatment systems (53,67). The presence and levels of such contaminants can vary based on data center designs, maintenance schedules, and operational parameters like cooling tower cycles of concentration (67).

As data centers grow in both numbers and size, the local environmental effects of such wastewater discharges may come under increasing scrutiny, as in recent news coverage of an Amazon data center in Oregon (111). Only one of the assessed studies quantified wastewater discharges (15), but none of these studies quantified wastewater discharge characteristics, downstream treatment needs, or potential ecological impacts. However, around one-sixth (17%) of the assessed studies offered at least a brief acknowledgement of discharge water quality as a potential data center water use issue and one that should be included in future studies.

## 5.3 Quantifying “peak” water demand

As shown in Figure 4, the direct water inputs of a data center can vary by hour, month, and season, with “peak” water inputs generally needed during the hottest hours of the year. Such peak water supply demand can often be substantially higher than a data center’s annual, monthly, or even daily average water supply demand. Understanding the timing and magnitude of such peaks can shed light on the potential water supply system stress posed by a data center, or on the need for onsite peak water or chilled water storage capacity as a mitigation approach (see Section 2.132), especially during hot hours when the supply system may already be oversubscribed (83). Understanding potential peaks can also shed light on the potential water supply infrastructure needs of future data center projects. However, only one of the assessed studies actually quantified peak water supply capacity needs (83), whereas four others (107,112–114) offered at least a brief acknowledgement of peak demand as a potential data center water use impact on local water supply systems.

## 5.4 Consideration of specific water source types

Data center water inputs generally originate from surface waters (i.e., rivers and lakes), underground aquifers, and reclaimed water sources (115). Studies that move beyond estimating withdrawal and input quantities alone by further associating those quantities with specific local sources can enable better assessments of local water stress, provide greater transparency to communities, and illuminate when potentially beneficial water reclamation sources are, or can be, utilized.  Similar to water stress (Section 5.1), analyzing specific water source dependencies typically requires geographically locating the water use quantities estimated by a study at the water supply system level.  Only four of the assessed studies (114,116–118) at least loosely connected water quantities with specific water sources, whereas another three (83,119,120) offered at least a brief acknowledgement of water supply source types as an important concept. Therefore, associating water withdrawal quantities with specific water sources remains an important opportunity area in future studies.

## 5.5 Embodied water use

As discussed in Section 2, embodied water use is included in the strict definition of the data center water footprint (16,27,92). However, it is rarely included in data center water use studies due to extensive requirements for data on the quantities, types, supply chains, and manufacturing water needs of myriad data center IT and infrastructure components (121).  As the water-intensive semiconductor hardware requirements of data centers grow over time, consideration of embodied water use may emerge as a key priority for future analysts.  To date, however, only five studies (16%) (92,108,112,117,122) have included quantitative information about embodied water use as part of broader direct and indirect water estimation studies, although one of these is limited to the “embodied” water in electricity for desalination of data center direct water use (108).

# 6. The path forward

The growing number of studies and the growing demand for water use estimates among the media, the public, and policymakers is creating a golden opportunity for the analyst community. Namely, this critical mass of interest among analysts and their stakeholders can help both accelerate and strengthen the scientific impacts of data center water use analysis as a bone fide field, and one with high scientific standards.  All of the proposed best practices have been exemplified by studies within the assessed collection; they are just not widespread or universal due to the nascence of the field.  However, all best practices are all achievable with concerted actions by analysts, their stakeholders, and policymakers.  Priority actions for each community are listed below.

## 6.1 Actions for analysts

1. Strengthen individual adherence to best practices, such as those proposed in this study. Particular attention should be paid to improving the representativeness of data inputs, providing meaningful scientific justification for analytical choices, and contributing to community building through meaningful discussions of limitations and actionable areas of future work.
2. Engage in research community building to “raise all boats,” including contributing to the development and dissemination of shared datasets, models, best practice standards, and scientific exchanges to propel the field forward. Importantly, identifying urgent data and standardization gaps is a pressing community priority, such as common standards for hydropower EWIF accounting bases.
3. Engage with stakeholders to better understand their uses of estimates and how analysts can better support their decision needs. Help educate them on the responsible interpretation and use of published estimates.
4. Consider analyzing emerging issues such as quantifications of water stress, peak water loads, water discharge quantities and quality, withdrawals from specific water sources, and embodied water use to broaden water use estimation studies to more decision-relevant impacts.

## 6.2 Actions for stakeholders

1. The analyst best practices here provide a recipe for high quality and transparent estimates. As such, they double as critical questions that any stakeholder should ask before using any given number in future research, outreach, or decision-making. Don’t use numbers blindly.
2. Engage with analysts to ask questions, understand their analyses, and to communicate needs for quality, transparency, and utility. This will help promote “virtuous cycles” of feedback to analysts to promote continuous improvement and greater adherence to best practices.
3. Practice responsible use of data center water use estimates. All estimates are confined to specific analysis scopes (direct, indirect, embodied), water use components (withdrawals, inputs, consumption, discharges), and regions and shouldn’t be generalized beyond those boundaries. Avoid imprecise terms when using estimates in your work and communications.

## 6.3 Actions for policymakers

1. There are many empirical data gaps that have led to the use of non-representive data in many studies. Such data gaps lead to uncertainties and blind spots in data center water use estimation. Improved and broadened reporting standards for data center operators coupled with public data portals can help, especially for the data and insights related to cooling technologies, total water withdrawals, inputs, consumption, and discharges and PUE and WUE at individual sites.

2. Public investments in shared data and modeling platforms can help ensure that all analysts use the best data and methods while being able to learn from, critique, and compare results from other researchers.
3. Scientific convenings can help build communities through knowledge sharing, network building, refinement and codification of best practice standards, and the development of research roadmaps that can identify multi-year research and public investment plans to improve datasets, models, and estimation techniques.
4. Policies should be based on the best science and numerical estimates. Actions 1-3 for stakeholders are critical for policymakers, too.

**Declaration of competing interest**

E.M. has no competing interests to declare.

**Acknowledgements**

E.M. gratefully acknowledges funding for this research from the ClimateWorks Foundation, the Mellichamp Academic Initiative in Sustainability at the University of California, Santa Barbara, and via the Energy Demand Changes Induced by Technological and Social Innovations (EDITS) network, an initiative coordinated by the Research Institute of Innovative Technology for the Earth (RITE) and the International Institute for Applied Systems Analysis (IIASA) and funded by the Ministry of Economy, Trade and Industry (METI), Japan. E.M also gratefully acknowledges data collection by Vrunda Tol and Karol Paya in the early stages of this study.

**Appendix. Supplementary materials**

Supplementary_Information.docx

Study_Assessments.xlsx

**Data availability**

All data associated with figures and study assessments are available as supplementary information.

# Estimating data center water use: Best practices and critical questions

Eric Masanet[1,2,*]

[1] Bren School of Environmental Science and Management, University of California, Santa Barbara, USA

[2] Department of Mechanical Engineering, University of California, Santa Barbara, USA

* emasanet@ucsb.edu

# Supplementary Information

# S1: Study selection

This review focuses on studies that present original estimates of data center water use at *macro-scales* (i.e., at global, regional, country, state, provincial, and/or city levels), whether in retrospective fashion, prospective fashion, or both. These are the types of studies that inform researchers, policymakers, the public, and the media about the scales, drivers, and trajectories of water use by data centers as a distinct sector.

Studies appearing in the literature as of September 15th, 2026, were included in this review. Studies posted to preprint servers were included, given that their results are often cited by both the media and other researchers prior to peer review. Also included are original research institution estimates that are published entirely in online articles, such as (1), which are also cited by various stakeholders.

The studies included in the best practice alignment assessment are summarized in Table S1. They are listed in chronological order based on the day, month, and year of publication. The Table S1 studies were identified through searches of scientific literature databases (Scopus and Google Scholar) and internet search engines using combinations of the following broad keywords: data center(s), data centre(s), datacenter(s), datacentre(s), and water. Forward- and reverse-citation searches of identified studies were also conducted to compile additional studies meeting this review's inclusion criteria.

The following types of studies were excluded from the best practice alignment assessment:

- studies that analyzed a single or small number of individual data centers, such as (2), (3), (4), or (5)
- studies focused solely on attributing water use to a specific digital application or a unit of computation, such as (6), (7), or (8)
- studies that review or present water evaluation methodologies without including macro-level water use estimates, such as (9), (10), (11), or (12), inclusive of studies that only estimate water usage effectiveness (WUE) (13,14);
- studies produced by market analysis firms or consultancies for which only sample or highlight results are made public and must be purchased for access to full data, methods, and results, such as (15), (16), or (17), and
- studies that simply repeated macro-level water use estimates sourced from other studies without presenting any original estimates of their own, such as (18) or (19).

However, excluded studies may have been cited elsewhere in the paper as relevant.

Further details on each of the studies in Table S1 can be found in the supplementary Excel file.

**Table S1:** Summary of assessed studies that were published as of September 15th, 2026

| Label | Citation | Title | Publication type | Reference |
|---|---|---|---|---|
| 2016-1 | Shehabi et al. (2016) | United States Data Center Energy Usage Report | Report | (20) |
| 2021-1 | Obringer et al. (2021) | The overlooked environmental footprint of increasing Internet use | Journal article | (7) |
| 2021-2 | Siddik et al. (2021) | The environmental footprint of data centers in the United States | Journal article | (21) |
| 2023-1 | Farfan and Lohrman (2023) | Gone with the clouds: Estimating the electricity and water footprint of digital data services in Europe | Journal article | (22) |
| 2024-1 | Tan, D., and C.T. Low (2024) | China ICT running dry? The rise of AI & climate risks amplify existing water risks faced by thirsty data centres | Report | (23) |
| 2024-2 | Jerleus et al. (2024) | Environmental footprints of the data center service sector in Sweden | Journal article | (24) |
| 2024-3 | Shehabi et al. (2024) | 2024 United States Data Center Energy Usage Report | Report | (25) |
| 2025-1 | Jha et al. (2025) | Forecasting US data center CO2 emissions using AI models: emissions reduction strategies and policy recommendations | Journal article | (26) |
| 2025-2 | Li et al. (2025)[1] | Making AI Less “Thirsty”: Uncovering and Addressing the SecretWater Footprint of AI Models | Preprint | (27) |
| 2025-3 | IEA (2025) | Energy and AI | Report | (29) |
| 2025-4 | Jiang et al. (2025) | Facility-level energy-driven water footprint and scarcity implications of Chinese data | Journal article | (30) |

[1] Li et al. (2025) originally appeared as a preprint (27) that was later published as a peer-reviewed journal article (28). Only the preprint version (dated March 26, 2026) contains the methodological details for the study’s national data center water use estimates, which was the subject of the best practice alignment assessment presented in this review.

| Label | Citation | Title | Publication type | Reference |
|---|---|---|---|---|
| | | centers: a bottom-up analysis and scenario-based projection | | |
| 2025-5 | Krishnamoorti, R., and A. Datta (2025) | The Future of the Electric Grid in Texas: Opportunities and Challenges in the Next Decade | Report | (31) |
| 2025-6 | Herrera et al. (2025) | Sustainable AI infrastructure: A scenario-based forecast of water footprint under uncertainty | Journal article | (32) |
| 2025-7 | James et al. (2025) | Drained by Data: The Cumulative Impact of Data Centers on Regional Water Stress | Report | (33) |
| 2025-8 | Xiao et al. (2025) | Environmental impact and net-zero pathways for sustainable artificial intelligence servers in the USA | Journal article | (34) |
| 2025-9 | Liu et al. (2025) | An Assessment of California Data Centers' Environmental and Public Health Impacts | Report | (12) |
| 2025-10 | Mandala (2025) | Data Centres as Enabling Infrastructure | Report | (35) |
| 2025-11 | Arzumanyan et al. (2025) | Water Use Requirements for Data Centers in Texas: A White Paper on the Evolving Demands of Water Use in Data Center Infrastructure in Texas | Report | (36) |
| 2026-1 | deVries (2026) | The carbon and water footprints of data centers and what this could mean for artificial intelligence | Journal article | (37) |
| 2026-2 | GWI and Xylem (2026) | Watering the New Economy: Managing the impacts of the AI revolution | Report | (38) |
| 2026-3 | Cook (2026) | Thirsty Data and the Lone Star State: The Impact of Data Center Growth on Texas' Water Supply | Report | (39) |
| 2026-4 | Henzl and McKenna (2026) | Data Center Water and Electricity Consumption in Nevada | Report | (40) |

| Label | Citation | Title | Publication type | Reference |
|---|---|---|---|---|
| 2026-5 | Han et al. (2026) | Small Bottle, Big Pipe: Quantifying and Addressing the Impact of Data Centers on PublicWater Systems | Preprint | (41) |
| 2026-6 | Liang (2026) | The cloud's thirst: Quantifying AI's water footprint and its impact on the water-energy nexus in São Paulo, Brazil | Journal article | (42) |
| 2026-7 | Attallah and Warner (2026) | Data Centers in Indiana and the Midwest: Assessment of Power and Water Demand in the AI Era | Conference Proceedings | (43) |
| 2026-8 | Aczel et al. (2026) | Environmental Cost of AI's Energy Use: Carbon, Water and Land Footprints | Report | (44) |
| 2026-9 | Guidi and Dominici (2026) | The hidden water geography of U.S. hyperscale data centers in the AI era | Preprint | (45) |
| 2026-10 | Awad et al. (2026) | Net sustainability assessment of AI-driven data centers in the GCC: an MCDA approach framework | Journal article | (46) |
| 2026-11 | Biedny and Luna (2026) | Data center water use and economic contributions in context: Benchmarking analysis across nine industrial sectors in the United States | Report | (47) |
| 2026-12 | James et al. (2026) | Water Behind the Watts: The Hidden Risk of Powering Data Centers | Report | (48) |
| 2026-13 | Khoi Le and Selvaraju (2026) | Tech Thirst: Data center water consumption could triple by 2030 without efficiency gains | Web article | (1) |

# S2: Study characteristics

Figure S1 summarizes key characteristics of the 31 assessed studies in Table S1.

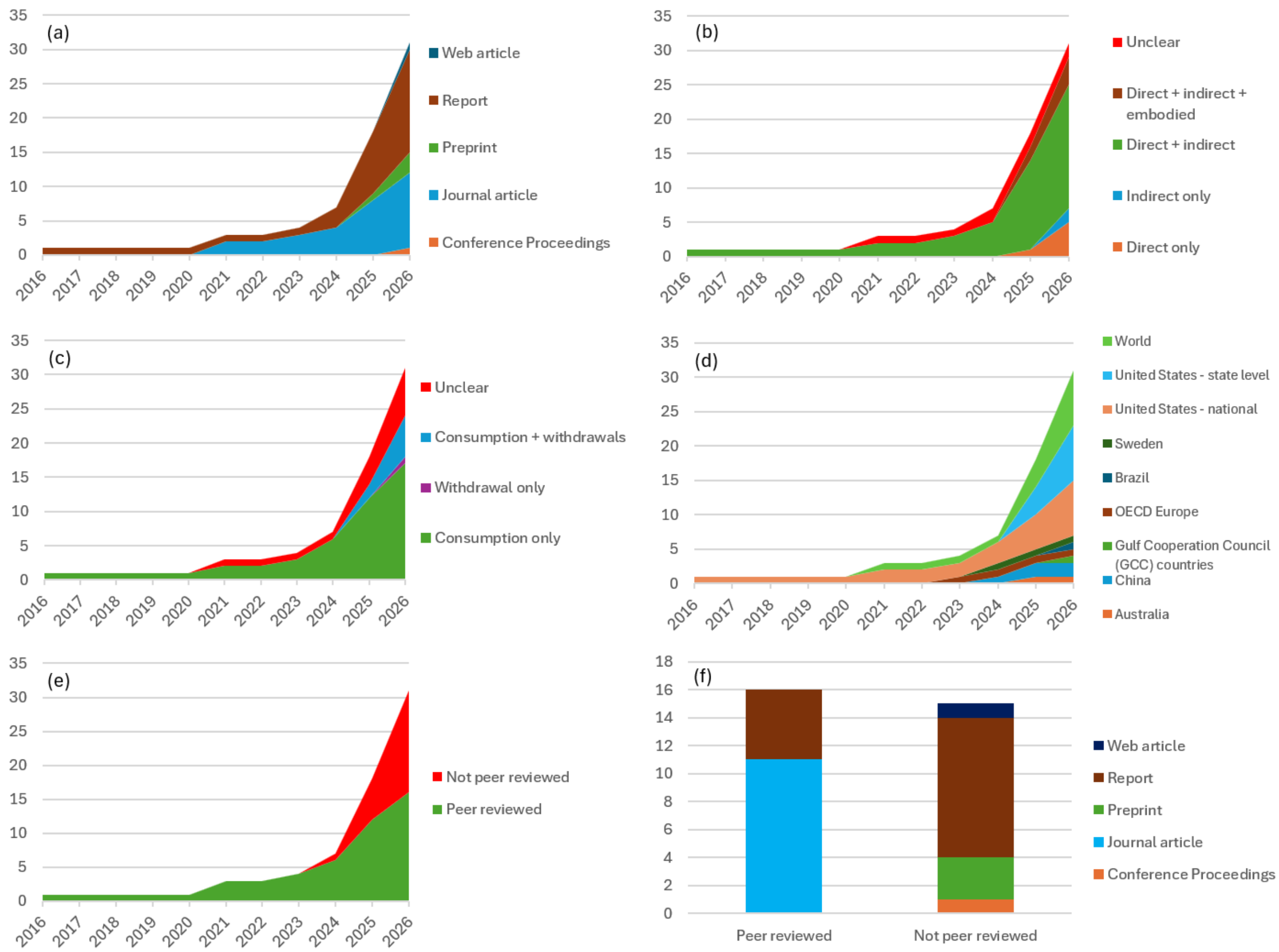

**Figure S1:** Cumulative number of studies according to: (a) type of publication; (b) water use stages analyzed; (c) water use components analyzed; (d) region of study; (e) peer review status; and (f) peer review status by publication type. Note: peer review status as of September 15th, 2026.

The number of macro-level studies has grown sharply since around 2023, coinciding with the “AI boom.” Notably, the number of journal articles and reports has increased dramatically since 2023 (Figure S1a), with the latter being published by a broad range of institutions, including research centers, think tanks, nonprofit organizations, and consulting firms.

The majority of the assessed studies included both direct and indirect water use in their analysis scopes (Figure S1b); the next most common scope was to quantify direct water use only. Of the assessed studies, most focused on water consumption, with a few studies including both water consumption and withdrawals; however, some studies did not explicitly declare the water use components considered (Figure S1c). Prior to 2023, the few published studies focused on either the United States at the national level or the world; since 2023, the number of regions of study has diversified substantially, including more studies that focus on specific U.S. states (Figure S1d).

Lastly, nearly half of the cumulative studies were not peer reviewed at the time of their public release (Figure S1e), with the majority of non peer-reviewed studies being reports issued by consulting firms, think tanks, and nonprofit organizations (Figure S1f).

# S3: History of news coverage

As one measure of media attention to the topic of data center water use, a Nexis Uni search was conducted with the following search parameters (49):

- **Search terms:** ("data center" OR "data centre" OR "data centers" OR "data centres" OR datacenter! OR datacentre!) AND water AND (use OR usage OR consumption OR withdrawal!)
- **Timeline:** January 1st, 2016 to June 30th, 2026
- **Source types:** Industry Trade Press, Web-based Publications, Newswires & Press Releases, Newspapers, News Transcripts, Magazines and Journals, Newsletters, Blogs, Aggregate News Sources
- **Search scope:** terms found anywhere in document

The goal of this search was not to establish the exact number of news stories in different media, but rather to establish rough senses of the scale of, and growth in, coverage in English-language news sources over the past decade. Figure S2 plots the number of “hits” based on these search results by source type. Source types are defined by Nexis Uni (49). Year 2026 data include all “hits” during the period 01-Jan-2026 to 30-Jun-2026.

Potential changes in Nexi Uni source type definitions and/or indexed media within each source type may occur over time. These changes may affect yearly trends in news item counts for each source type. Such changes were not investigated in this study. The annual source type totals in Figure S2 are solely intended to illustrate general trends toward more search term "hits" by year and source type; the quality and relevance of individual "hits" were not investigated in this study.

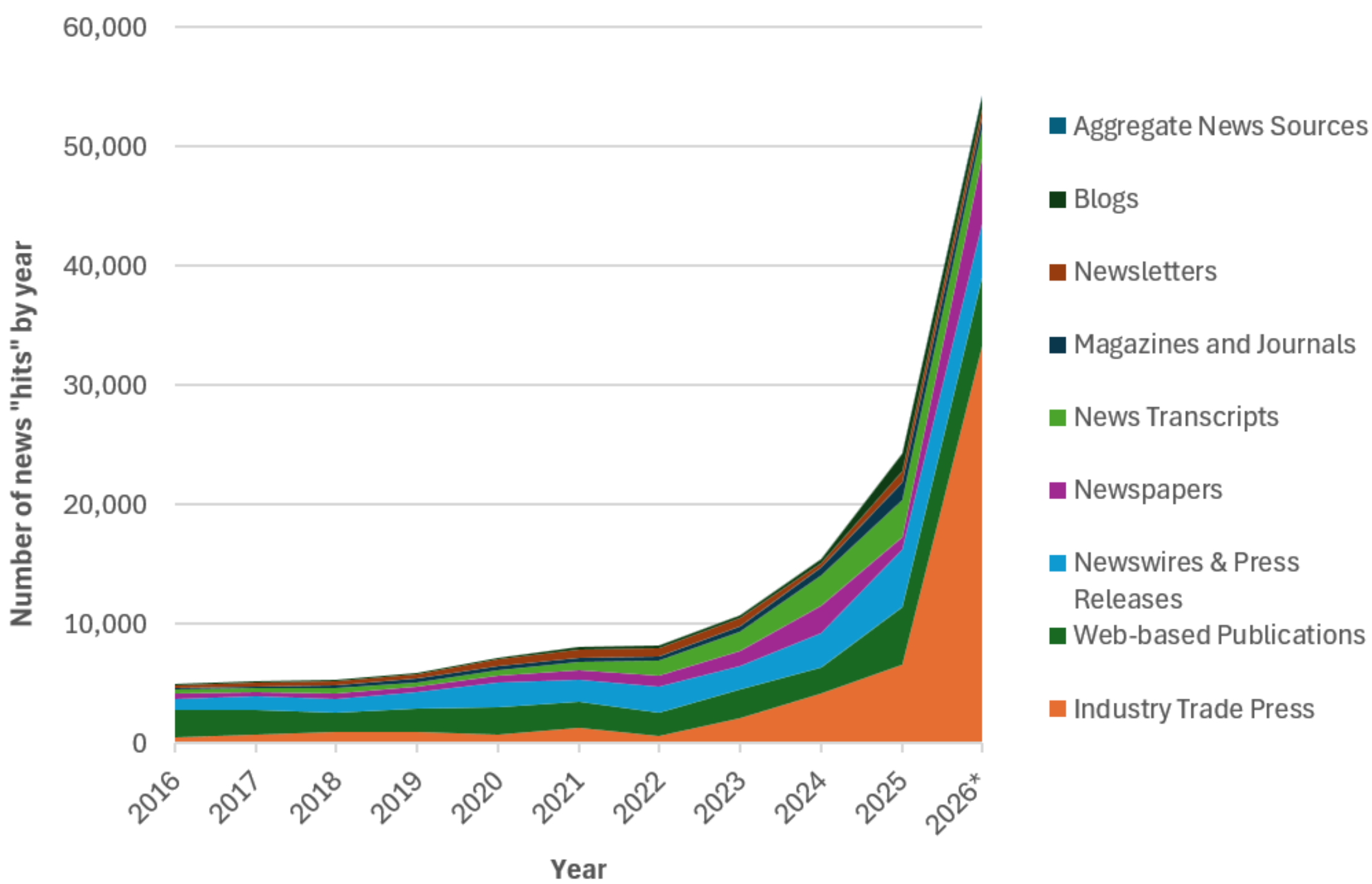

**Figure S2:** Number of “hits” by English language news source type and year. Year 2026 data refer to the period 01-Jan-2026 to 30-Jun-2026.

# S4: Follow the heat: cooling system variant schematics

To aid in visualizing how different cooling system configurations rely on water and electric power, simplified schematics of 14 important configurations that analysts will encounter are provided in Figures S4-S22. The depicted configurations are summarized in Table S3 for convenience. All figures share a common legend (Figure S3).

The simplified “follow the heat” schematics are meant to identify the key fluid loops involved, the heat exchanges between them, and how heat travels from the IT rack to outside of the data center. In reality, the detailed equipment designs for implementing these systems can be more complex.

The depicted configurations for liquid-cooled racks apply to direct-to-chip/cold plate, single-phase immersion, and two-phase immersion systems. Explanations for differences between these liquid-cooling approaches are provided under the figures.

The depicted configurations are not always mutually exclusive, since data centers often utilize multiple configurations synergistically. For example, a data center with air-cooled IT racks may utilize airside economizers during hours when exterior ambient conditions enable “free cooling” (Figure S14) and deploy “back-up” water-cooled mechanical chillers (Figure S4) during hours when free cooling isn’t available. In such cases, parallel configurations can also be thought of as different cooling “modes” within the same data center and multiple options from Figures S4-S22 will apply to the same site.

Similarly, some data centers can also use combinations of air-cooled racks and liquid-cooled racks, with shared or independent heat rejection technologies. In such cases, different configurations below may apply to different data halls or to different rack clusters within the same data center.

For simplicity, data center airflows related to data hall air change exhaust and makeup air intake (required at all data centers) are not depicted, nor are dedicated conventional air humidification devices within the data hall. Various pumps, fans, blowers, and other fluid movement equipment are omitted for simplicity. Interior power chain components, such as switchgear, transformers, uninterruptible power supplies, and power distribution units are also omitted for simplicity. However, power chain equipment also gives off heat that must be continuously removed from the data center in all cooling modes.

Direct-expansion cooling systems with computer room air conditioners are excluded, given that they are mostly used by small data centers. Emerging data center designs, such as orbital or undersea data centers, are also excluded due to their early development stage. Finally, the basic pathway from IT rack heat generation to ultimate outdoor heat rejection into an exterior heat sink is described after each figure so that readers can easily “follow the heat” as indicated by red flows in the diagrams.

**Table S2:** Summary of “follow the heat” cooling system configuration schematics

| Figure | Cooling system configuration |
|---|---|
| **S4** | Air-cooled racks with a water-cooled chiller and an open-loop cooling tower |
| **S5** | Liquid-cooled racks with a water-cooled chiller and an open-loop cooling tower |
| **S6** | Air-cooled racks with a water-cooled chiller and a closed-loop cooling tower |
| **S7** | Liquid-cooled racks with a water-cooled chiller and a closed-loop cooling tower |
| **S8** | Air-cooled racks using “free cooling” via a waterside economizer (open-loop cooling tower) |
| **S9** | Liquid-cooled racks using “free cooling” via a waterside economizer (open-loop cooling tower) |
| **S10** | Air-cooled racks using “free cooling” via a waterside economizer (closed-loop cooling tower) |
| **S11** | Liquid-cooled racks using “free cooling” via a waterside economizer (closed-loop cooling tower) |
| **S12** | Air-cooled racks with an air-cooled chiller |
| **S13** | Liquid-cooled racks with an air-cooled chiller |
| **S14** | Air-cooled racks using “free cooling” via an airside economizer |
| **S15** | Air-cooled racks using “free cooling” via an airside economizer with direct adiabatic cooling |
| **S16** | Air-cooled racks using “free cooling” via an airside economizer with indirect adiabatic cooling |
| **S17** | Air-cooled racks with chillerless, once-through water cooling |
| **S18** | Liquid-cooled racks with chillerless, once-through water cooling |
| **S19** | High-temperature liquid-cooled racks with dry coolers |
| **S20** | High-temperature liquid-cooled racks with dry coolers and adiabatic assist |
| **S21** | Air-cooled racks with dry coolers |
| **S22** | Air-cooled racks with dry coolers and adiabatic assist |

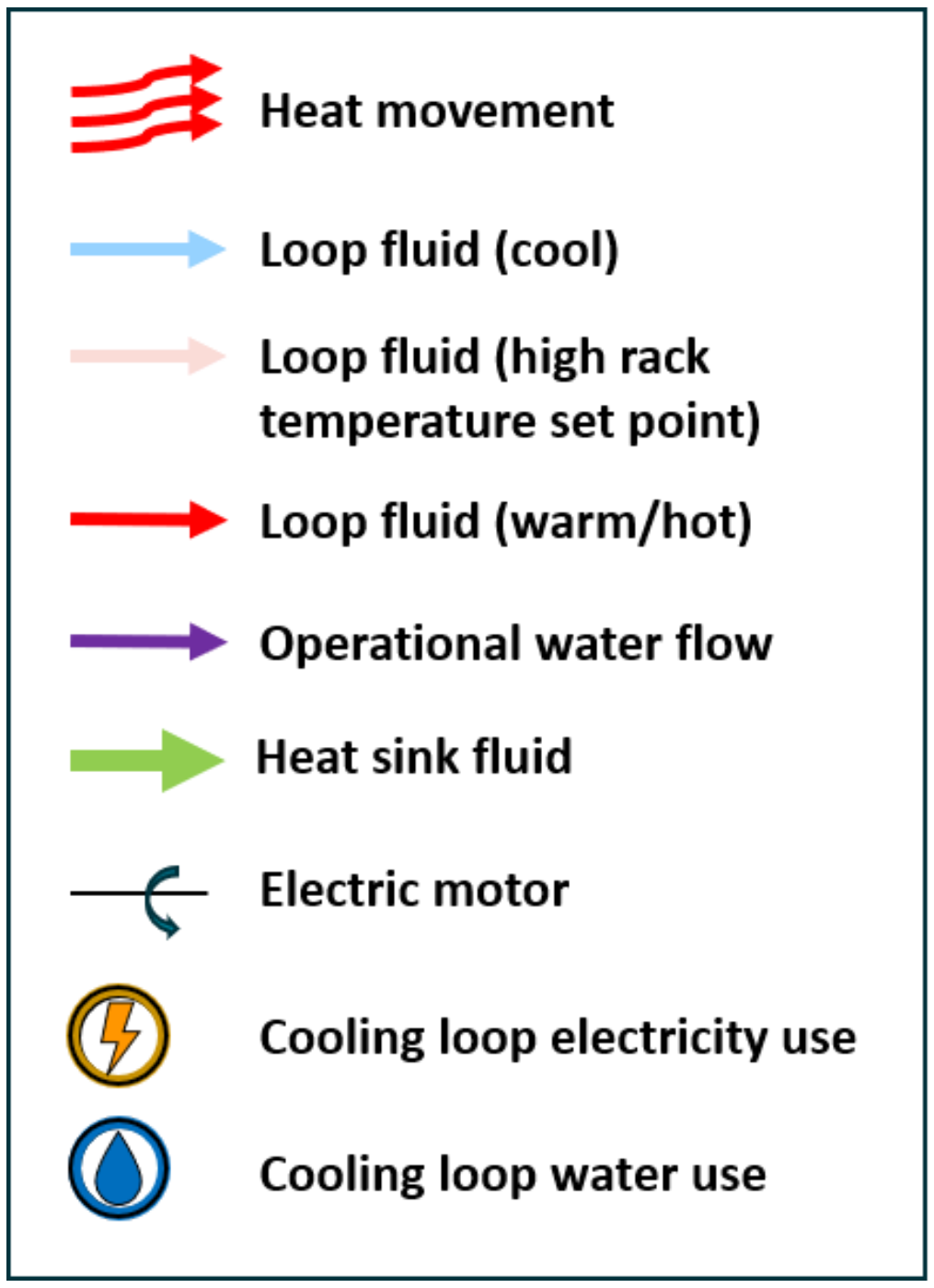

**Figure S3:** Cooling system configuration schematic legend

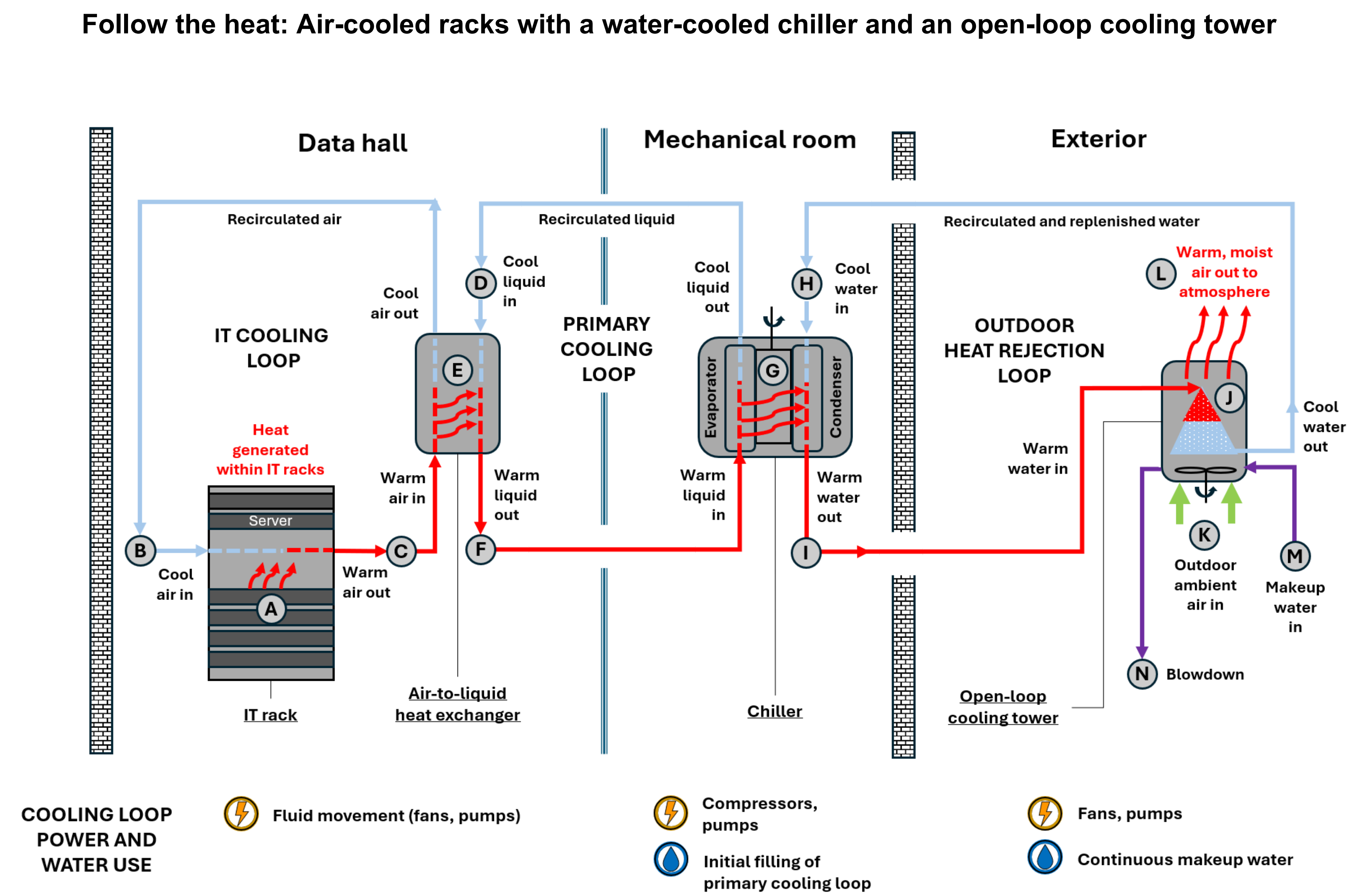

**Figure S4:** Air-cooled racks with a water-cooled chiller and an open-loop cooling tower

This configuration utilizes air in the IT cooling loop, chilled water (or a water/glycol mixture) in the primary cooling loop, and water in the building heat rejection loop. The primary cooling loop is closed and the building heat rejection loop is open. Cool incoming air (point B) absorbs the heat generated within the racks (point A), becoming warm air upon exiting the rack (point C). The warm air flows to an air-to-liquid heat exchanger, which is fed by chilled liquid (point D). Heat moves from the warm air to the chilled liquid (point E), the latter becoming warm as it exits the heat exchanger (point F). The warm liquid returns to the chiller, wherein it becomes chilled again (point D) as its heat is moved within the chiller to the chiller's condenser (point G). Cool water flows across the condenser (point H), becoming warm (point I) as it absorbs the heat from the condenser. This warm water then enters the open-loop cooling tower, where it is sprayed downward (point J) as exterior ambient air (which acts as the ultimate heat sink) is drawn upward against it (point K). Some of the warm water gets evaporated into the ambient air, which becomes warm and humid (point L), taking the heat that originated at the IT rack (point A) with it. The remaining water at the bottom of the cooling tower is now cool and flows back to the chiller condenser (point H) for another heat removal cycle. "Makeup" water is continuously added to the cooling tower (point M) to replace the water that was lost in evaporation and due to "blowdown" water that is discharged (point N) for wastewater treatment. Blowdown is necessary to remove concentrated minerals and water treatment chemicals that build up in the cooling tower.

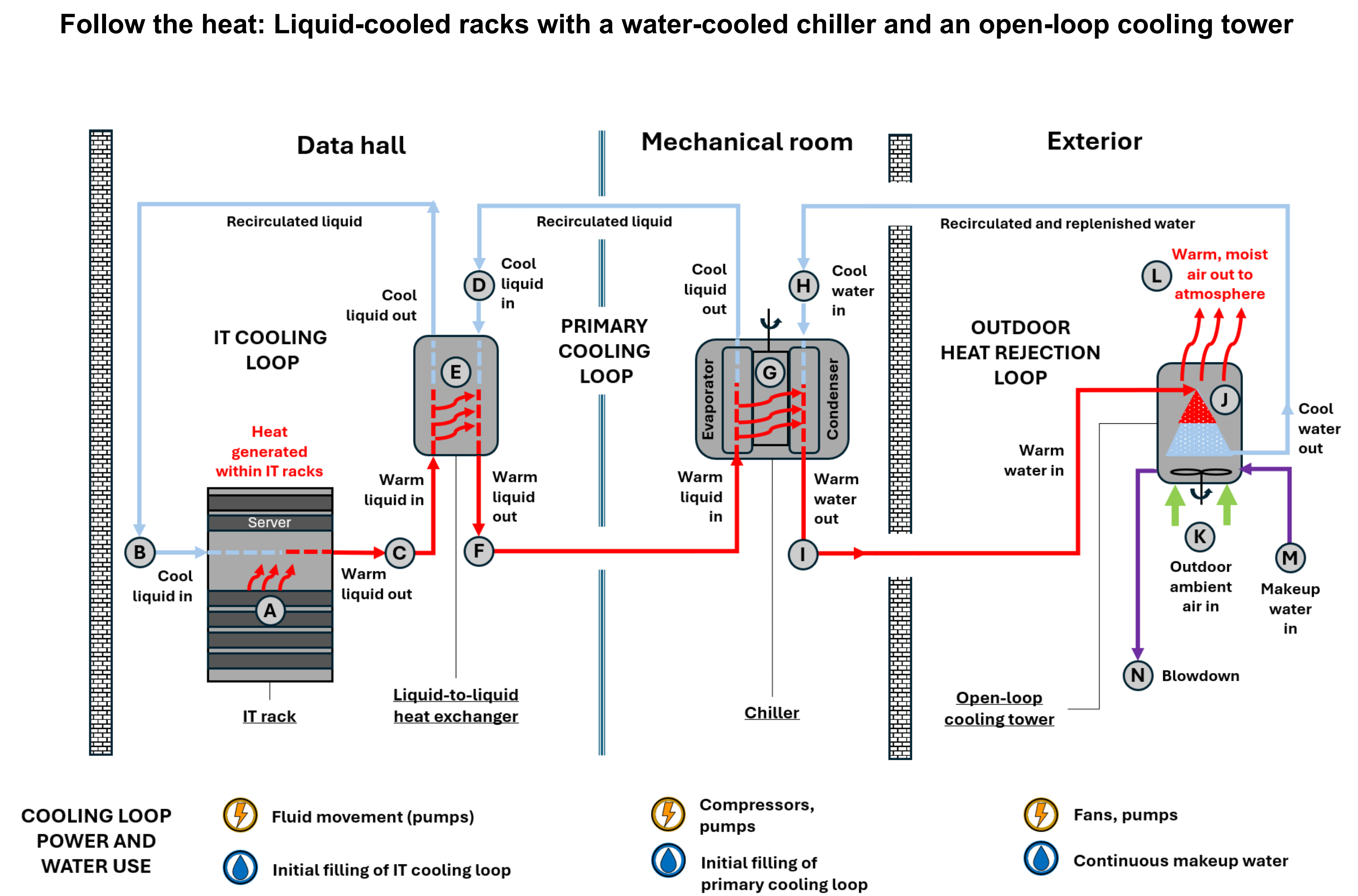

**Figure S5:** Liquid-cooled racks with a water-cooled chiller and an open-loop cooling tower

This configuration utilizes liquid in the IT cooling loop (water, water/glycol mixture, or a dielectric fluid), chilled water (or a water/glycol mixture) in the primary cooling loop, and water in the building heat rejection loop. The IT cooling loop is closed, the primary cooling loop is closed, and the building heat rejection loop is open. Cool incoming liquid (point B) absorbs the heat generated within the racks (point A), becoming warm liquid upon exiting the rack (point C). The warm liquid flows to a liquid-to-liquid heat exchanger, which is fed by chilled liquid (point D). Heat moves from the warm air to the chilled liquid (point E), the latter becoming warm as it exits the heat exchanger (point F). The warm liquid returns to the chiller, wherein it becomes chilled again (point D) as its heat is moved within the chiller to the chiller's condenser (point G). Cool water flows across the condenser (point H), becoming warm (point I) as it absorbs the heat from the condenser. This warm water then enters the open-loop cooling tower, where it is sprayed downward (point J) as exterior ambient air (which acts as the ultimate heat sink) is drawn upward against it (point K). Some of the warm water gets evaporated into the ambient air, which becomes warm and humid (point L), taking the heat that originated at the IT rack (point A) with it. The remaining water at the bottom of the cooling tower is now cool and flows back to the chiller condenser (point H) for another heat removal cycle. "Makeup" water is continuously added to the cooling tower (point M) to replace the water that was lost in evaporation and due to "blowdown" water that is discharged (point N) for wastewater treatment. Blowdown is necessary to remove concentrated minerals and water treatment chemicals that build up in the cooling tower.

For single-phase immersion cooling systems, the IT equipment is contained in a tank instead of a traditional rack. The IT cooling loop contains a single-phase dielectric fluid that is pumped out of the tank at point C for heat removal at point E, after which point it is pumped back to the tank at point B for another cooling cycle.

For two-phase immersion cooling systems, the IT equipment is contained in a tank instead of a traditional rack. The tank contains a two-phase dielectric fluid that transitions from a liquid to a vapor when it absorbs heat, rising to the top of the tank. The heat exchanger depicted at point E is situated at the top of the tank and cools the vapor back into a liquid state, causing it to fall back down in the tank for more cooling. Thus, the "loop" depicted between point C and point B occurs entirely within the two-phase immersion cooling tank.

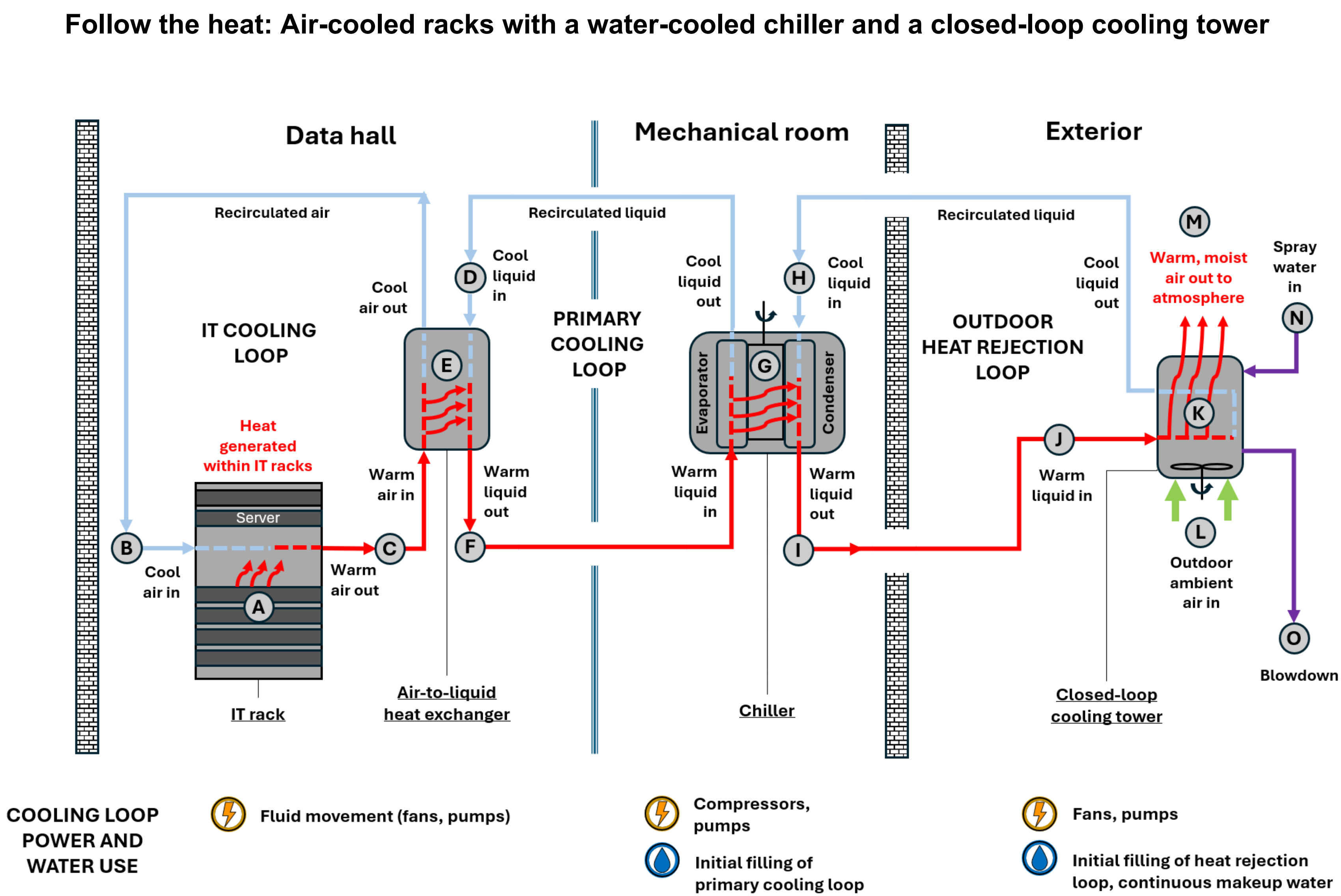

**Figure S6:** Air-cooled racks with a water-cooled chiller and a closed-loop cooling tower

This configuration utilizes air in the IT cooling loop, chilled water (or a water/glycol mixture) in the primary cooling loop, and water (or a water/glycol mixture) in the building heat rejection loop. The primary cooling loop is closed and the building heat rejection loop is closed. However, indirect "open loop" water use still occurs due to a water spray within the cooling tower, as explained below. Cool incoming air (point B) absorbs the heat generated within the racks (point A), becoming warm air upon exiting the rack (point C). The warm air flows to an air-to-liquid heat exchanger, which is fed by chilled liquid (point D). Heat moves from the warm air to the chilled liquid (point E), the latter becoming warm as it exits the heat exchanger (point F). The warm liquid returns to the chiller, wherein it becomes chilled again (point D) as its heat is moved within the chiller to the chiller's condenser (point G). Cool liquid flows across the condenser (point H), becoming warm (point I) as it absorbs the heat from the condenser. This warm liquid then enters the closed-loop cooling tower (point J), where it flows through a closed liquid-to-air heat exchanger (point K) within the cooling tower. Water is sprayed downward on the heat exchanger within the cooling tower as exterior ambient air (which acts as the ultimate heat sink) is drawn upward (point L) across the heat exchanger. As the incoming air flows across the heat exchanger, some of the spray water evaporates into it, making the air warm and humid (point M) as it exits the cooling tower, taking the heat that originated at the IT rack (point A) with it. The liquid exiting the closed-loop heat exchanger is now cool and flows back to the chiller condenser (point H) for another heat removal cycle. Additional spray water is continuously added to the cooling tower (point N) to replace the water that was lost to evaporation and due to "blowdown" water that is discharged (point O) for wastewater treatment. Blowdown is necessary to remove concentrated minerals and water treatment chemicals that build up within the cooling tower. The spray water is "open loop" even though the outdoor heat rejection water is "closed loop." This is known as "indirect" cooling of the heat rejection water loop.

## Follow the heat: Liquid-cooled racks with a water-cooled chiller and a closed-loop cooling tower

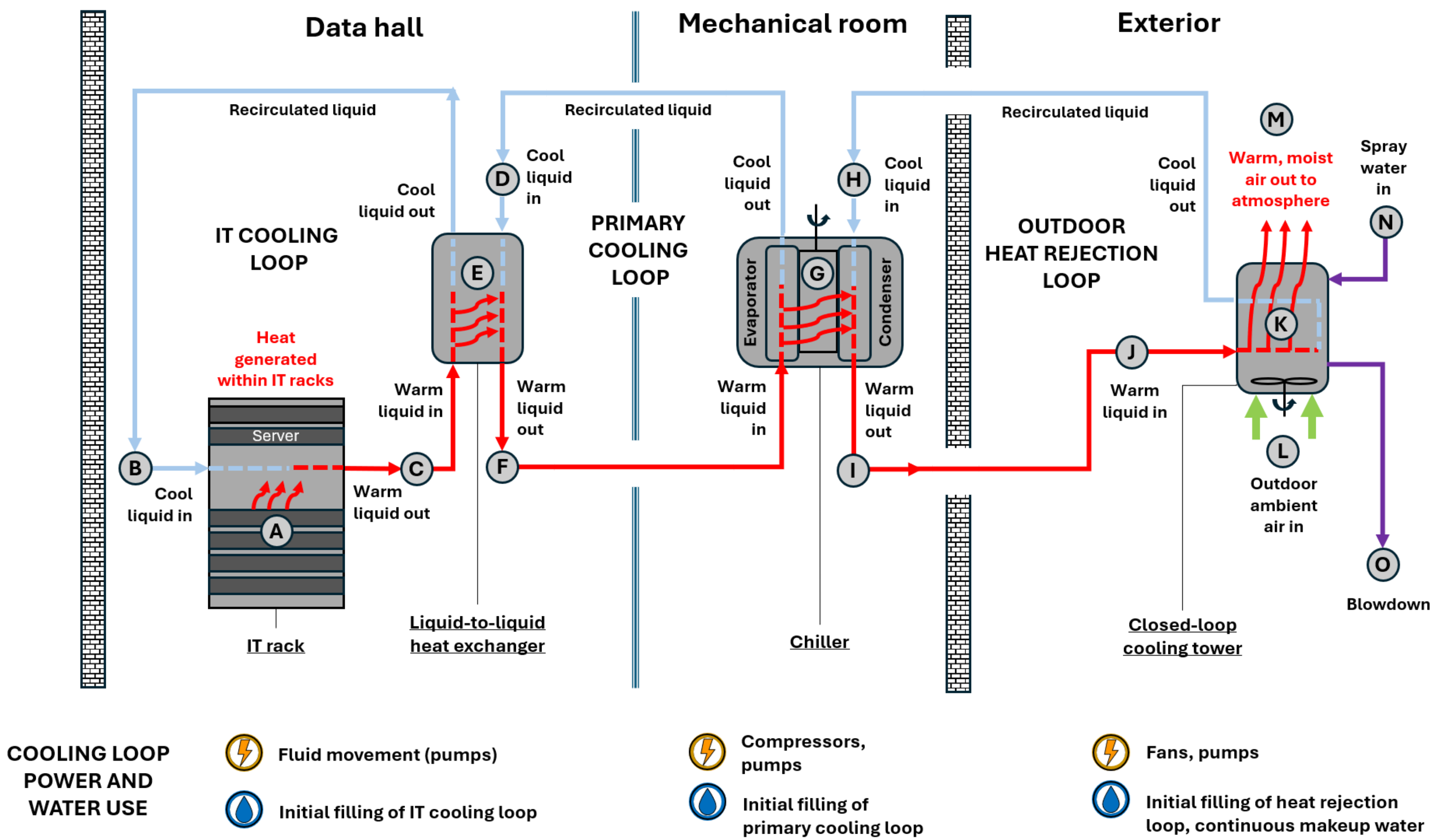

**Figure S7:** Liquid-cooled racks with a water-cooled chiller and a closed-loop cooling tower

This configuration utilizes liquid in the IT cooling loop (water, water/glycol mixture, or a dielectric fluid), chilled water (or a water/glycol mixture) in the primary cooling loop, and water (or a water/glycol mixture) in the building heat rejection loop. The IT cooling loop is closed, the primary cooling loop is closed, and the building heat rejection loop is closed. However, indirect "open loop" water use still occurs due to a water spray within the cooling tower, as explained below. Cool incoming liquid (point B) absorbs the heat generated within the racks (point A), becoming warm liquid upon exiting the rack (point C). The warm liquid flows to a liquid-to-liquid heat exchanger, which is fed by chilled liquid (point D). Heat moves from the warm liquid to the chilled liquid (point E), the latter becoming warm as it exits the heat exchanger (point F). The warm liquid returns to the chiller, wherein it becomes chilled again (point D) as its heat is moved within the chiller to the chiller's condenser (point G). Cool liquid flows across the condenser (point H), becoming warm (point I) as it absorbs the heat from the condenser. This warm liquid then enters the closed-loop cooling tower (point J), where it flows through a closed liquid-to-air heat exchanger (point K) within the cooling tower. Water is sprayed downward on the heat exchanger within the cooling tower as exterior ambient air (which acts as the ultimate heat sink) is drawn upward (point L) across the heat exchanger. As the incoming air flows across the heat exchanger, some of the spray water evaporates into it, making the air warm and humid (point M) as it exits the cooling tower, taking the heat that originated at the IT rack (point A) with it. The liquid exiting the closed-loop heat exchanger is now cool and flows back to the chiller condenser (point H) for another heat removal cycle. Additional spray water is continuously added to the cooling tower (point N) to replace the water that was lost to evaporation and due to "blowdown" water that is discharged (point O) for wastewater treatment. Blowdown is necessary to remove concentrated minerals and water treatment chemicals that build up within the cooling tower. The spray water is "open loop" even though the outdoor heat rejection water is "closed loop." This is known as "indirect" cooling of the heat rejection water loop.

For single-phase immersion cooling systems, the IT equipment is contained in a tank instead of a traditional rack. The IT cooling loop contains a single-phase dielectric fluid that is pumped out of the tank at point C for heat removal at point E, after which point it is pumped back to the tank at point B for another cooling cycle.

For two-phase immersion cooling systems, the IT equipment is contained in a tank instead of a traditional rack. The tank contains a two-phase dielectric fluid that transitions from a liquid to a vapor when it absorbs heat, rising to the top of the tank. The heat exchanger depicted at point E is situated at the top of the tank and cools the vapor back into a liquid state, causing it to fall back down in the tank for more cooling. Thus, the "loop" depicted between point C and point B occurs entirely within the two-phase immersion cooling tank.

**Follow the heat: Air-cooled racks using "free cooling" via a waterside economizer (open-loop cooling tower)**

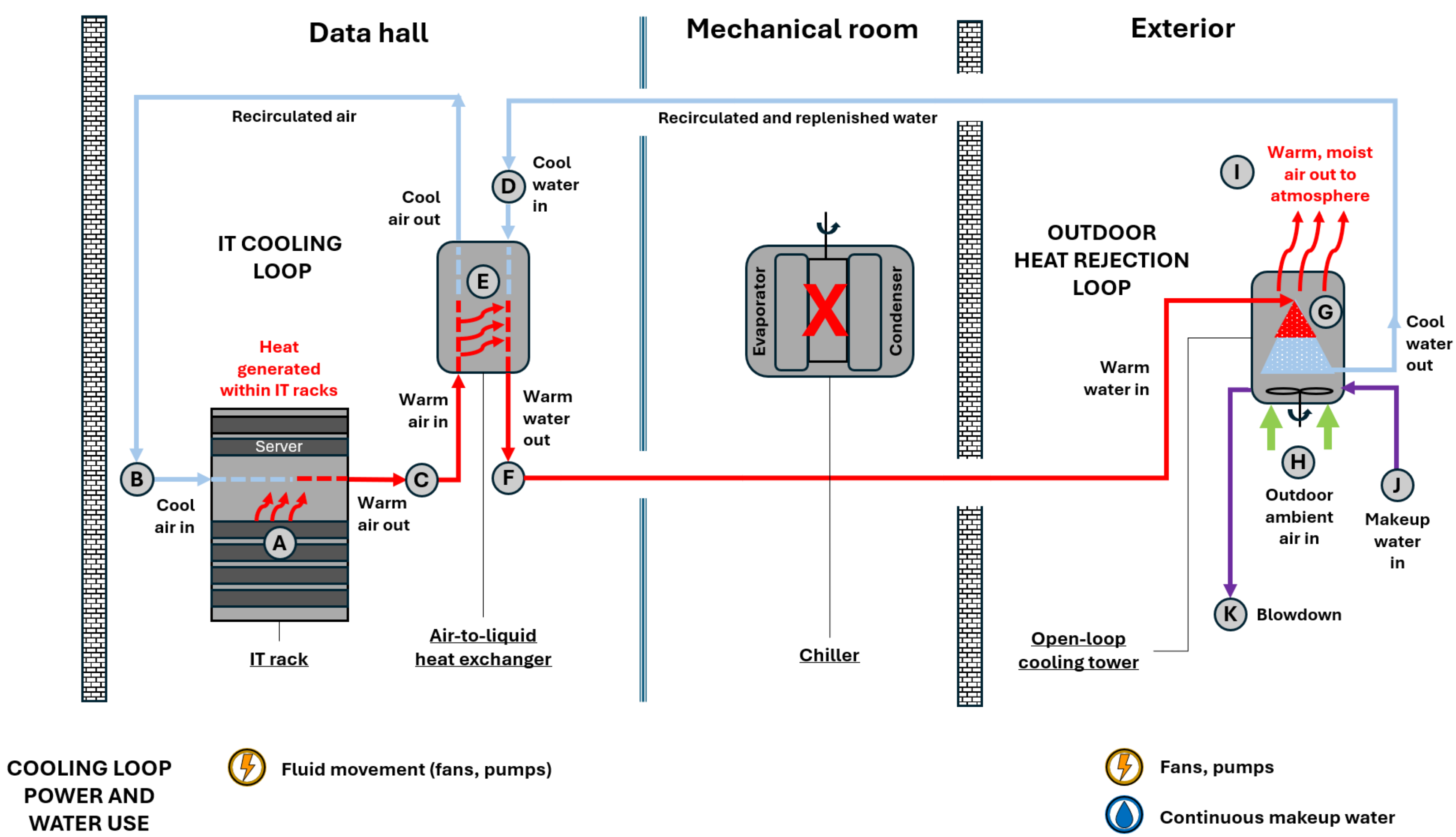

**Figure S8:** Air-cooled racks using "free cooling" via a waterside economizer (open-loop cooling tower)

This configuration utilizes air in the IT cooling loop and water in the building heat rejection loop. The building heat rejection loop is open. Cool incoming air (point B) absorbs the heat generated within the racks (point A), becoming warm air upon exiting the rack (point C). The warm air flows to an air-to-liquid heat exchanger, which is fed by cooled water (point D) provided directly from the open-loop cooling tower outside the building. Heat moves from the warm air to the cool water (point E), the latter becoming warm as it exits the heat exchanger (point F). This warm water then returns to the open-loop cooling tower, where it is sprayed downward (point G) as exterior ambient air (which acts as the ultimate heat sink) is drawn upward against it (point H). Some of the warm water gets evaporated into the ambient air, which becomes warm and humid (point I), taking the heat that originated at the IT rack (point A) with it. The remaining water at the bottom of the cooling tower is now cool and flows back to the data hall heat exchanger (point D) for another heat removal cycle. “Makeup” water is continuously added to the cooling tower (point J) to replace the water that was lost in evaporation and due to “blowdown” water that is discharged (point K) for wastewater treatment. Blowdown is necessary to remove concentrated minerals and water treatment chemicals that build up in the cooling tower.

This configuration is known as “free cooling,” since the operation of the chiller and its primary cooling loop (see Figure S4) is eliminated. Free cooling is possible during periods when outdoor ambient air conditions allow the cooling tower exit water to be sufficiently cold at point D such that chiller operation isn’t necessary. Free cooling via waterside economizers can be a viable cooling mode at many data centers for many hours of the year in suitable climates.

**Follow the heat: Liquid-cooled racks using "free cooling" via a waterside economizer (open-loop cooling tower)**

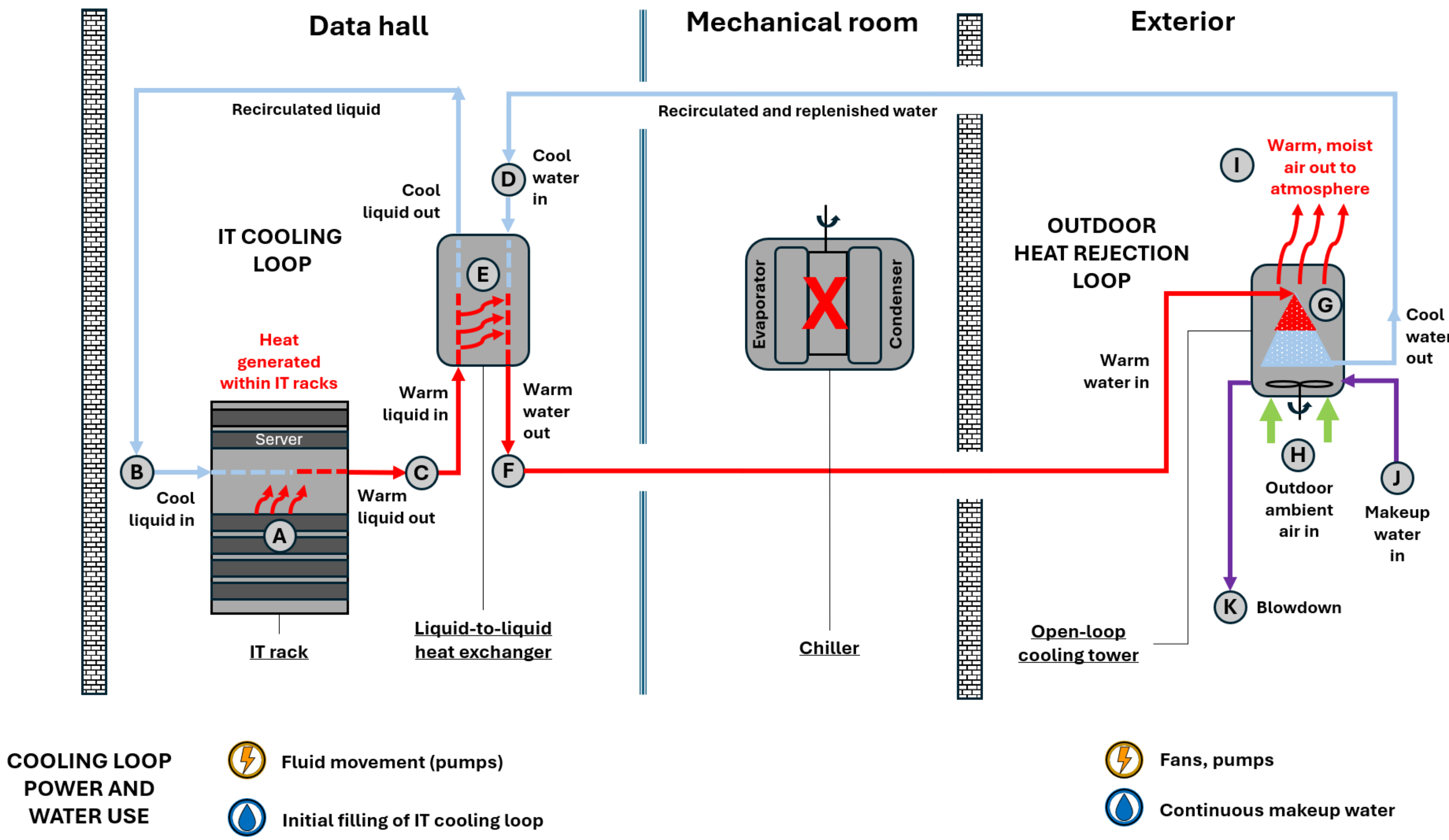

**Figure S9:** Liquid-cooled racks using "free cooling" via a waterside economizer (open-loop cooling tower)

This configuration utilizes liquid in the IT cooling loop (water, water/glycol mixture, or dielectric fluid) and water in the building heat rejection loop. The IT cooling loop is closed and the building heat rejection loop is open. Cool incoming liquid (point B) absorbs the heat generated within the racks (point A), becoming warm liquid upon exiting the rack (point C). The warm liquid flows to an air-to-liquid heat exchanger, which is fed by cooled water (point D) provided directly from the open-loop cooling tower outside the building. Heat moves from the warm liquid to the cool water (point E), the latter becoming warm as it exits the heat exchanger (point F). This warm water then returns to the open-loop cooling tower, where it is sprayed downward (point G) as exterior ambient air (which acts as the ultimate heat sink) is drawn upward against it (point H). Some of the warm water gets evaporated into the ambient air, which becomes warm and humid (point I), taking the heat that originated at the IT rack (point A) with it. The remaining water at the bottom of the cooling tower is now cool and flows back to the data hall heat exchanger (point D) for another heat removal cycle. "Makeup" water is continuously added to the cooling tower (point J) to replace the water that was lost in evaporation and due to "blowdown" water that is discharged (point K) for wastewater treatment. Blowdown is necessary to remove concentrated minerals and water treatment chemicals that build up in the cooling tower.

This configuration is known as "free cooling," since the operation of the chiller and its primary cooling loop (see Figure S5) is eliminated. Free cooling is possible during periods when outdoor ambient air conditions allow the cooling tower exit water to be sufficiently cold at point D such that chiller operation isn't necessary. Free cooling via waterside economizers can be a viable cooling mode at many data centers for many hours of the year in suitable climates.

For single-phase immersion cooling systems, the IT equipment is contained in a tank instead of a traditional rack. The IT cooling loop contains a single-phase dielectric fluid that is pumped out of the tank at point C for heat removal at point E, after which point it is pumped back to the tank at point B for another cooling cycle.

For two-phase immersion cooling systems, the IT equipment is contained in a tank instead of a traditional rack. The tank contains a two-phase dielectric fluid that transitions from a liquid to a vapor when it absorbs heat, rising to the top of the tank. The heat exchanger depicted at point E is situated at the top of the tank and cools the vapor back into a liquid state, causing it to fall back down in the tank for more cooling. Thus, the "loop" depicted between point C and point B occurs entirely within the two-phase immersion cooling tank.

**Follow the heat: Air-cooled racks using "free cooling" via a waterside economizer (closed-loop cooling tower)**

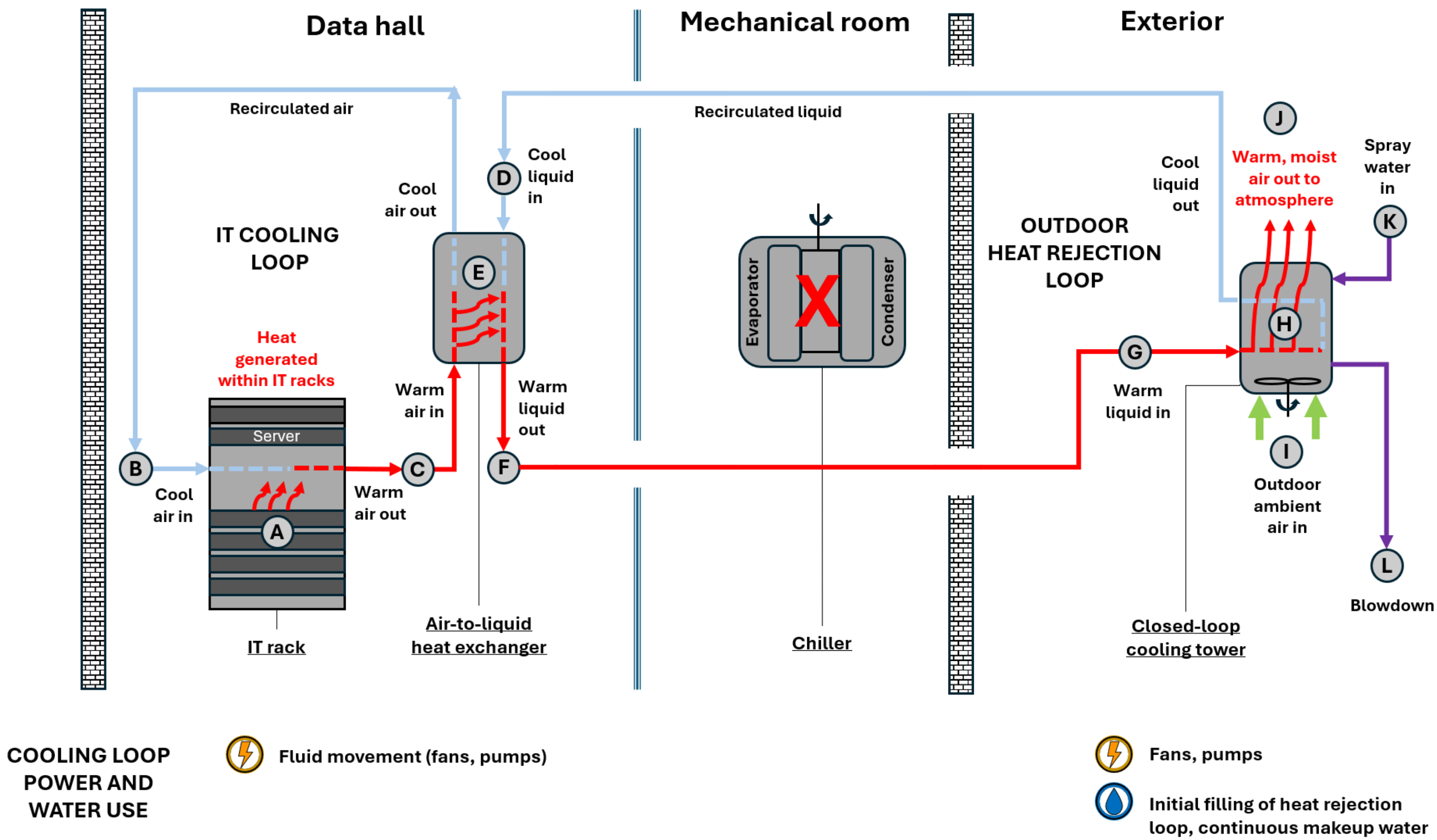

**Figure S10:** Air-cooled racks using "free cooling" via a waterside economizer (closed-loop cooling tower)

This configuration utilizes air in the IT cooling loop and water (or a water/glycol mixture) in the building heat rejection loop.  The building heat rejection loop is closed.  However, indirect “open loop” water use still occurs due to a water spray within the cooling tower, as explained below.  Cool incoming air (point B) absorbs the heat generated within the racks (point A), becoming warm air upon exiting the rack (point C). The warm air flows to an air-to-liquid heat exchanger, which is fed by cooled liquid (point D) provided directly from the closed-loop cooling tower outside the building. Heat moves from the warm air to the cool liquid (point E), the latter becoming warm as it exits the heat exchanger (point F). This warm liquid then returns to the closed-loop cooling tower (point J), where it flows through a closed liquid-to-air heat exchanger (point H) within the cooling tower.  Water is sprayed downward on the heat exchanger within the cooling tower as exterior ambient air (which acts as the ultimate heat sink) is drawn upward (point I) across the heat exchanger.  As the incoming air flows across the heat exchanger, some of the spray water evaporates into it, making the air warm and humid (point J) as it exits the cooling tower, taking the heat that originated at the IT rack (point A) with it. The water exiting the closed-loop heat exchanger is now cool and flows back to the data hall heat exchanger (point D) for another heat removal cycle. Additional spray water is continuously added to the cooling tower (point K) to replace the water that was lost in evaporation and due to “blowdown” water that is discharged (point L) for wastewater treatment.  Blowdown is necessary to remove concentrated minerals and water treatment chemicals that build up in the cooling tower. The spray water is “open loop” even though the outdoor heat rejection water is “closed loop.” This is known as “indirect” cooling of the heat rejection water loop.

This configuration is known as “free cooling,” since the operation of the chiller and its primary cooling loop (see Figure S6) is eliminated.  Free cooling is possible during periods when outdoor ambient air conditions allow the cooling tower exit liquid to be sufficiently cold at point D such that chiller operation isn’t necessary.  Free cooling via waterside economizers can be a viable cooling mode at many data centers for many hours of the year in suitable climates.

**Follow the heat: Liquid-cooled racks using “free cooling” via a waterside economizer (closed-loop cooling tower)**

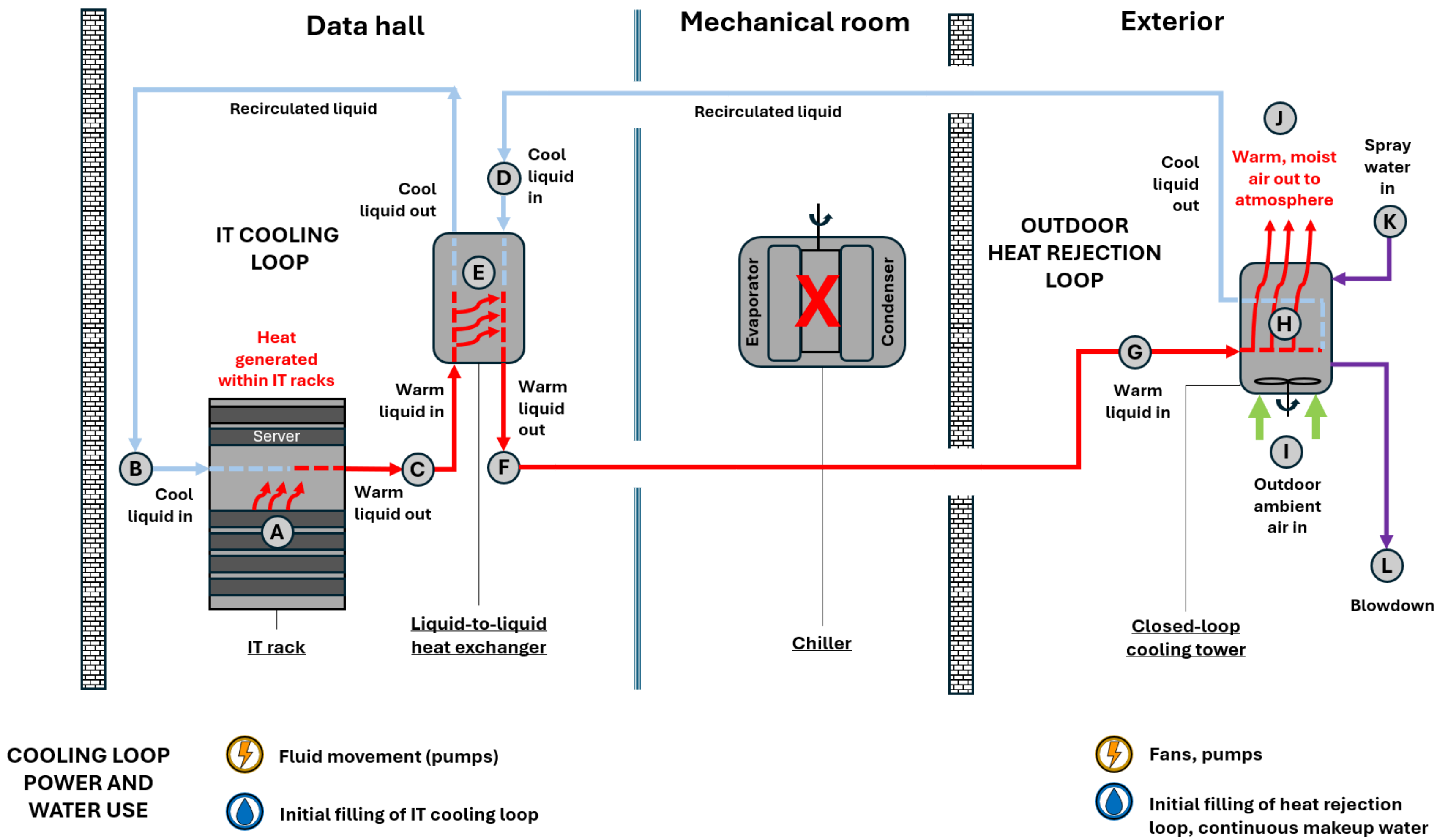

**Figure S11:** Liquid-cooled racks using “free cooling” via a waterside economizer (closed-loop cooling tower)

This configuration utilizes liquid in the IT cooling loop (water, water/glycol mixture, or a dielectric fluid), chilled water (or a water/glycol mixture) in the primary cooling loop, and water (or a water/glycol mixture) in the building heat rejection loop. The IT cooling loop is closed and the building heat rejection loop is closed. However, indirect “open loop” water use still occurs due to a water spray within the cooling tower, as explained below. Cool incoming liquid (point B) absorbs the heat generated within the racks (point A), becoming warm liquid upon exiting the rack (point C). The warm liquid flows to a liquid-to-liquid heat exchanger, which is fed by cooled liquid (point D) provided directly from the closed-loop cooling tower outside the building. Heat moves from the warm air to the cool liquid (point E), the latter becoming warm as it exits the heat exchanger (point F). This warm liquid then returns to the closed-loop cooling tower (point J), where it flows through a closed liquid-to-air heat exchanger (point H) within the cooling tower. Water is sprayed downward on the heat exchanger within the cooling tower as exterior ambient air (which acts as the ultimate heat sink) is drawn upward (point I) across the heat exchanger. As the incoming air flows across the heat exchanger, some of the spray water evaporates into it, making the air warm and humid (point J) as it exits the cooling tower, taking the heat that originated at the IT rack (point A) with it. The water exiting the closed-loop heat exchanger is now cool and flows back to the data hall heat exchanger (point D) for another heat removal cycle. Additional spray water is continuously added to the cooling tower (point K) to replace the water that was lost in evaporation and due to “blowdown” water that is discharged (point L) for wastewater treatment. Blowdown is necessary to remove concentrated minerals and water treatment chemicals that build up in the cooling tower. The spray water is “open loop” even though the outdoor heat rejection water is “closed loop.” This is known as “indirect” cooling of the heat rejection water loop.

This configuration is known as “free cooling,” since the operation of the chiller and its primary cooling loop (see Figure S7) is eliminated. Free cooling is possible during periods when outdoor ambient air conditions allow the cooling tower exit liquid to be sufficiently cold at point D such that chiller operation isn’t necessary. Free cooling via waterside economizers can be a viable cooling mode at many data centers for many hours of the year in suitable climates.

For single-phase immersion cooling systems, the IT equipment is contained in a tank instead of a traditional rack. The IT cooling loop contains a single-phase dielectric fluid that is pumped out of the tank at point C for heat removal at point E, after which point it is pumped back to the tank at point B for another cooling cycle.

For two-phase immersion cooling systems, the IT equipment is contained in a tank instead of a traditional rack. The tank contains a two-phase dielectric fluid that transitions from a liquid to a vapor when it absorbs heat, rising to the top of the tank. The heat exchanger depicted at point E is situated at the top of the tank and cools the vapor back into a liquid state, causing it to fall back down in the tank for more cooling. Thus, the “loop” depicted between point C and point B occurs entirely within the two-phase immersion cooling tank.

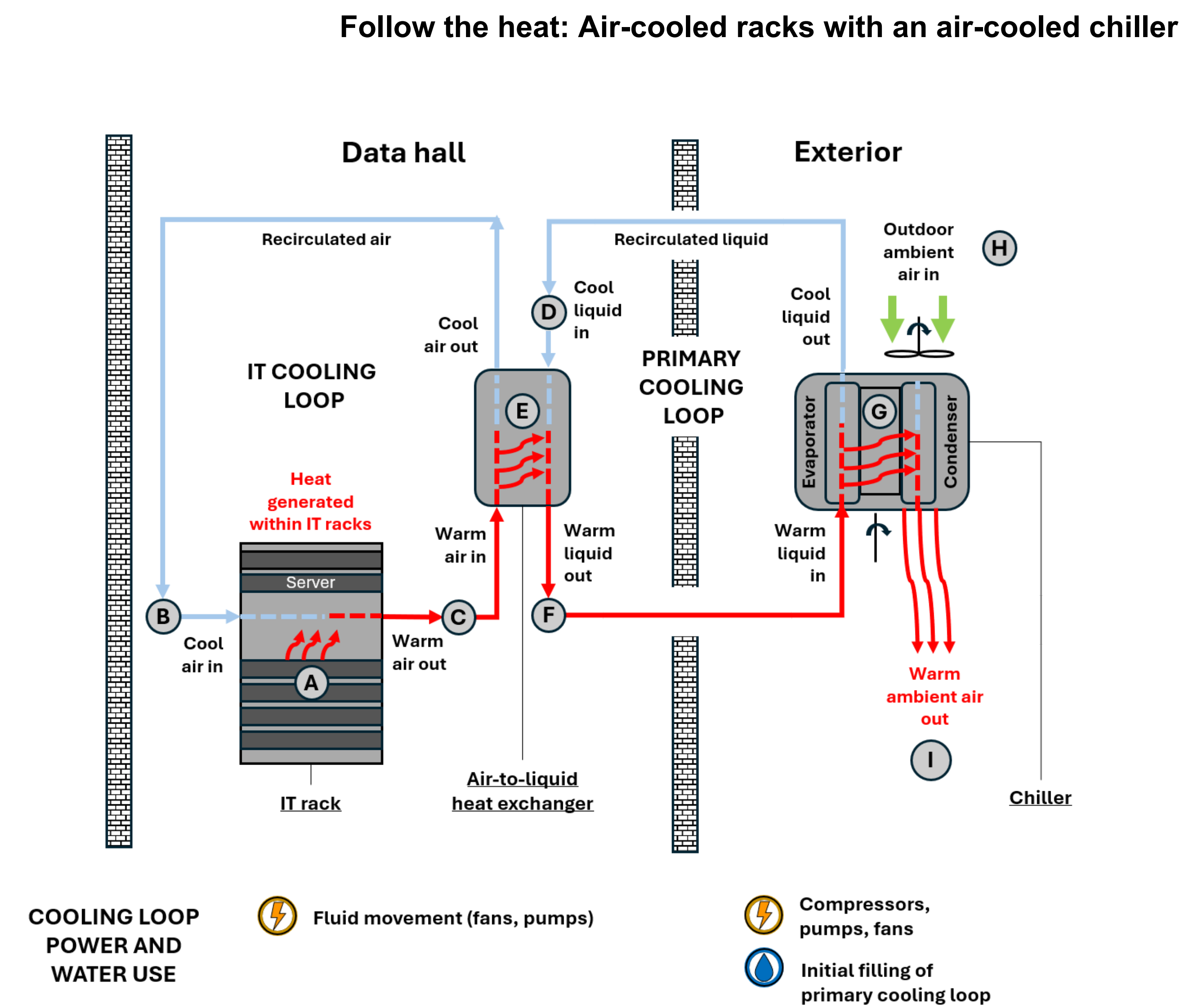

**Figure S12:** Air-cooled racks with an air-cooled chiller

This configuration utilizes air in the IT cooling loop and chilled water (or a water/glycol mixture) in the primary cooling loop. There is no building heat rejection loop, since the air-cooled chiller itself is located outside of the building.  The primary cooling loop is closed. Cool incoming air (point B) absorbs the heat generated within the racks (point A), becoming warm air upon exiting the rack (point C). The warm air flows to an air-to-liquid heat exchanger, which is fed by chilled liquid (point D). Heat moves from the warm air to the chilled liquid (point E), the latter becoming warm as it exits the heat exchanger (point F). The warm liquid returns to the chiller, wherein it becomes chilled again (point D) as its heat is moved within the chiller to the chiller's condenser (point G).  Exterior ambient air (which acts as the ultimate heat sink) is drawn using a fan (point H) across the chiller's condenser coils, absorbing heat from the condenser in the process.  The air exits the condenser coils as warm air (point I), taking the heat that originated at the IT rack (point A) with it.

## Follow the heat: Liquid-cooled racks with an air-cooled chiller

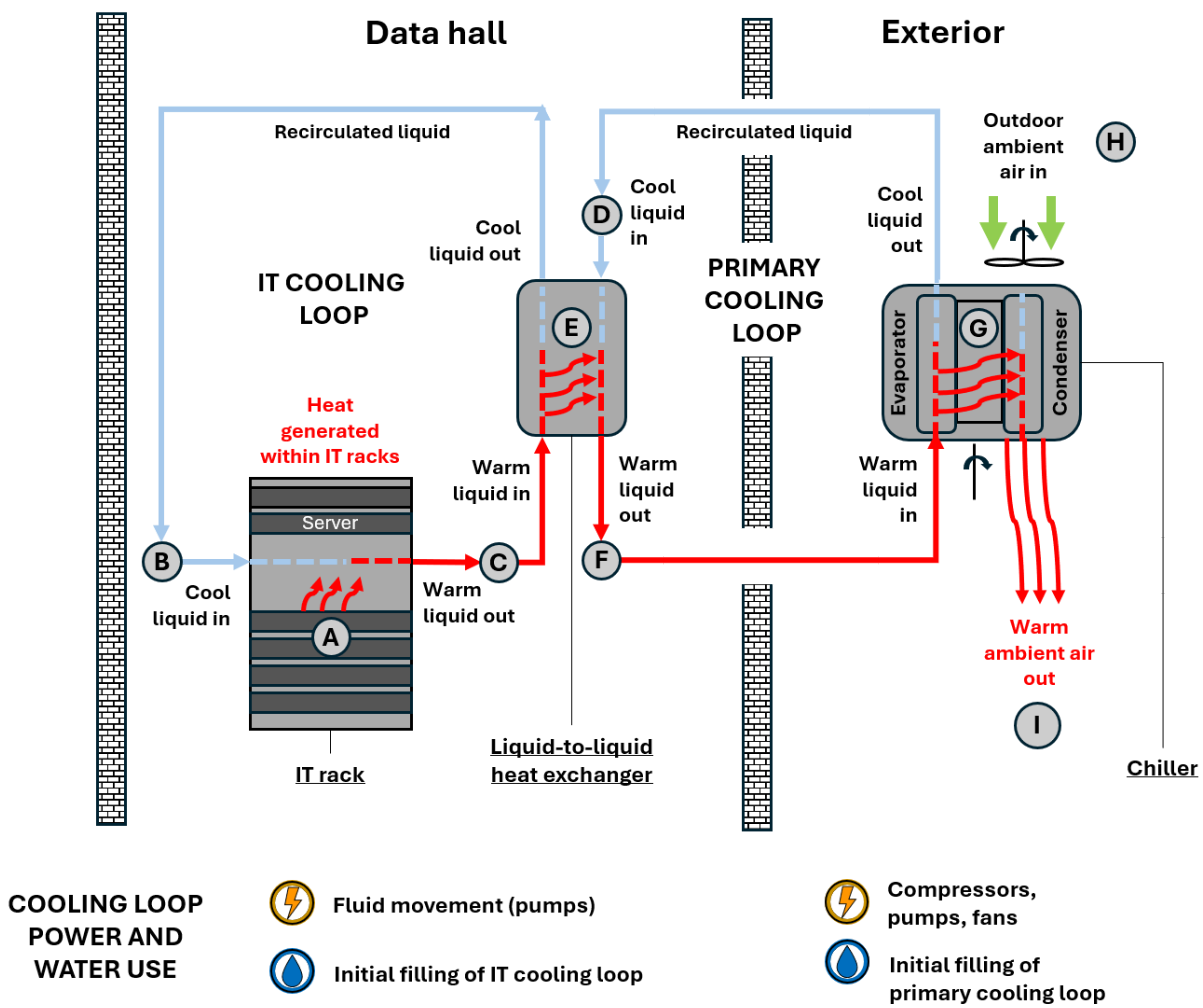

**Figure S13:** Liquid-cooled racks with an air-cooled chiller

This configuration utilizes liquid in the IT cooling loop (water, water/glycol mixture, or a dielectric fluid) and chilled water (or a water/glycol mixture) in the primary cooling loop. There is no building heat rejection loop, since the air-cooled chiller itself is located outside of the building. The primary cooling loop is closed. Cool incoming liquid (point B) absorbs the heat generated within the racks (point A), becoming warm liquid upon exiting the rack (point C). The warm liquid flows to a liquid-to-liquid heat exchanger, which is fed by chilled liquid (point D). Heat moves from the warm liquid to the chilled liquid (point E), the latter becoming warm as it exits the heat exchanger (point F). The warm liquid returns to the chiller, wherein it becomes chilled again (point D) as its heat is moved within the chiller to the chiller’s condenser (point G). Exterior ambient air (which acts as the ultimate heat sink) is drawn using a fan (point H) across the chiller’s condenser coils, absorbing heat from the condenser in the process. The air exits the condenser coils as warm air (point I), taking the heat that originated at the IT rack (point A) with it.

For single-phase immersion cooling systems, the IT equipment is contained in a tank instead of a traditional rack. The IT cooling loop contains a single-phase dielectric fluid that is pumped out of the tank at point C for heat removal at point E, after which point it is pumped back to the tank at point B for another cooling cycle.

For two-phase immersion cooling systems, the IT equipment is contained in a tank instead of a traditional rack. The tank contains a two-phase dielectric fluid that transitions from a liquid to a vapor when it absorbs heat, rising to the top of the tank. The heat exchanger depicted at point E is situated at the top of the tank and cools the vapor back into a liquid state, causing it to fall back down in the tank for more cooling. Thus, the “loop” depicted between point C and point B occurs entirely within the two-phase immersion cooling tank.

## Follow the heat: Air-cooled racks using "free cooling" via an airside economizer

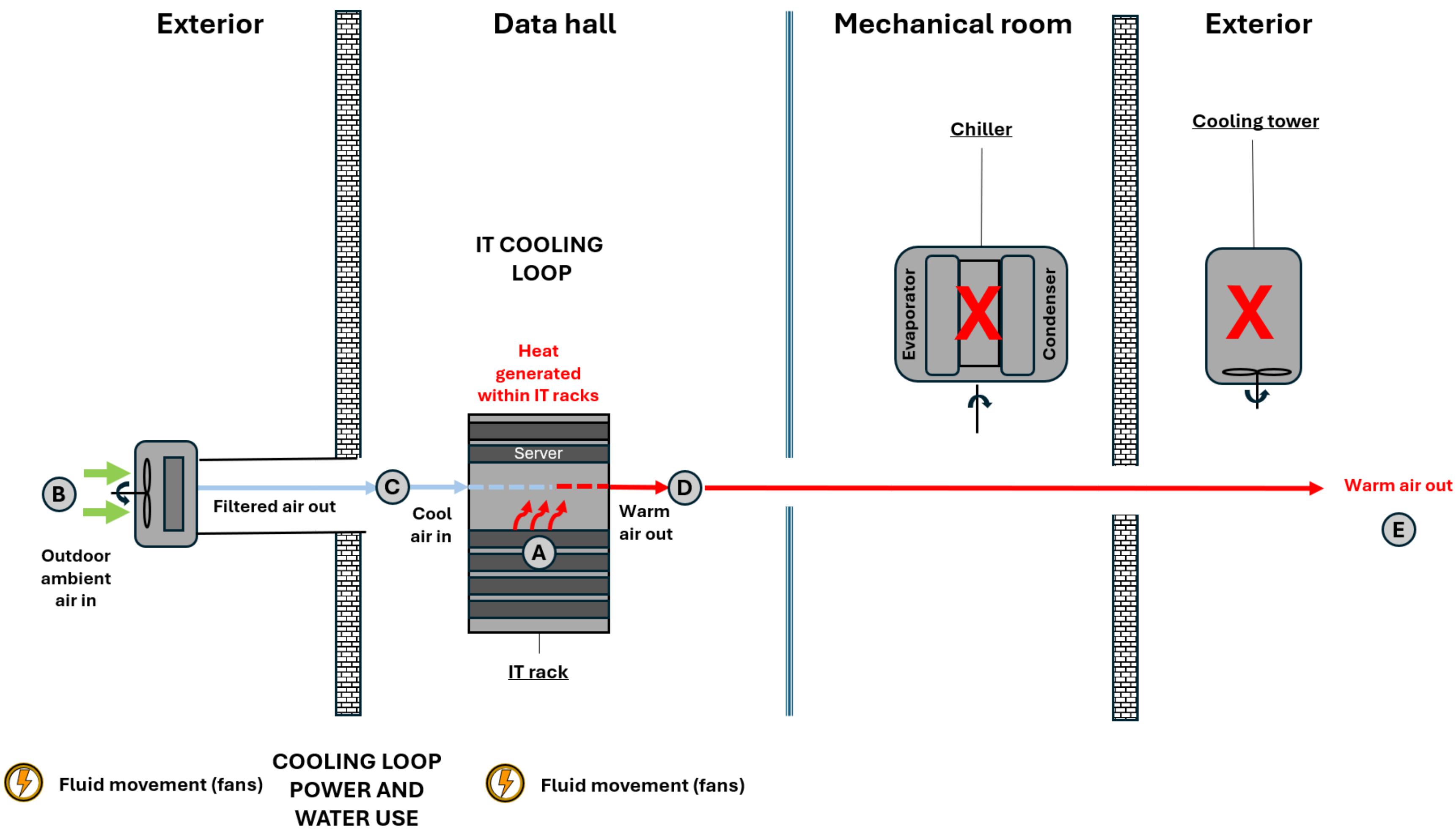

**Figure S14:** Air-cooled racks using "free cooling" via an airside economizer

This configuration utilizes air in the IT cooling loop. Cool and filtered outside air (which acts as the ultimate heat sink) is drawn into the data hall (point B), where it directly enters the racks as cool air (point C). The cool incoming air (point C) absorbs the heat generated within the racks (point A), becoming warm air upon exiting the rack (point D). This warm air is then directly exhausted to the exterior of the data center (point E), taking the heat that originated at the IT rack (point A) with it.

This configuration is known as “free cooling,” since the operation of the chiller, the primary cooling loop, the cooling tower, and the building heat rejection loop (see figure S5) are eliminated. Free cooling is possible during periods when outdoor ambient air conditions allow the incoming rack air at point C to be sufficiently cool such that chiller operation isn’t necessary. Free cooling via airside economizers can be a viable cooling mode at many data centers for many hours of the year in suitable climates.

**Follow the heat: Air-cooled racks using "free cooling" via an airside economizer with direct adiabatic cooling**

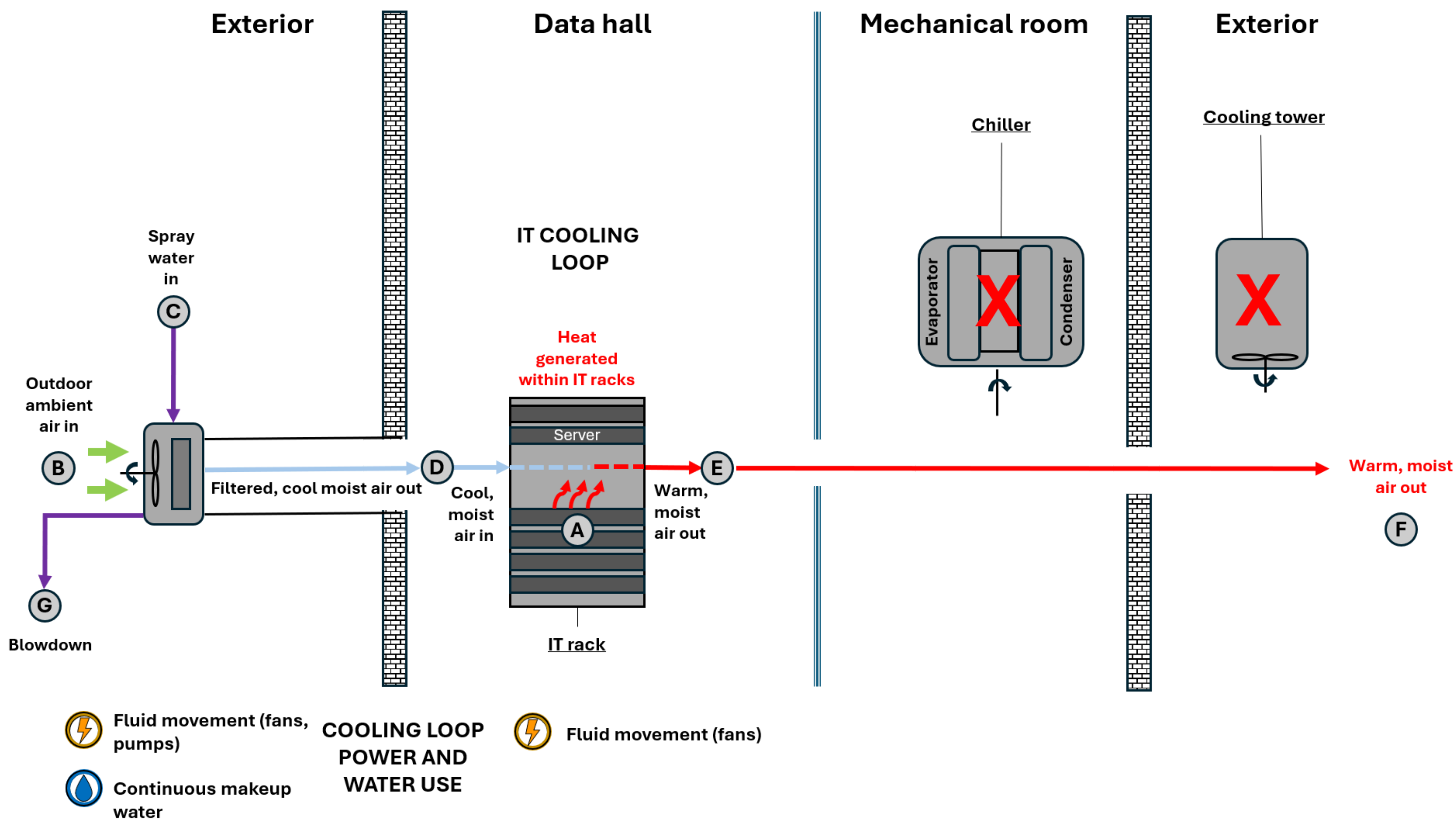

**Figure S15:** Air-cooled racks using "free cooling" via an airside economizer with direct adiabatic cooling

This configuration utilizes air in the IT cooling loop.  Outside air is drawn into an adiabatic cooler (point B), which contains a wetted media subject to a continuous water spray (point C).  The addition of water both cools and humidifies the incoming rack air (point D), which acts as the ultimate heat sink.  The cool incoming rack air (point D) absorbs the heat generated within the racks (point A), becoming warm air upon exiting the rack (point E).  This warm, moist air is then directly exhausted to the exterior of the data center (point E), taking the heat that originated at the IT rack (point A) with it. When required, “blowdown” water is also discharged (point G) for wastewater treatment to remove concentrated minerals and water treatment chemicals that build up in the adiabatic cooler.

This configuration is known as “free cooling” with “adiabatic assist,” since the operation of the chiller, the primary cooling loop, the cooling tower, and the building heat rejection loop (see figure S5) are eliminated.  Free cooling is possible during periods when outdoor ambient air conditions, with the addition of injected water, allow the incoming rack air at point D to be sufficiently cool such that chiller operation isn’t necessary.  Free cooling via airside economizers with adiabatic assist works best in climates with warm, dry air and can be a viable cooling mode at many data centers for many hours of the year in such climates.

**Follow the heat: Air-cooled racks using "free cooling" via an airside economizer with indirect adiabatic cooling**

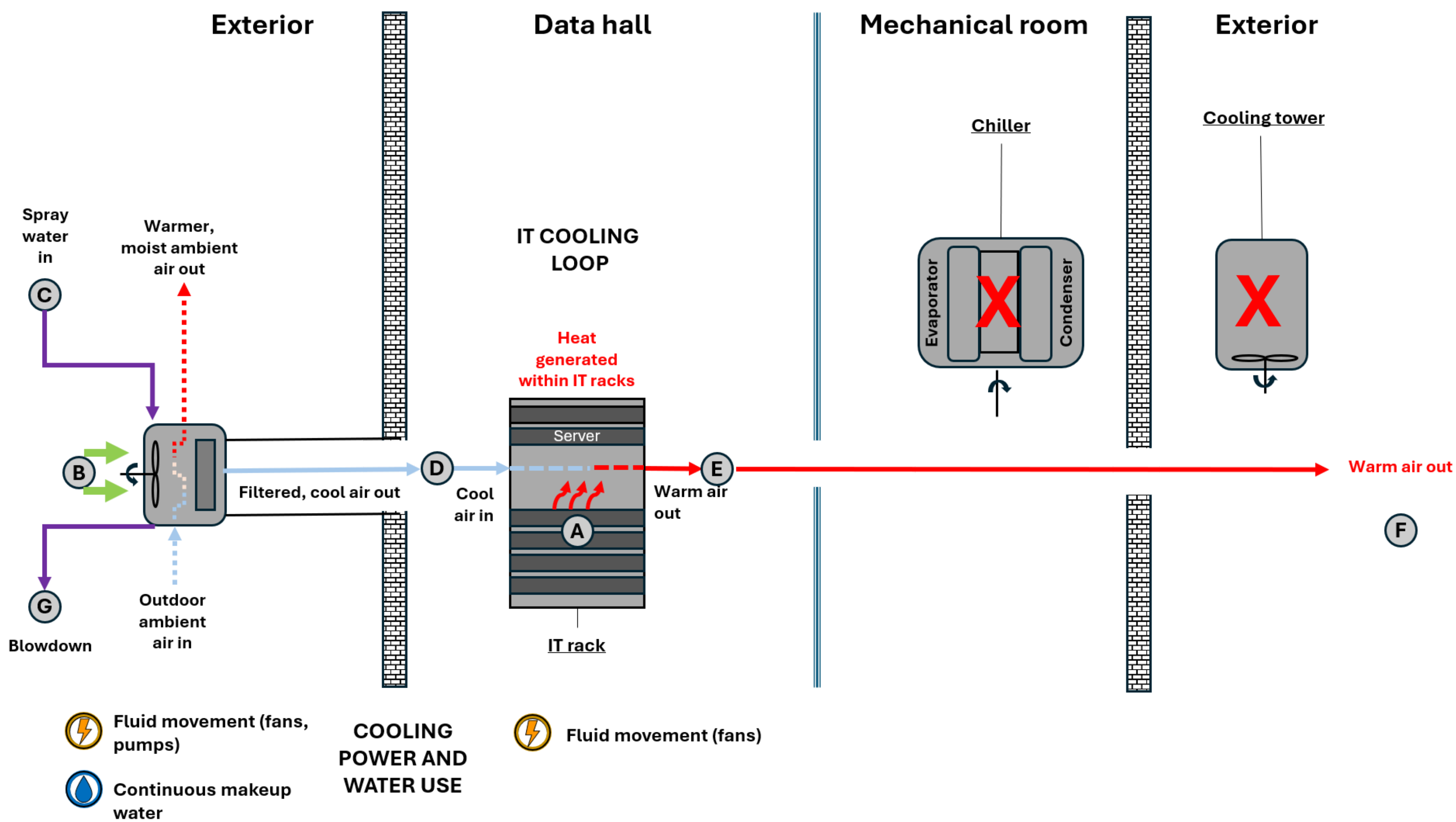

**Figure S16:** Air-cooled racks using "free cooling" via an airside economizer with indirect adiabatic cooling

This configuration utilizes air in the IT cooling loop.  Outside air is drawn into an indirect adiabatic cooler (point B), which contains a separate airflow channel that is sprayed with a wetted media (point C), cooling the airflow in that channel.  This separate airflow channel cools indirectly the incoming rack air (point D), which acts as the ultimate heat sink.  The cool incoming rack air (point D) absorbs the heat generated within the racks (point A), becoming warm air upon exiting the rack (point E).  This warm air is then directly exhausted to the exterior of the data center (point E), taking the heat that originated at the IT rack (point A) with it. When required, “blowdown” water is also discharged (point G) for wastewater treatment to remove concentrated minerals and water treatment chemicals that build up in the adiabatic cooler.

This configuration is known as “free cooling” with indirect “adiabatic assist,” since the operation of the chiller, the primary cooling loop, the cooling tower, and the building heat rejection loop (see figure S5) are eliminated.  Free cooling is possible during periods when outdoor ambient air conditions, with the addition of injected water, allow the incoming rack air at point D to be sufficiently cool such that chiller operation isn’t necessary.  Free cooling via airside economizers with adiabatic assist works best in climates with warm, dry air and can be a viable cooling mode at many data centers for many hours of the year in such climates.

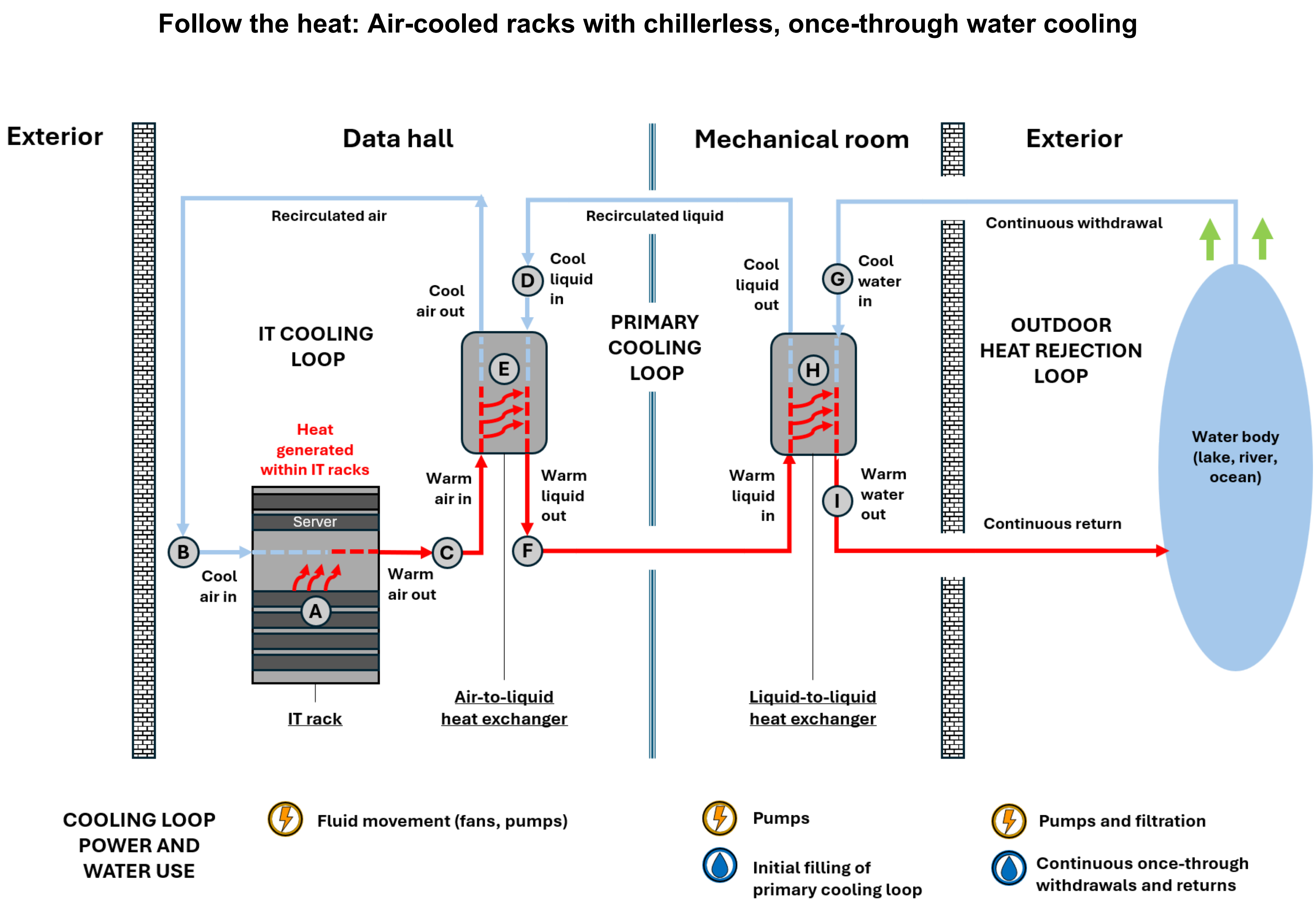

**S17:** Air-cooled racks with chillerless, once-through water cooling

This configuration utilizes air in the IT cooling loop, water (or a water/glycol mixture) in the primary cooling loop, and water in the building heat rejection loop. The primary cooling loop is closed and the building heat rejection loop is open. Cool incoming air (point B) absorbs the heat generated within the racks (point A), becoming warm air upon exiting the rack (point C). The warm air flows to an air-to-liquid heat exchanger, which is fed by cool liquid (point D). Heat moves from the warm air to the cool liquid (point E), the latter becoming warm as it exits the heat exchanger (point F). The warm liquid enters a liquid-to-liquid heat exchanger, which is fed by cool water (point G) that is withdrawn with filtration from an external water body, such as a river, lake, or ocean. This cool inlet water (point G) serves as the ultimate heat sink. Heat moves from the warm liquid to the cool water (point H), the latter becoming warm as it exits the heat exchanger (point I), taking the heat that originated at the IT rack (point A) with it. Because all of the water that is withdrawn (point G) is immediately returned to the water body (point I), this system is known as a “once through” cooling system. Since discharge quantities equal withdrawal quantities, no water consumption occurs.

Such configurations are associated with floating data centers or terrestrial data centers that use “once-through” cooling sourced from adjacent water bodies. This conceptualization may also apply to underwater data centers that cool the IT equipment with recirculated air, except the entire data center is submerged in the waterbody within a controlled-atmosphere, waterproof container.

## Follow the heat: Liquid-cooled racks with chillerless, once-through water cooling

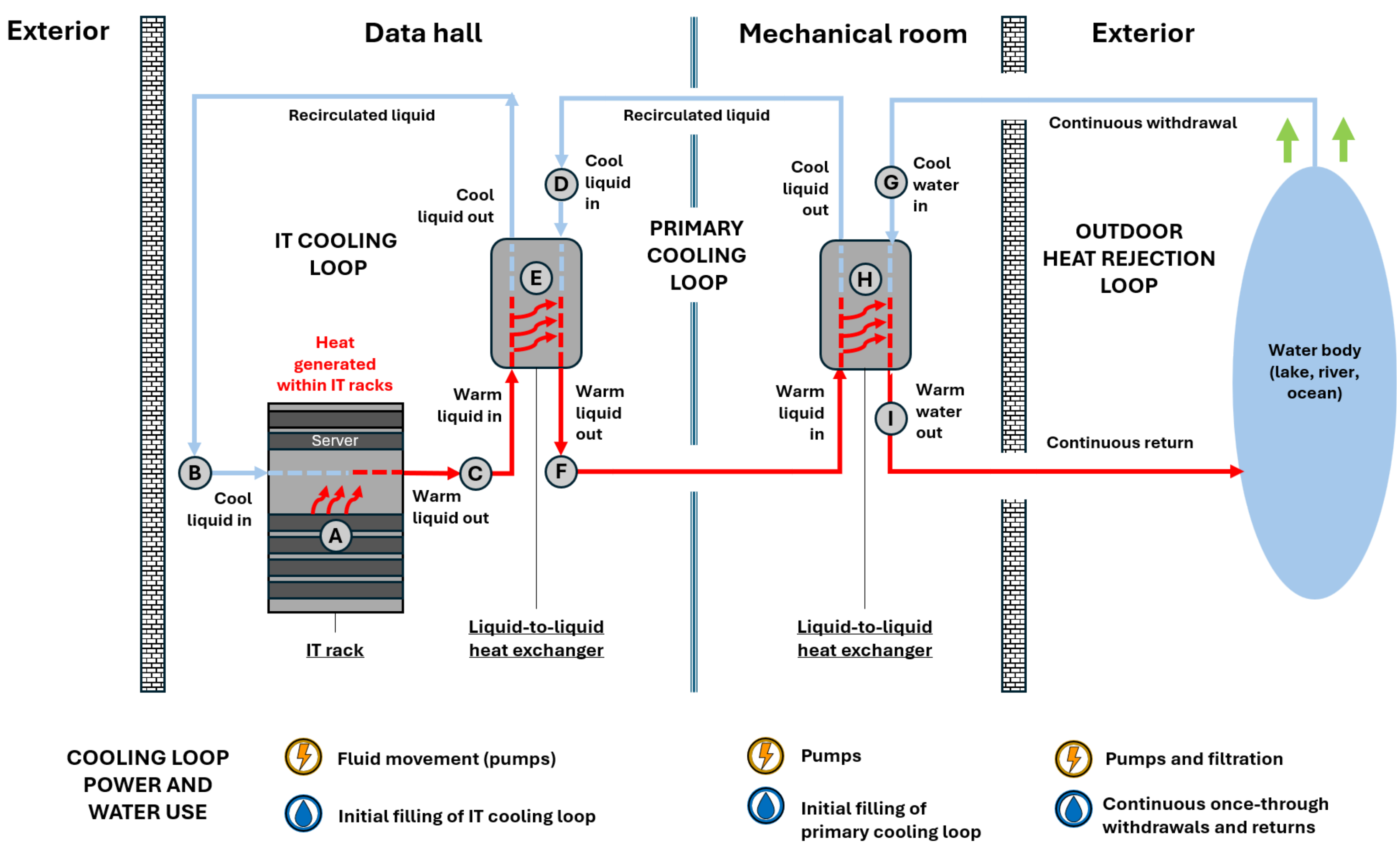

**S18:** Liquid-cooled racks with chillerless, once-through water cooling

This configuration utilizes liquid in the IT cooling loop, water (or a water/glycol mixture) in the primary cooling loop, and water in the building heat rejection loop. The IT cooling loop is closed, the primary cooling loop is closed, and the building heat rejection loop is open. Cool incoming liquid (point B) absorbs the heat generated within the racks (point A), becoming warm liquid upon exiting the rack (point C). The warm liquid flows to a liquid-to-liquid heat exchanger, which is fed by cool liquid (point D). Heat moves from the warm liquid to the cool liquid (point E), the latter becoming warm as it exits the heat exchanger (point F). The warm liquid enters a liquid-to-liquid heat exchanger, which is fed by cool water (point G) that is withdrawn with filtration from an external water body, such as a river, lake, or ocean. This cool inlet water (point G) serves as the ultimate heat sink. Heat moves from the warm liquid to the cool water (point H), the latter becoming warm as it exits the heat exchanger (point I), taking the heat that originated at the IT rack (point A) with it. Because all of the water that is withdrawn (point G) is immediately returned to the water body (point I), this system is known as a “once through” cooling system. Since discharge quantities equal withdrawal quantities, no water consumption occurs.

Such configurations are associated with floating data centers or terrestrial data centers that use “once-through” cooling sourced from adjacent water bodies. This conceptualization may also apply to underwater data centers that use liquid-cooling of IT equipment, except the entire data center is submerged in the waterbody within a controlled-atmosphere, waterproof container.

For single-phase immersion cooling systems, the IT equipment is contained in a tank instead of a traditional rack. The IT cooling loop contains a single-phase dielectric fluid that is pumped out of the tank at point C for heat removal at point E, after which point it is pumped back to the tank at point B for another cooling cycle.

For two-phase immersion cooling systems, the IT equipment is contained in a tank instead of a traditional rack. The tank contains a two-phase dielectric fluid that transitions from a liquid to a vapor when it absorbs heat, rising to the top of the tank. The heat exchanger depicted at point E is situated at the top of the tank and cools the vapor back into a liquid state, causing it to fall back down in the tank for more cooling. Thus, the “loop” depicted between point C and point B occurs entirely within the two-phase immersion cooling tank.

## Follow the heat: High-temperature liquid-cooled racks with dry cooling

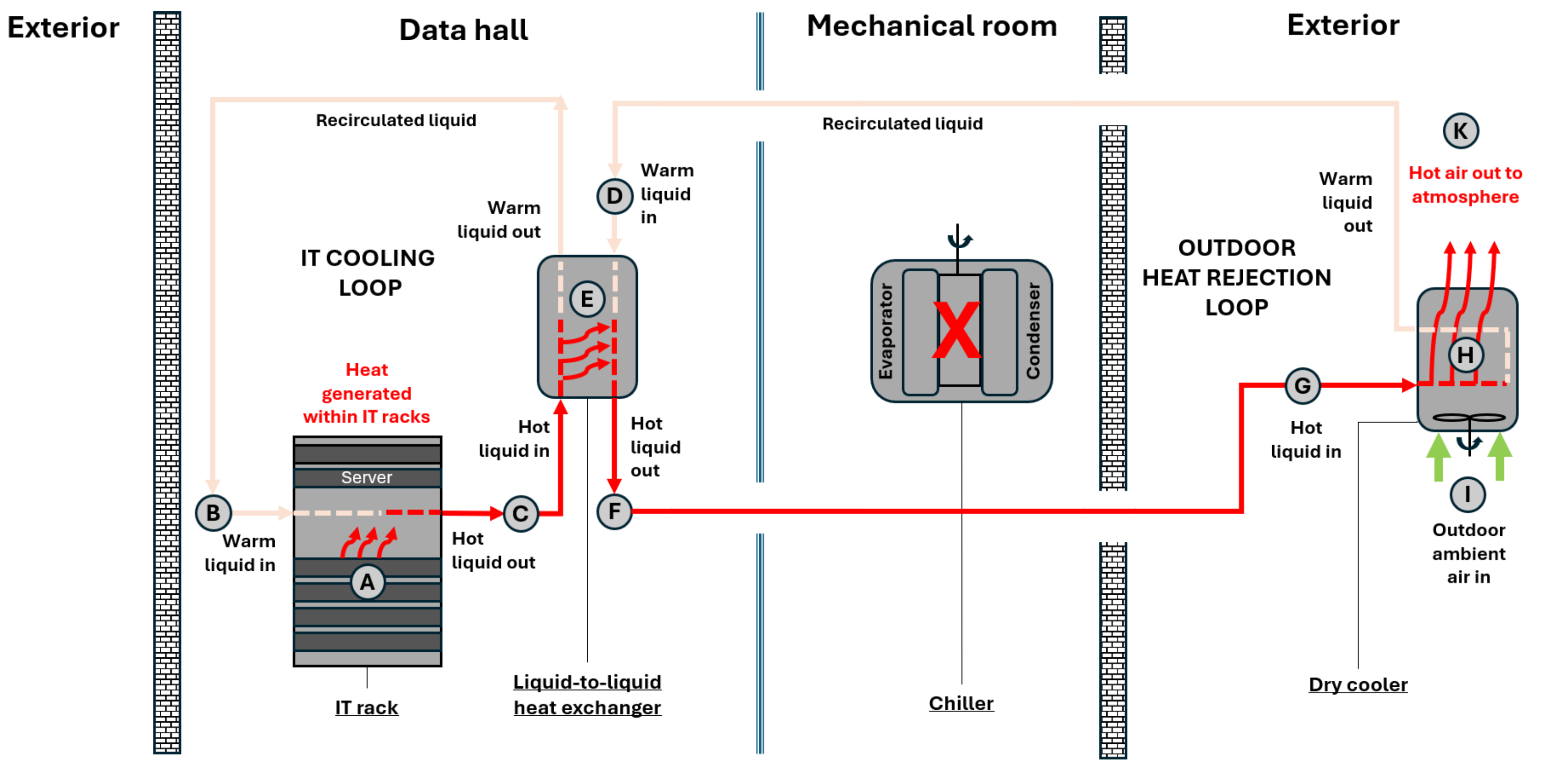

**COOLING LOOP POWER AND WATER USE**

- Fluid movement (pumps)
- Initial filling of IT cooling loop
- Fans, pumps
- Initial filling of heat rejection loop

**S19:** High-temperature liquid-cooled racks with dry cooling

This configuration utilizes liquid in the IT cooling loop (water, water/glycol mixture, or a dielectric fluid) and liquid (water or a water/glycol mixture) in the building heat rejection loop. The IT cooling loop is closed and the building heat rejection loop is closed. Additionally, high (e.g., 40-50 C) rack inlet temperature set points are utilized, allowing warm water to be used as the incoming rack heat removal liquid (point B). This warm incoming liquid (point B) absorbs the heat generated within the racks (point A), becoming hot liquid upon exiting the rack (point C). The hot liquid flows to a liquid-to-liquid heat exchanger, which is fed by warm liquid (point D) provided directly from the closed-loop “dry” cooler outside the building. Heat moves from the hot liquid to the warm liquid (point E), the latter becoming hot as it exits the heat exchanger (point F). The hot liquid returns to the closed-loop “dry” cooler (point G), where it flows through a closed liquid-to-air heat exchanger (point H). Ambient air (point I), which is the ultimate heat sink, is drawn across the “dry”cooler’s heat exchanger. The liquid exiting the heat exchanger becomes cooler, returning as warm liquid to the data hall heat exchanger (point D) for another cooling cycle. As the ambient air exits the heat exchanger (point K), it becomes hot, taking the heat that originated at the IT rack (point A) with it.

This configuration is referred to as “chillerless,” since the operation of the chiller and its primary cooling loop (see Figure S7) is eliminated. Chillerless operation is possible when outdoor dry bulb temperatures are sufficiently lower than the warm water temperature required at point D, which can be viable at many data centers for many hours of the year in suitable climates.

Immersion cooling systems can also operate at higher temperature set points.

For single-phase immersion cooling systems, the IT equipment is contained in a tank instead of a traditional rack. The IT cooling loop contains a single-phase dielectric fluid that is pumped out of the tank at point C for heat removal at point E, after which point it is pumped back to the tank at point B for another cooling cycle.

For two-phase immersion cooling systems, the IT equipment is contained in a tank instead of a traditional rack. The tank contains a two-phase dielectric fluid that transitions from a liquid to a vapor when it absorbs heat, rising to the top of the tank. The heat exchanger depicted at point E is situated at the top of the tank and cools the vapor back into a liquid state, causing it to fall back down in the tank for more cooling. Thus, the “loop” depicted between point C and point B occurs entirely within the two-phase immersion cooling tank.

## Follow the heat: High-temperature liquid-cooled racks with dry cooling and adiabatic assist

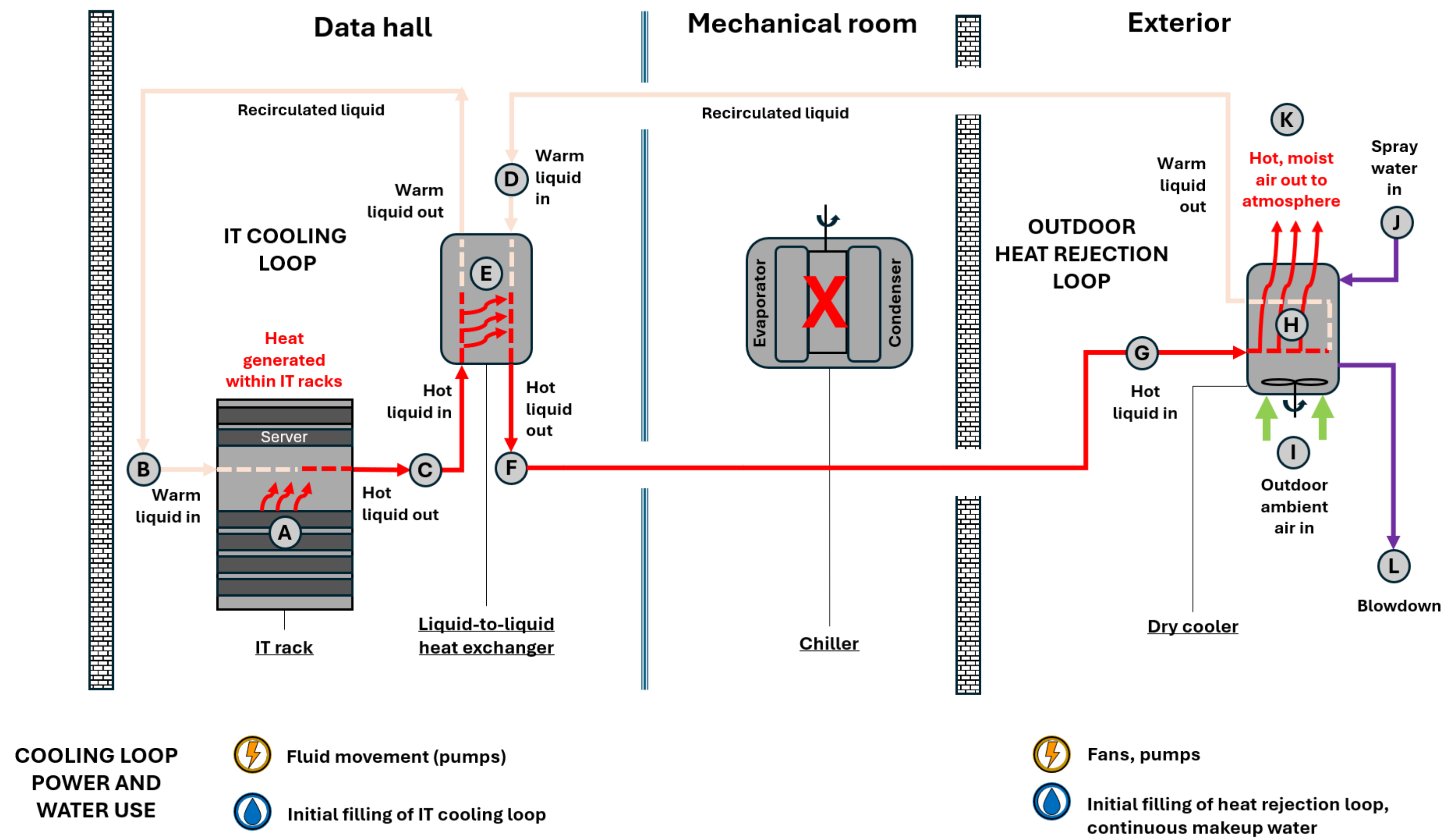

**S20:** High-temperature liquid-cooled racks with dry cooling and adiabatic assist

This configuration utilizes liquid in the IT cooling loop (water, water/glycol mixture, or a dielectric fluid) and liquid (water or a water/glycol mixture) in the building heat rejection loop.  The IT cooling loop is closed and the building heat rejection loop is closed. However, indirect “open loop” water use still occurs due to adiabatic assist in the “dry” cooler, as explained below.  Additionally, high (e.g., 40-50 C) rack inlet temperature set points are utilized, allowing warm water to be used as the incoming rack heat removal liquid (point B).  This warm incoming liquid (point B) absorbs the heat generated within the racks (point A), becoming hot liquid upon exiting the rack (point C). The hot liquid flows to a liquid-to-liquid heat exchanger, which is fed by warm liquid (point D) provided directly from the closed-loop “dry” cooler with adiabatic assist outside the building. Heat moves from the hot liquid to the warm liquid (point E), the latter becoming hot as it exits the heat exchanger (point F). The hot liquid returns to the closed-loop “dry” cooler with adiabatic assist (point G), where it flows through a closed liquid-to-air heat exchanger (point H).  Ambient air (point I), which is the ultimate heat sink, is drawn across the “dry”cooler’s heat exchanger.  The liquid exiting the heat exchanger becomes cooler, returning as warm liquid to the data hall heat exchanger (point D) for another cooling cycle.  Before the ambient air is drawn across the heat exchanger, it passes through a wetted media that is kept damp via a continuous water spray (point J). The introduction of water lowers the incoming air temperature, which increases its cooling ability, while also adding humidity.  As the cooled and humidified ambient air exits the heat exchanger (point K), it becomes hot, taking the heat that originated at the IT rack (point A) with it.

Additional spray water is continuously added to the cooling tower (point J) to replace the water that was lost to evaporation and due to any required “blowdown” water that is discharged (point L) to a wastewater treatment system.  Blowdown is necessary to remove concentrated minerals and water treatment chemicals that build up in the adiabatic cooler. The spray water is “open loop” even though the “dry” cooler’s outdoor heat rejection water is “closed loop.” This is known as “indirect” cooling of the heat rejection water loop.

This configuration is referred to as “chillerless,” since the operation of the chiller and its primary cooling loop (see Figure S8) is eliminated.  Chillerless operation is possible when outdoor wet bulb temperatures are sufficiently lower than the warm water temperature required at point D, which can be viable at many data centers for many hours of the year in suitable climates.

Immersion cooling systems can also operate at higher temperature set points. For single-phase immersion cooling systems, the IT equipment is contained in a tank instead of a traditional rack. The IT cooling loop contains a single-phase dielectric fluid that is pumped out of the tank at point C for heat removal at point E, after which point it is pumped back to the tank at point B for another cooling cycle. For two-phase immersion cooling systems,  the IT equipment is contained in a tank instead of a traditional rack.  The tank contains a two-phase dielectric fluid that transitions from a liquid to a vapor when it absorbs heat, rising to the top of the tank. The heat exchanger depicted at point E is situated at the top of the tank and cools the vapor back into a liquid state, causing it to fall back down in the tank for more cooling.  Thus, the “loop” depicted between point C and point B occurs entirely within the two-phase immersion cooling tank.

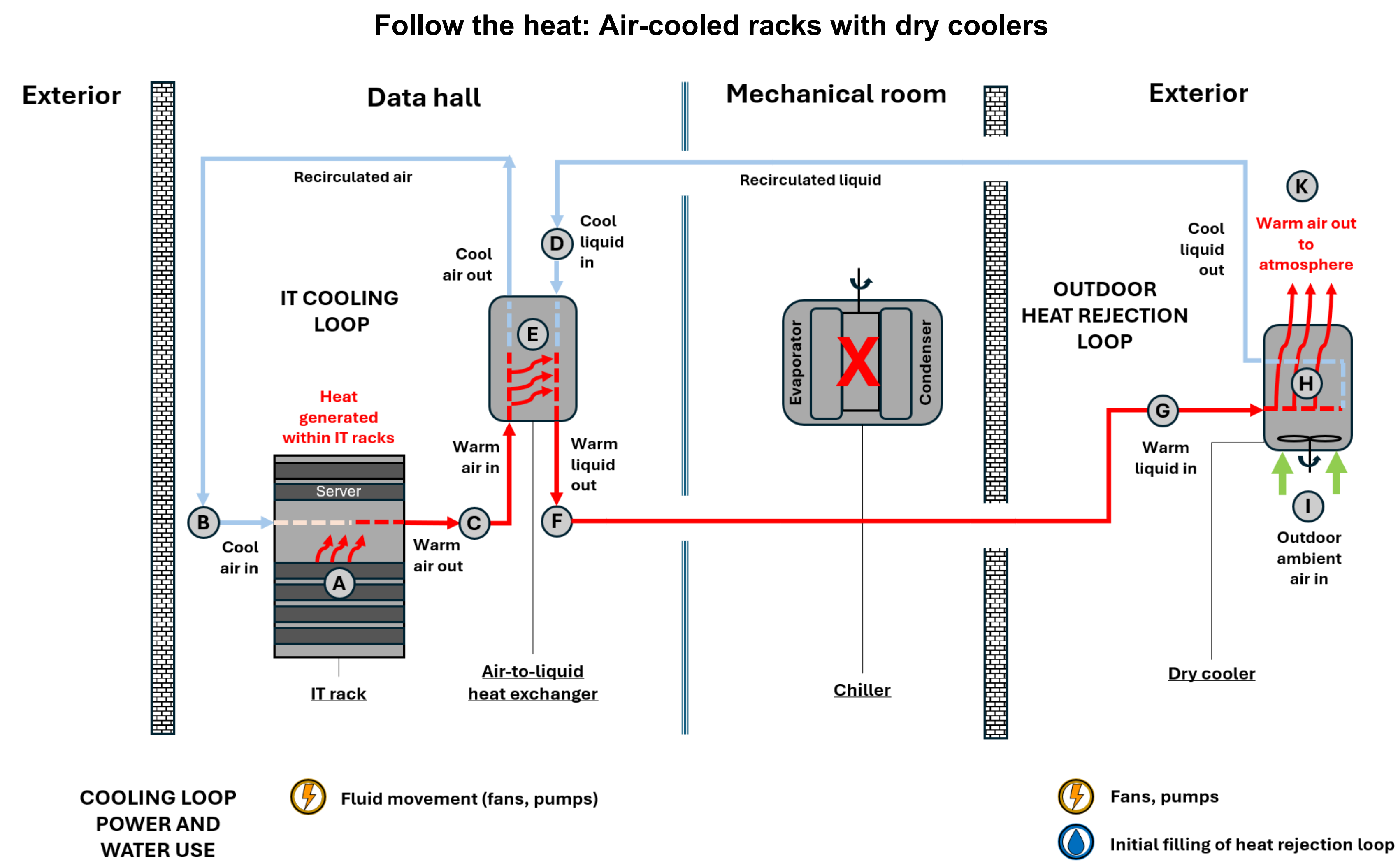

**S21:** Air-cooled racks with dry coolers

This configuration utilizes air in the IT cooling loop and liquid (water or a water/glycol mixture) in the building heat rejection loop. The building heat rejection loop is closed. The cool incoming air (point B) absorbs the heat generated within the racks (point A), becoming warm air upon exiting the rack (point C). The warm air flows to an air-to-liquid heat exchanger, which is fed by cool liquid (point D) provided directly from the closed-loop “dry” cooler outside the building. Heat moves from the warm air to the cool liquid (point E), the latter becoming warm as it exits the heat exchanger (point F). The warm liquid returns to the closed-loop “dry” cooler (point G), where it flows through a closed liquid-to-air heat exchanger (point H). Ambient air (point I), which is the ultimate heat sink, is drawn across the “dry”cooler’s heat exchanger. The liquid exiting the heat exchanger becomes cooler, returning as cool liquid to the data hall heat exchanger (point D) for another cooling cycle. As the ambient air exits the heat exchanger (point K), it becomes warm, taking the heat that originated at the IT rack (point A) with it.

This configuration is referred to as “chillerless,” since the operation of the chiller and its primary cooling loop (see Figure S8) is eliminated. Chillerless operation is possible when outdoor dry bulb temperatures are sufficiently lower than the cool water temperature required at point D, which can be viable at many data centers for many hours of the year in suitable climates.

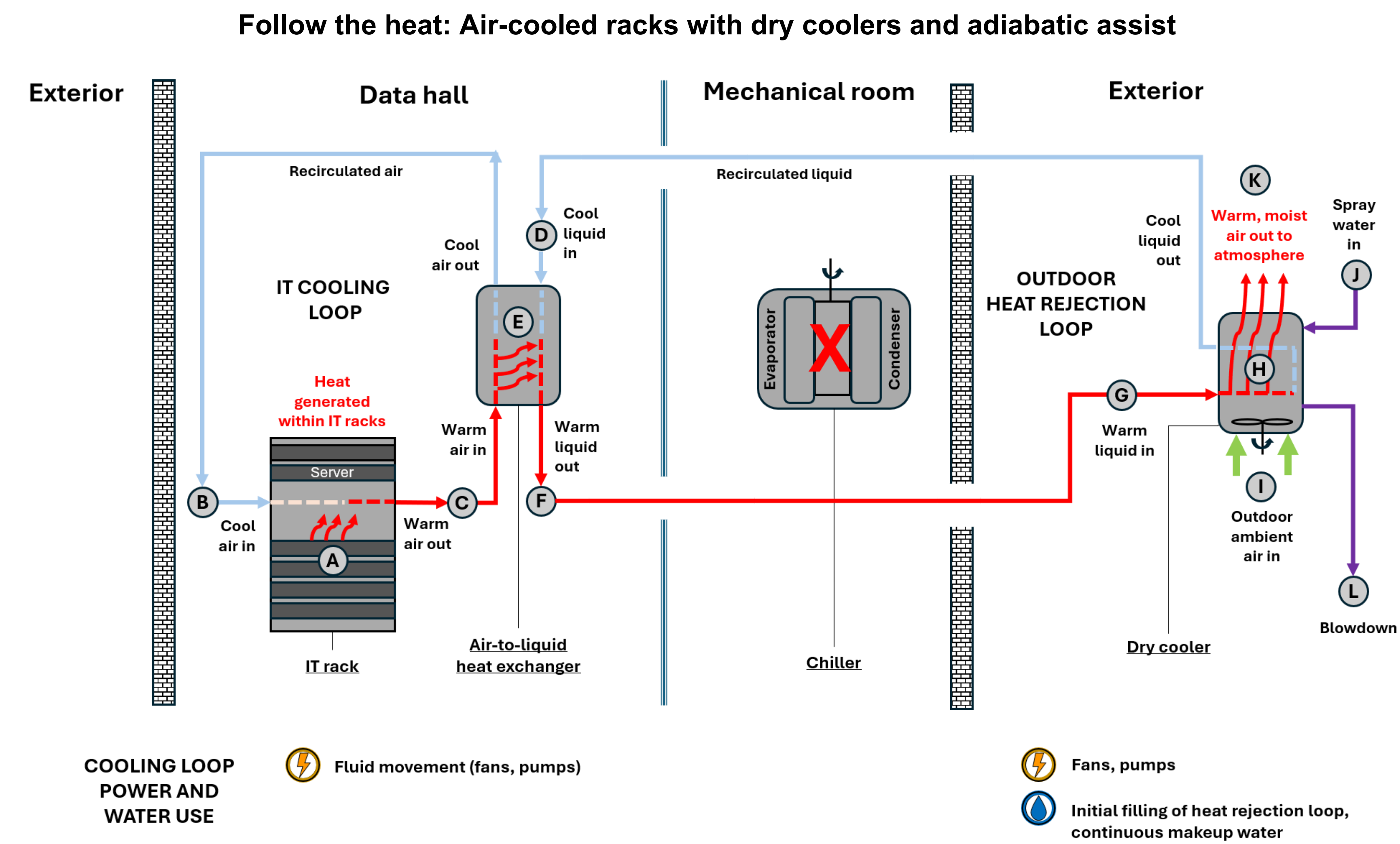

**S22:** Air-cooled racks with dry coolers and adiabatic assist

This configuration utilizes air in the IT cooling loop and liquid (water or a water/glycol mixture) in the building heat rejection loop. The building heat rejection loop is closed. The cool incoming air (point B) absorbs the heat generated within the racks (point A), becoming warm air upon exiting the rack (point C). The warm air flows to an air-to-liquid heat exchanger, which is fed by cool liquid (point D) provided directly from the closed-loop “dry” cooler with adiabatic assist outside the building. Heat moves from the warm air to the cool liquid (point E), the latter becoming warm as it exits the heat exchanger (point F). The warm liquid returns to the closed-loop “dry” cooler with adiabatic assist (point G), where it flows through a closed liquid-to-air heat exchanger (point H). Ambient air (point I), which is the ultimate heat sink, is drawn across the “dry”cooler’s heat exchanger. The liquid exiting the heat exchanger becomes cooler, returning as cool liquid to the data hall heat exchanger (point D) for another cooling cycle. Before the ambient air is drawn across the heat exchanger, it passes through a wetted media that is kept damp via a continuous water spray (point J). The introduction of water lowers the incoming air temperature, which increases its cooling ability, while also adding humidity. As the cooled and humidified ambient air exits the heat exchanger (point K), it becomes warm, taking the heat that originated at the IT rack (point A) with it.

Additional spray water is continuously added to the cooling tower (point J) to replace the water that was lost to evaporation and due to any required “blowdown” water that is discharged (point L) to a wastewater treatment system. Blowdown is necessary to remove concentrated minerals and water treatment chemicals that build up in the adiabatic cooler. The spray water is “open loop” even though the “dry” cooler’s outdoor heat rejection water is “closed loop.” This is known as “indirect” cooling of the heat rejection water loop.

This configuration is referred to as “chillerless,” since the operation of the chiller and its primary cooling loop (see Figure S8) is eliminated. Chillerless operation is possible when outdoor dry bulb temperatures are sufficiently lower than the cool water temperature required at point D, which can be viable at many data centers for many hours of the year in suitable climates.

# S5: Best practice alignment evaluation rubric

Table S3 presents a simple three-point scoring rubric to assess study adherence to each water use estimation best practice in Table 3 of the main paper. Scores of 3, 2, and 1 corresponding to strong, moderate, and weak/no adherence to a given criterion, respectively. The scoring rubric was designed to be as objective and replicable as possible when measuring the adherence of each study in Table S1 to each best practice. However, the goal of the study is to identify collective trends across studies rather than to isolate or focus on the strengths or weaknesses of any particular study, since all studies have improvement potential. The rationale for each score assigned to each study can be reviewed in the Supplementary Information Excel file.

**Table S3:** Best practice alignment evaluation rubric applied to studies in Table S1

| | | | **Study assessment score** | | |
|---|---|---|---|---|---|
| **Category** | **Criterion** | **#** | **1 (weak/no adherence)** | **2 (moderate adherence)** | **3 (strong adherence)** |
| **Clarity** | Scope | BP1 | The study does not clearly state the data center water use stages (Scope 1/direct, Scope 2/indirect, Scope 3/embodied) and components (withdrawals, inputs, consumption, discharges) included in the study | The study clearly states some, but not all, of the data center water use stages (Scope 1/direct, Scope 2/indirect, Scope 3/embodied) and components (withdrawals, inputs, consumption, discharges) included in the study or the scope can be reasonably inferred from the contents | The study clearly states the data center water use stages (Scope 1/direct, Scope 2/indirect, Scope 3/embodied) and components (withdrawals, inputs, consumption, discharges) included in the study |
| | Definitions | BP2 | Doesn't clearly define all included water use components, or analogous terms for these concepts | Doesn't clearly define all included water use components, or analogous terms for these concepts, but their definitions can be reasonably inferred from the contents | Clearly defines all included water use components, especially withdrawals and/or consumption, or analogous terms for these concepts |

| Category | Criterion | # | Study assessment score<br>1 (weak/no adherence) | 2 (moderate adherence) | 3 (strong adherence) |
|---|---|---|---|---|---|
| | Terminological consistency | BP3 | Mixes and uses multiple competing terms that may be misinterpreted, with many instances of variants or instances of variants seen in results graphs/tables, which can confuse readers and enable misuse of results | Mixes some terms, but not to the point where a layreader would reasonably misinterpret the study, with instances of variants being substantial but not overwhelming the instances of defined terms and no instances of variants in results graphs/tables, which can confuse readers and enable misuse of results | Uses consistent terminology throughout the study when describing the included water use components, with fewer instances of variants compared to defined terms and no instances of variants in results graphs/tables, which can confuse readers and enable misuse of results |
| **Scientific due diligence** | Electricity demand transparency | BP4a | Does not state which, or the type of, electricity demand value/outlook that is used, whether total or IT electricity demand and irrespective of units (watts of power capacity or watt-hours of electricity demand) | Clearly states which, and the type of (e.g., low, medium, or high), total electricity demand value(s)/outlook(s) that is/are used, but not IT electricity demand value(s) or the reported values are in units of watts of power capacity, not watt-hours of electricity demand | Clearly states which, and the type of (e.g., low, medium, or high), IT electricity demand value(s)/outlook(s) that is/are used and in units of watt-hours |
| | Electricity demand justification | BP5a | Does not offer a clear, scientific justification for the chosen electricity values/outlooks | Offers only limited/partial justification of the chosen electricity values/outlooks | Offers clear, scientific justification for the chosen electricity values/outlooks considering the goals of the study |
| | WUE transparency | BP4b | Does not state the WUE value(s) utilized | Does not state WUE value(s) directly, but refers to other sources containing the WUE value(s) or the WUE value(s) can be reasonably inferred from the content or the stated WUE value(s) use non-conventional units, such as liters per kWh of total data center electricity use | Clearly states the WUE value(s) utilized in units of l/kWh-IT |
| | WUE justification | BP5b | Does not offer a clear, scientific justification for the chosen WUE value(s) | Offers only limited/partial justification of the chosen WUE value(s) | Offers clear, scientific justification for the chosen WUE value(s) considering the goals of the study |

| Category | Criterion | # | Study assessment score<br>1 (weak/no adherence) | 2 (moderate adherence) | 3 (strong adherence) |
|---|---|---|---|---|---|
| | PUE transparency | BP4c | Does not state the PUE value(s) utilized | Does not state PUE value(s) directly, but refers to other sources containing the PUE value(s) or the PUE value(s) can be reasonably inferred from the content | Clearly states the PUE value(s) utilized in l/kWh-IT or comparable units |
| | PUE justification | BP5c | Does not offer a clear, scientific justification for the chosen PUE value(s) OR no demonstration of congruence with chosen WUE value(s) | Offers only limited/partial justification of the chosen PUE value(s) or demonstration of congruence with the chosen WUE value(s) | Offers clear, scientific justification for the chosen PUE value(s) considering the goals of the study and demonstrates congruence with the chosen WUE value(s) |
| | EWIF transparency | BP4d | Does not state the specific EWIF value(s) utilized and these cannot be easily inferred from the content | Does not state EWIF value(s) directly, but refers to other sources or EWIF value(s) can be easily inferred from the content | Clearly states the EWIF value(s) utilized in l/kWh or comparable units |
| | EWIF justification | BP5d | Does not offer a clear, scientific justification for the chosen EWIF value(s) considering the goals of the study | Offers only limited/partial justification of the chosen EWIF value(s) considering the goals of the study | Offers clear, scientific justification for the chosen EWIF value(s) considering the goals of the study |
| | Hydroelectricity EWIF transparency | BP6 | No description of the evaporation basis (gross or net) and allocation basis for any chosen EWIF(s) that contain or are specific to hydroelectric power sources in the study's geography | Describes either the evaporation basis (gross or net) or allocation basis for any chosen EWIF(s) that contain or are specific to hydroelectric power sources, but not both | Clearly describes the evaporation basis (gross or net) and allocation basis for any chosen EWIF(s) that contain or are specific to hydroelectric power sources or clearly indicates the absence of any hydroelectric power in its assumed power mixes. |
| | Uncertainty analysis | BP7 | No quantitative or qualitative treatment of uncertainty; only point values presented for results. | Uncertainty quantified for some, but not all, results and/or uncertainty is meaningfully discussed qualitatively. | Quantitative uncertainty analysis is provided, inclusive of upper/lower values and/or scenarios, possibly in addition to qualitative uncertainty discussion. |

| Category | Criterion | # | Study assessment score<br>1 (weak/no adherence) | 2 (moderate adherence) | 3 (strong adherence) |
|---|---|---|---|---|---|
| **Scientific quality and replicability** | WUE representativeness | BP8a | Chosen WUE value(s) do not correspond to expected or modeled cooling technologies and practices in the study's geographical scope, inclusive of expected changes over time, or this cannot be established based on provided information | Chosen WUE value(s) correspond to expected or modeled cooling technologies and practices, but not consistent with the study's geographical scope, inclusive of expected changes over time | Chosen WUE value(s) correspond to expected or modeled cooling technologies and practices in the study's geographical scope, inclusive of expected changes over time |
| | PUE representativeness | BP8b | Chosen PUE value(s) do not correspond to expected or modeled cooling technologies and practices in the study's geographical scope, inclusive of expected changes over time and are not congruent with the study's chosen WUE value(s), or this cannot be established based on provided information | Chosen PUE value(s) correspond to expected or modeled cooling technologies and practices, including consistency with the study's geographical scope and expected changes over time, but are not congruent with the study's chosen WUE value(s) | Chosen PUE value(s) correspond to expected or modeled cooling technologies and practices in the study's geographical scope, inclusive of expected changes over time AND are congruent with the study's chosen WUE value(s) |
| | EWIF representativeness | BP8c | Chosen EWIF value(s) do not correspond to expected or modeled power technologies, cooling configurations, and climate conditions in the study's geographical scope, and considering the study time period, inclusive of expected changes over time, or this cannot be established based on provided information. | Chosen EWIF value(s) correspond to expected or modeled power technologies, but not consistent with the study's geographical scope and/or cooling configurations, and/or considering the study time period, inclusive of expected changes over time | Chosen EWIF value(s) correspond to expected or modeled power technologies, cooling configurations, and climate conditions in the study's geographical scope, and considering the study time period, inclusive of expected changes over time |
| | Replicability | BP9 | Replication is not possible due to lack of data, equations, assumptions, and/or results presented in the study. | Replication is possible, but not easily, and only through backing out assumptions from provided information and/or consulting study references. | All water use estimates can be replicated easily using data, equations, assumptions, and results presented in the study. |

| Category | Criterion | # | Study assessment score<br>1 (weak/no adherence) | 2 (moderate adherence) | 3 (strong adherence) |
|---|---|---|---|---|---|
| | Reporting granularity | BP10 | Reports only aggregate results that combine indirect and direct water use values together, and data center types, and/or into cumulative temporal totals only, depending on the scope | Does not fully report water analysis results by water use stage, water use component, key data center market segment, and year (or other timestep), but some can be reasonably inferred from the document | Reports water analysis results by water use stage, water use component, key data center market segment, and year (or other timestep) depending on the scope |
| | Peer review | BP11 | No peer review or acknowledgement of external feedback is evident. | Inputs and/or comments from external experts are acknowledged, but these are not framed as reviewers or constituting a review. | Peer-reviewed journal, conference proceeding, or acknowledgement of peer/external expert reviewers, either by name or anonymously. |
| **Community knowledge building** | Limitations | BP12 | No limitations related to specific methods, data sources, or findings related to water analysis are discussed. | Limitations related to specific methods, data, and/or results associated with water analysis are acknowledged, but with limited/no discussion. | Limitations related to specific methods, data, and/or results associated with water analysis are thoroughly discussed. |
| | Future work | BP13 | No future work suggestions associated with water analysis are offered. | Future work is suggested associated with water analysis, but in brief, limited, and/or generic fashion. | Multiple and specific future work suggestions associated with water analysis are offered to address limitations. |
| **Communications** | Valid comparisons | BP14 | Comparisons of results to other types of water use are not made on "apples to apples" bases (e.g., water consumption is compared with water withdrawal) and/or the comparative data aren't accompanied by any cited sources or explanations. | Some, but not all, comparisons of results to other types of water use are made on "apples to apples" basis (e.g., water consumption is not compared with water withdrawal) and those have comparative data sources are cited and explained. | All comparisons of results to other types of water use are made on "apples to apples" basis (e.g., water consumption is not compared with water withdrawal), as determined via citing and explaining the activity data to which data center water use are compared. |

# S6: Terminology frequency assessment

Table S4 summarizes the estimated frequency of different water use-related terms found in each of the studies listed in Table S1. Terms are divided into two categories: (1) the standard, precise water use components that are the specific targets of analysis within the study (i.e., withdrawals, inputs, consumption, and/or discharges), and (2) other “competing” terms that are used within the same study when referring to methods, results, or definitions. The latter category might sow confusion about which water use component is being referred to in a given statement, table, or figure.

As discussed in the main paper, studies that strictly refer to specific water use components with minimal use of competing terms can minimize reader confusion and potential misinterpretations of results. Conversely, studies that use multiple and many competing terms increase the likelihoods of reader confusion and subsequent misinterpretations and misuses of results (50).

The results in Table S4 were obtained by keyword searches of both target and competing terms within each study; competing terms include all of those found within the studies listed in Table S1. Each keyword search “hit” was then reviewed to establish if it was used in reference to the study’s water use methods, results, definitions, or other context that could affect study interpretation. When a “hit” wasn’t used as a competing term, it was excluded from the counts in Table S4. For example, “water usage” was included when it was used to describe context, methods, or results in a potentially confusing way, but “water usage” was excluded when referring to the metric “water usage effectiveness,” or WUE.

Only the main study documents were evaluated; examples include the main paper for peer-reviewed journal articles or the main report for technical reports. Supporting files (e.g., supporting information, appendix files, or code) that must be downloaded separately were not included in the keyword searches, since they may not be read by casual audiences. The estimated totals are listed in Table S4. Estimates are meant only as a rough indication of the potential for confusion and divergence from BP3. Future work may be warranted for more thorough analysis of the specific context of each competing term’s use, which may change the totals in Table S4.

Note that the usage of competing terms doesn’t always connote lack of clarity. For example, some studies use competing terms for background and context statements and then focus on target terms in their analyses. Some studies define competing terms clearly in terms of their target terms, making the use of competing terms less confusing. In such instances, studies weren’t necessarily assigned a low alignment score for BP3. Conversely, some studies with low instances of competing terms may have received low alignment scores for BP3 if the competing terms were used in results figures, tables, and or explanations.

| | Instances of target water use component terms* | | | | | Instances of competing terms, i.e., "water ______" | | | | | | | | |
|---|---|---|---|---|---|---|---|---|---|---|---|---|---|---|
| **Label** | **Withdrawals** | **Inputs** | **Consumption** | **Discharges** | **Total** | **Demand** | **Footprint** | **Impacts** | **Intensity** | **Losses** | **Needs** | **Use/usage** | **Requirements** | **Total** |
| 2016-1 | | | 10 | | **10** | 0 | 0 | 0 | 0 | 3 | 0 | 3 | 1 | **7** |
| 2021-1 | | | | | **0** | | 14 | | | | 1 | | | **15** |
| 2021-2 | | | 27 | | **27** | 6 | 43 | | 5 | | | 7 | 4 | **65** |
| 2023-1 | | | 33 | | **33** | 16 | 7 | 6 | 2 | 0 | 2 | 6 | 6 | **45** |
| 2024-1 | | | 13 | | **13** | 7 | 3 | | | | 1 | 24 | 3 | **38** |
| 2024-2 | | | 30 | | **30** | | 2 | | | | 1 | 6 | 1 | **10** |
| 2024-3 | | | 33 | | **33** | 2 | 3 | 1 | 3 | 0 | 0 | 7 | 1 | **17** |
| 2025-1 | | | 12 | | **12** | 0 | 0 | 0 | 0 | 0 | 1 | 4 | 0 | **5** |
| 2025-10 | | | 62 | | **62** | 4 | | 2 | | | 1 | 47 | 7 | **61** |
| 2025-11 | | | 29 | | **29** | 14 | 9 | 3 | 9 | 5 | 3 | 67 | | **110** |
| 2025-2 | 36 | | 69 | | **105** | 2 | 55 | 1 | 1 | 0 | 3 | 27 | 1 | **90** |
| 2025-3 | 7 | | 9 | | **16** | 2 | 0 | 0 | 2 | 0 | 1 | 7 | 3 | **15** |
| 2025-4 | 4 | | 34 | | **38** | 8 | 42 | | | | | 42 | | **92** |
| 2025-5 | | | | | **0** | 6 | 1 | 4 | 0 | 0 | 3 | 2 | 2 | **18** |
| 2025-6 | | | 65 | | **65** | 13 | 27 | | 8 | 5 | | 28 | 3 | **84** |
| 2025-7 | | | 9 | 13 | **22** | 10 | 1 | 4 | 5 | 0 | 1 | 55 | 2 | **78** |
| 2025-8 | | | 2 | | **2** | | 60 | | | 1 | | 2 | | **63** |
| 2025-9 | | | 98 | | **98** | 2 | 11 | 1 | 1 | 1 | 0 | 17 | 8 | **41** |
| 2026-1 | | | 66 | | **66** | | 24 | | | | | 1 | | **25** |
| 2026-2 | 59 | | 31 | | **90** | 11 | 19 | 13 | 16 | 15 | 2 | 115 | 4 | **195** |
| 2026-3 | 10 | | 37 | | **47** | 48 | 0 | 1 | 2 | 1 | 7 | 50 | 6 | **115** |
| 2026-4 | 6 | | 23 | | **29** | 53 | 2 | | 2 | | 38 | 19 | 3 | **117** |
| 2026-5 | 152 | | 94 | | **246** | 59 | 16 | | | 5 | 2 | 146 | 3 | **231** |
| 2026-6 | | | 14 | | **14** | 3 | 31 | 2 | 6 | 1 | 2 | | 2 | **47** |
| 2026-7 | 5 | | 7 | | **12** | 6 | 3 | 3 | | | | 9 | 3 | **24** |
| 2026-8 | | | | | **0** | 2 | 20 | 0 | 4 | 0 | 16 | 2 | 1 | **45** |
| 2026-9 | 2 | | 28 | | **30** | | 7 | | 1 | | | 10 | | **18** |
| 2026-10 | | | 17 | | **17** | 3 | 12 | | | | | 12 | 1 | **28** |
| 2026-11 | 73 | | 58 | | **131** | 6 | | 5 | | | | 95 | | **106** |
| 2026-12 | 22 | | | | **22** | 1 | 0 | 2 | 0 | 0 | 3 | 69 | 3 | **78** |
| 2026-13 | | | 15 | | **15** | 2 | 3 | 1 | | | | 3 | 1 | **10** |
| **Total** | 376 | 0 | 925 | 13 | **1314** | 286 | 415 | 49 | 67 | 37 | 88 | 882 | 69 | **1893** |

Note: based on main text keyword search counts, which may not be exact, and exclusions of instances that are not directly related to water use
* includes instances in text, tables, and figures and common variants (e.g., water "consumed" is counted as water "consumption")

**Table S4:** Keyword search "hits" for target terms and common competing terms

# S7: Origins of WUE and PUE values

During the evaluation of each study, its sources of WUE and/or PUE values were recorded in order to assess the representativeness of these values for the study's stated data center type(s), region(s), and time horizon(s). All studies including direct water use estimates were assessed for WUE value(s), whereas all studies including indirect water use estimates were assessed for PUE value(s) (Figure S1). Occasionally, a study estimated only direct water use, but utilized PUE assumptions to derive the IT electricity values associated with its direct water use estimates. In such cases, PUE sources were recorded.

If no WUE or PUE sources were explicitly declared or otherwise discernible in a study, its sources were considered undisclosed. When specific sources were declared, they were further assessed to identify any dependence on preceding studies for the WUE and/or PUE values contained in those specific sources. The goal was to trace back to all original sources of WUE and PUE values utilized in the 31 assessed studies.

The origins of declared WUE and PUE values are depicted in network diagrams in Figures S23 and S24, respectively. Original sources are labeled as providing reported values, advertised values, simulated values, or estimated values. Reported values are those that were published in audited company reports, whereas advertised values are those that appear in data center directories. Simulated values are those that are generated from PUE and/or WUE simulation models. Estimated values are those that appear as original estimates without citations or further information.

Additionally, studies that used onsite water intensity values (see main paper, Section 4.16) instead of WUE values were labeled.

Figure S23 maps the 27 studies that included direct water use estimation, wherein their citations to previous studies for WUE or onsite water intensity values are indicated with leftward arrows. Studies appear chronologically from left to right.

While most studies disclosed the sources of their WUE assumptions, four studies did not. Furthermore, more than one-third (10) of the 27 studies did not use an ISO 30134-compliant WUE value in their direct water use estimations.  Rather, these studies used an onsite water intensity value that was expressed in liters per kWh of total data center electricity use, precluding easy comparability to studies that used WUE values.

Three major clusters can be observed in Figure S23 that shed light on which types of WUE (or onsite water intensity) values were used for direct water use estimation.  The first cluster of ten studies utilized simulated WUE values.  This cluster of studies traces back to simulations run in 2025, 2024, or 2022 by Lei et al. (8), Shehabi et al. (25), or Lei and Masanet (51), respectively, all of which trace back to the Lei and Masanet (52) model originally developed in 2020.  The second cluster of ten studies utilized onsite water intensity values that ultimately trace back to two estimated values, from Shehabi et al. (20) and Heslin (53), both of which were published in 2016.  The third cluster of three studies utilized recently reported WUE values from major data center operators (Amazon, Equinix, Google, and Meta) or advertised WUE values from data center market directories. Thus, the use of recently operator-reported/provided WUE data remains rare, especially those that have been verified, with most studies relying on simulated or estimated values from previous studies, some of which may be (much) older than study time horizons.

Figure S24 maps the 27 studies that included indirect water use estimation, wherein their citations to previous studies for PUE values are indicated with leftward arrows. Studies appear chronologically from left to right.

Of these 27 studies, two-fifths (11) didn't disclose the PUE values associated with their indirect water use estimates, which makes verification of congruence with any utilized WUE values impossible. Non-disclosure of PUE values also obscures assumptions related to the cooling technologies and efficiencies underlying the analyses, which is a barrier to stakeholder transparency.

Similar to Figure S23 for WUE, three distinct clusters emerge for cited PUE sources. The first cluster of eight studies utilized simulated PUE values. This cluster of studies traces back to simulations run in 2024 or 2022 by Shehabi et al. (25) or Lei and Masanet (51), respectively, both of which trace back to the Lei and Masanet (52) model originally developed in 2020. The second cluster of four studies utilized PUE values that ultimately trace back to PUE values published by Shehabi et al. in 2011 (54), which themselves were based on previously published values, expert elicitation, and energy modeling results (20).

Finally, the third cluster of three studies utilized recently reported PUE values from data center operators (Google) or within industry surveys (Uptime Institute) or advertised WUE values from data center market directories (Datacentermap.com, idcquan.com). Thus, the use of recently operator-reported/provided PUE data also remains rare, with most studies relying on simulated or estimated values from previous studies, some of which may be (much) older than study time horizons.

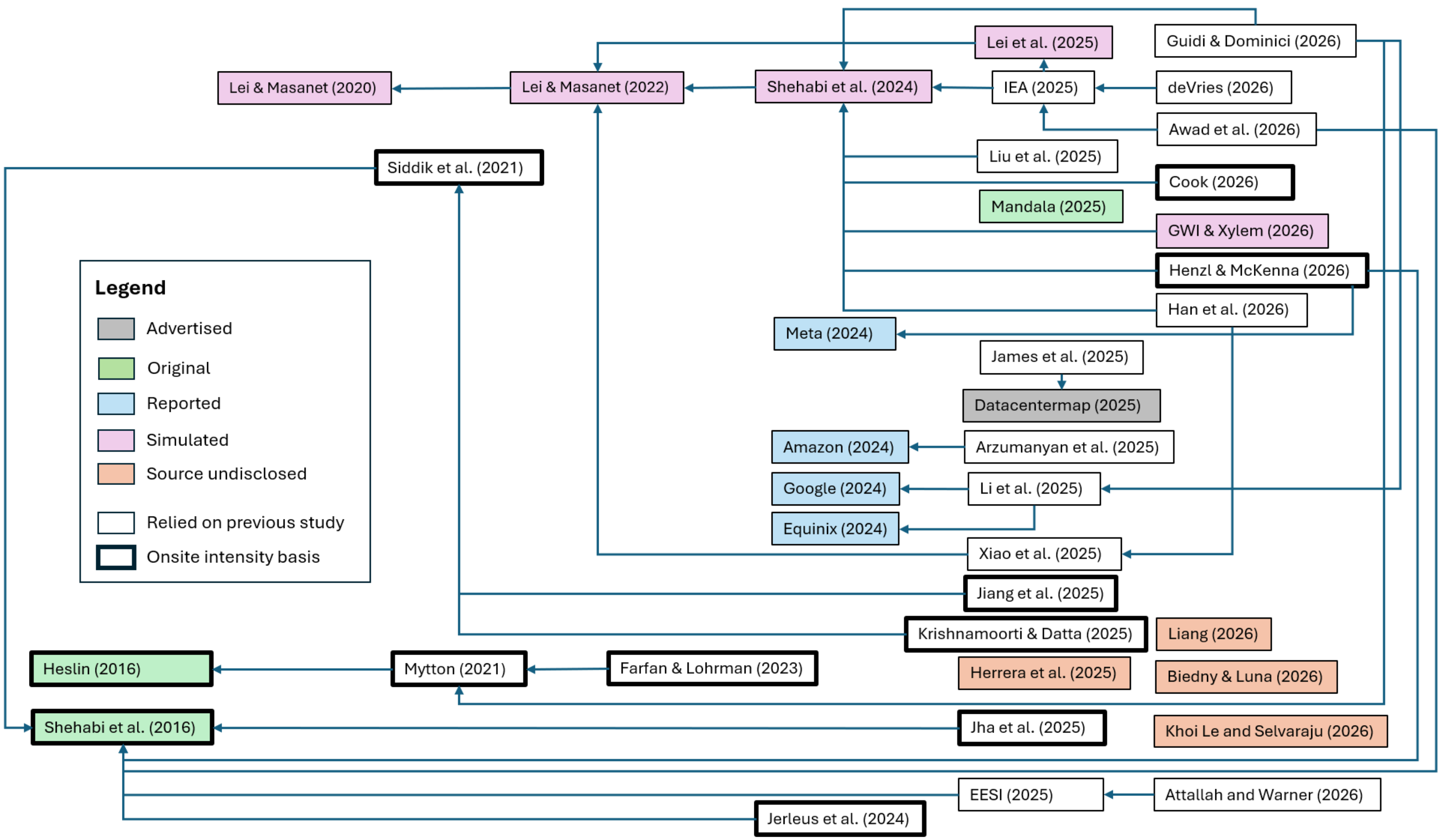

**Figure S23:** Origins of WUE values in 27 studies that estimated direct water use.

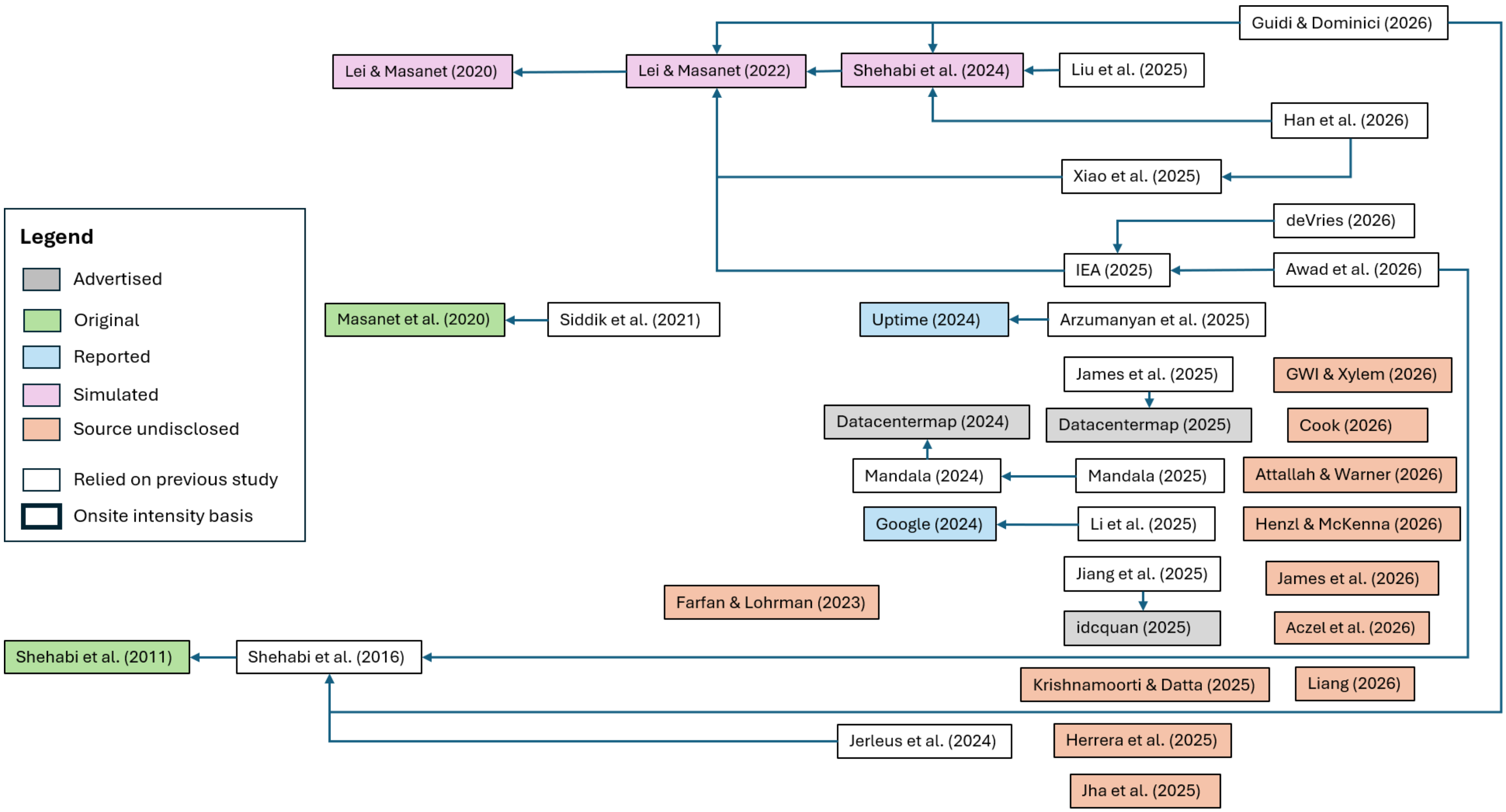

**Figure S24:** Origins of PUE values in 27 studies that estimated indirect water use

# S8: Temporal analysis of chosen EWIF factors

In addition to extracting the EWIF values utilized by each study (when available), this review also assessed the ages of selected EWIF values. This was necessary to assess the temporal representativeness of EWIF values compared to study time horizons, but it also shines a light on the need for updated EWIF research in general. As shown in Table S5, many of the EWIF values used in the assessed studies originated long before the time horizons to which they were applied in some studies.

The min and max values in Table S5 indicate the time difference between the data years associated with a study's selected EWIF value(s) and the years within a study's analysis time horizon. Only the studies that included Scope 2 water use estimates are included in the table.

A data year is defined as the year associated with the data used to derive a given EWIF value, which is always earlier than, or equal to, the year in which that EWIF value was published. A given EWIF value can often have a range of data years based on the use of multiple sources in its derivation, such as the EWIF values in Macknick et al (55). Also, because different data years can be associated with different power technology EWIF values, even within the same source study. Lastly, many sources of EWIF values are based on previously-published EWIF values; therefore, the estimates in Table S5 trace back to the original sources to establish data ages. Table S5 includes common classes of power generation technologies to visualize how time differences may vary by power technology.

Estimated time differences are shown as ranges for each study. The minimum value is calculated by subtracting the latest potential data year associated with an EWIF from the earliest year in a water estimation study's time horizon. For example, if the latest potential data year associated with an EWIF is 2020 and the earliest year in a study's time horizon is 2025, the minimum time difference is calculated as 5 years.

The maximum value is calculated by subtracting the earliest potential data year associated with an EWIF from the latest year in a water estimation study's time horizon. Therefore, if the earliest data year associated with an EWIF is 2005 and the latest year in a study's time horizon is 2030, the maximum time difference is calculated as 25 years. If an EWIF only has a single data year, its latest and earliest data years are equal.

Finally, all data and details supporting Table S5 are provided in the Supplementary Information Excel file.

| Study | Fossil/nuclear | | Hydroelectric | | Geothermal | | Biopower | | Solar CSP | | Solar PV | | Wind | | National grids | |
|---|---|---|---|---|---|---|---|---|---|---|---|---|---|---|---|---|
| | Min | Max | Min | Max | Min | Max | Min | Max | Min | Max | Min | Max | Min | Max | Min | Max |
| 2024-3 | 1 | 5 | 10 | 23 | | | | | | | 13 | 27 | 20 | 21 | | |
| 2016-1 | 11 | 25 | 5 | 21 | | | | | | | | | | | | |
| 2025-1 | 27 | 35 | 21 | 31 | | | | | | | | | | | | |
| 2025-9 | -3 | 10 | 6 | 28 | | | | | | | 9 | 32 | 16 | 26 | | |
| 2026-1 | 3 | 7 | 12 | 25 | | | | | | | 15 | 29 | 22 | 23 | | |
| 2026-4 | 2 | 10 | 2 | 10 | 2 | 10 | | | | | 2 | 10 | 2 | 10 | | |
| 2023-1 | 11 | 37 | 19 | 38 | 11 | 28 | 14 | 22 | 11 | 37 | 10 | 19 | 16 | 26 | | |
| 2024-2 | 11 | 39 | 18 | 18 | 11 | 42 | 0 | 0 | 11 | 28 | 12 | 26 | 19 | 20 | | |
| 2026-9 | 15 | 43 | 23 | 34 | 15 | 46 | 18 | 18 | 15 | 33 | 14 | 30 | 20 | 24 | | |
| 2021-2 | 0 | 0 | 5 | 18 | | | | | 8 | 25 | 16 | 29 | 14 | 22 | | |
| 2025-8 | 13 | 37 | 10 | 20 | 13 | 28 | 16 | 22 | 13 | 37 | | | | | | |
| 2025-4 | 7 | 41 | | | | | | | | | 16 | 42 | | | | |
| 2026-3 | 2 | 7 | | | | | 2 | 7 | | | | | | | | |
| 2025-11 | 1 | 6 | | | | | 1 | 6 | | | | | | | | |
| 2025-7 | 2 | 8 | 9 | 15 | | | 9 | 15 | 2 | 8 | 9 | 15 | 9 | 15 | | |
| 2025-2 | | | | | | | | | | | | | | | 11 | 11 |
| 2025-2 | | | | | | | | | | | | | | | 6 | 6 |
| 2025-5 | 16 | 26 | | | | | | | | | | | | | | |
| 2025-6 | 7 | 56 | 7 | 39 | 8 | 39 | 7 | 32 | 7 | 56 | 7 | 56 | 7 | 40 | | |
| 2026-6 | 13 | 47 | 20 | 26 | | | 2 | 8 | | | 14 | 34 | 21 | 28 | | |
| 2026-12 | 0 | 6 | 8 | 14 | | | 8 | 14 | | | 8 | 14 | 8 | 14 | | |

| Years of difference | | | | | | | | | | | | |
|---|---|---|---|---|---|---|---|---|---|---|---|---|
| | 5 | 10 | 15 | 20 | 25 | 30 | 35 | 40 | 45 | 50 | 55 | 60 |

**Table S5**: Potential minimum and maximum time discrepancies between a given study's time horizon and the data origin age (years) of the EWIFs used in that study by generation type. Color codes correspond to the scale below the table.

# Supplementary Information References